\documentclass{aa}
\usepackage{array}
\usepackage{color}
\usepackage{graphics}
\usepackage{hyperref}
\usepackage[varg]{txfonts}

\begin{document}

\title{Multiply-imaged time-varying sources behind galaxy clusters Comparing FRBs to QSOs, SNe, and GRBs}
\titlerunning{Multiply-imaged, time-varying sources behind galaxy clusters}
\author{Jenny Wagner\inst{1}  \and Jori Liesenborgs\inst{2} \and David Eichler\inst{3}}
\institute{Universit\"at Heidelberg, Zentrum f\"ur Astronomie, Astron. Rechen-Institut, M\"onchhofstr. 12--14, 69120 Heidelberg, Germany\\
\email{j.wagner@uni-heidelberg.de}
\and
Expertisecentrum voor Digitale Media, Universiteit Hasselt, Wetenschapspark 2, B-3590, Diepenbeek, Belgium \\
\email{jori.liesenborgs@uhasselt.be}
\and
Department of Physics, Ben-Gurion University, P.O.Box 653, Beer-Sheva 84105 Israel \\
\email{eichler@bgumail.bgu.ac.il}}
\date{Received XX; accepted XX}

\abstract{
With upcoming (continuum) surveys of high-resolution radio telescopes, detection rates of fast radio bursts (FRBs) might approach $10^5$ per sky per day by future extremely large observatories, such as the possible extension of the Square Kilometer Array (SKA) to a phase 2 array. Depending on the redshift distribution of FRBs and using the repeating FRB121102 as a model, we calculate a detection rate of multiply-imaged FRBs with their multiply-imaged hosts caused by the distribution of galaxy-cluster scale gravitational lenses of the order of $10^{-4}$ per square degree per year for a minimum total flux of the host of 10 $\mu$Jy at 1.4 GHz for SKA phase 2. 
Our comparison of estimated detection rates for quasars, supernovae, gamma ray bursts, and FRBs shows that multiple images of FRBs could be more numerous than those of gamma ray bursts and supernovae and as numerous as multiple images of quasars. 
%If the host galaxy of the multiply-imaged FRB cannot be resolved, gravitational lens models can estimate the redshift of the FRB. 
Time delays between the multiple images of an FRB break degeneracies in model-based and model-independent lens reconstructions as other time-varying sources do, yet without a microlensing bias as FRBs are more point-like and have shorter duration times. We estimate the relative imprecision of FRB time-delay measurements to be $10^{-10}$ for time delays on the order of 100 days for galaxy-cluster scale lenses, yielding more precise (local) lens properties than time delays from the other time-varying sources. Using the lens modelling software Grale, we show the increase in accuracy and precision of the reconstructed scaled surface mass density map of a simulated cluster-scale lens when adding time delays for one set of multiple images to the set of observational constraints.
%\com{As the understanding of FRBs is still limited, our results heavily rely on FRB-model assumptions. Yet, we hope our analyses contribute to the design specifications for sky surveys as SKA phase 2, and show that, compared to other transient sources, FRBs could be more suitable to increase the precision and accuracy of lens mass reconstructions to map the dark matter distribution in gravitational lenses to a greater detail.}
%We also derive a general equation to determine fluctuations of the column density of electrons in the cluster gas between the positions of the multiple images of the FRB from the differences in the dispersion measures of these images. Hence, we obtain differences in the lensing potential and in the density fluctuations of the cluster gas between the same positions, which may help to calibrate the gravitational potential derived from X-ray and lensing measurements against each other. 
%\textit{Context.} 
%\\
%\textit{Aims.} 
%\\
%\textit{Methods.} 
%\\
%\textit{Results.} 
%\\
%\textit{Conclusions.} 
}

\maketitle

%%%%%%%%%%%%%%%%%
\section{Introduction}
\label{sec:introduction}

Fast radio bursts (FRBs) are transients of a few milliseconds duration. Serendipitously, 60 of such bursts have been observed in the frequency ranges of about 0.4 to 6 GHz, \cite{bib:Petroff}\footnote{\url{http://www.frbcat.org}}. Due to the low angular resolution of continuum sky radio surveys, FRBs have only been observed in the time domain so far with best localisation precisions on the order of arcminutes. Yet, one FRB, FRB121102, showed repeating bursts, so that its angular position could be determined to milliarcsecond precision and the redshift of its host galaxy as $z = 0.193$, \cite{bib:Bassa, bib:Chatterjee, bib:Marcote, bib:Tendulkar}.

Given upcoming radio surveys with increased resolution, sky coverage and field of view, like those made possible by the SKA in phase 1 and 2, the origin and nature of FRBs can be investigated in more detail, for instance as detailed in \cite{bib:Dai, bib:Katz, bib:Waxman}, if the estimated detection rates of FRBs are matched by the observations of future sky surveys, as, for instance, discussed in \cite{bib:Bannister,bib:Fialkov1,bib:Jarvis,bib:Johnston, bib:Norris}.

Assuming that some of these FRBs are located behind masses that act as gravitational lenses, two applications can be envisioned: First, as point-like transients, they are highly suitable to break degeneracies in the reconstructions of gravitational lensing mass distributions in the same way as quasars do, see e.g.\ \cite{bib:Dahle1, bib:Dahle2}. Second, given a precise mass model of the lens, they could be used like quasars multiply-imaged by galaxies to determine the Hubble constant, as detailed, e.g. in \cite{bib:Suyu} and \cite{bib:Li}. 
Alternatively, multiply-imaged supernovae could be used for these tasks: \cite{bib:Goobar} detected a multiply-imaged supernova Type Ia behind a foreground galaxy and \cite{bib:Kelly} found a supernova Type II behind the Hubble Frontier Field cluster MACS1149. Yet, observational data are still rare. FRBs could greatly improve the statistics and the reconstruction precision, if they can be detected more frequently than quasars and supernovae. 

\cite{bib:Meneghetti2} show state-of-the-art confidence bounds of the reconstructed scaled mass density distribution and of the lensing magnification without time delay information for various reconstruction approaches for two elaborately simulated galaxy clusters. \cite{bib:Liesenborgs2} treat the case including time-delay constraints from a set of three multiple images for the free-form reconstruction algorithm Grale for a less detailed galaxy cluster simulation. They find that including highly precise time-delay measurements, reconstructions of the cluster lensing potential or the cluster mass become more accurate and precise.
FRBs have the advantage of very short pulse widths compared to the time delay between the multiple images of the FRB caused by the gravitational lensing effect (on the order of much more than days). 
While previous works mostly focused on constraining dark matter on small scales, \cite{bib:Cordes, bib:Dai, bib:Eichler, bib:Li2, bib:Munoz, bib:Zheng}, we address the usage of FRBs for gravitational lens reconstructions on galaxy-cluster scale. 
Without a serendipitous detection analogous to \cite{bib:Kelly} and \cite{bib:Goobar}, the practical application may lie in the far future. However, despite the high uncertainties and model assumptions, our investigations show that multiply-imaged FRBs are at least as worthy of consideration as probes for mapping the lensing mass distribution as multiply-imaged supernovae are.

In Section~\ref{sec:occurrence}, we derive the constraints to detect multiply-imaged FRBs and estimate the detection rates for SKA phase 2 using the properties of the repeating FRB121102 as a model for the FRBs to be observed. In addition, we investigate the rate of spurious detections and discuss the advantages of galaxy clusters as lenses compared to galaxy-scale lenses. We show the contribution of multiply-imaged FRBs to the reconstructions of the galaxy-cluster scale lensing mass and lensing potential in Section~\ref{sec:usage}. 
Then, in Section~\ref{sec:comparison}, we compare the detection rates and the usage of FRBs with the detection rates and the usage of quasars, gamma-ray bursts (GRBs) or supernovae in the applications that are introduced in Section~\ref{sec:usage}.
At last, in Section~\ref{sec:redshifts}, we investigate how well the distance, and thus the redshift, of a multiply-imaged FRB
can be constrained by gravitational lens models. With a summary of our results, we conclude in Section~\ref{sec:conclusion}.

%The repetition of the burst can be attributed to a single source or several independent sources coming from a region in the sky smaller than the angular resolution can resolve. 

%A second explanation is  to a gravitational lensing effect Among them can be due to a spatially unresolved gravitational lensing effect  because state-of-the-art continuum survey telescopes only have angular resolutions of several arcminutes or even degrees \textbf{(references)}. Hence, a doublet of images that is caused by gravitational lensing of a source FGRB or FRB is detected as a repeated signal because the separation between the two images is on the order of some ten arcseconds for a galaxy-cluster scale gravitational lens \textbf{(references)}. 
%\\[0.2cm]
%Fast radio bursts (FRBs)
%\textbf{Related work / possibly useful references}
%\begin{itemize}
%\item \textbf{Cosmology with SKA} \\ \url{http://cdsads.u-strasbg.fr/abs/2015aska.confE..18J}
%\item \textbf{Radio Continuum Surveys with SKA Pathfinders, technical specifications} \\
%\url{http://cdsads.u-strasbg.fr/abs/2013PASA...30...20N}
%\item \textbf{MACHO test with multiply-imaged FRBs} \\
%\url{http://cdsads.u-strasbg.fr/abs/2016PhRvL.117i1301M}
%\item \textbf{Constraining $H_0$ from galaxy-scale lensing of FRBs} \\ \url{http://cdsads.u-strasbg.fr/abs/2017arXiv170806357L}
%\item \textbf{Nature of FRBs and motion of their sources} \\ \url{http://cdsads.u-strasbg.fr/abs/2017ApJ...847...19D}
%\end{itemize}

%%%%%%%%%%%%%%%%%
\section{Properties of multiply-imaged FRBs}
\label{sec:occurrence}

%Given typical angular and temporal resolutions of current and future radio telescopes, we first summarise the specifications of those surveys that are capable of detecting multiply-imaged FRBs in Section~\ref{sec:detectability}.
In Section~\ref{sec:FRBs}, we summarise the findings of previous works and observations on the estimated occurrence rate of FRBs by already existing and by future observatories. The resulting distribution of FRBs in redshift is subsequently employed in Section~\ref{sec:lensed_FRBs} to determine the number of multiply-imaged FRBs behind galaxy clusters that we expect to observe per square degree per year. To arrive at this result, we first determine the probability to detect multiply-imaged FRBs from galaxy-cluster scale lenses. For this, we select a distribution function of galaxy clusters and their cross sections in which multiple images can occur from existing approaches. 

In Section~\ref{sec:props_of_interest}, we estimate expected time delays and image separations caused by typical galaxy clusters as gravitational lenses. They support the possible way to detect multiply-imaged FRBs that is outlined in Section~\ref{sec:props_of_interest}. Galaxy clusters contain smaller-scale structures that can cause additional lensing effects. We show how they can be distinguished and disentangled from the deflections and magnifications caused by the cluster as a whole in Section~\ref{sec:microlensing}. Finally, we show how to distinguish multiple images of FRBs from spurious detections due to spatially close, independent FRBs coming from different sources in Section~\ref{sec:confusion}.

\subsection{Occurrence rate of FRBs}
\label{sec:FRBs}

The rate of observable FRBs strongly depends on their astrophysical origins and their redshift distribution, \cite{bib:Fialkov1, bib:Fialkov2, bib:Law, bib:Li2}. As \cite{bib:Fialkov1} state, the detectability of FRBs depends on three factors: the spectral shape of the individual FRBs, the FRB luminosity function, and the population of host galaxies.  

Concerning the spectral shape distribution, the observed FRBs show a bell-shaped spectral profile usually fitted by a Gaussian in the observed band. Central frequencies are in the range of 2.8 to 3.2~GHz with peak fluxes between 130 and 3340~mJy, and the full-width-half-maximum of the Gaussian is in the range of 290 to 690~MHz, \cite{bib:Fialkov1}. If the small ensemble of FRBs observed so far turns out to be representative, FRBs may not be detectable outside the narrow band of frequencies around the peak frequency. Consequently, redshifted FRB signals will require surveys at lower frequencies to detect them. As also shown in \cite{bib:Fialkov1}, future surveys as SKA phase 1 and 2 will be able to distinguish between a Gaussian-shaped and a flat spectral profile from the observed number counts. Focusing on the Gaussian-shaped profiles, SKA phase 1 may not yield many detections due to an insufficient sensitivity below 0.95 GHz, as shown in \cite{bib:Fialkov1}. As the number of observable, multiply-imaged FRBs will be even lower, only SKA phase 2 could potentially detect them.
%In \cite{bib:Law}, a joint analysis from several observations of FRB 121102 indicates that its entire spectrum cannot be modelled with a single spectral index. 

The second factor, the FRB luminosity function, affects the detection rate of FRBs and their usage. If FRBs have the same intrinsic brightness and can be calibrated as standard candles, distances to the observed FRBs can be measured. In \cite{bib:Fialkov1} two luminosity functions are considered: the first one assumes that all FRBs have the same intrinsic luminosity $L_\mathrm{int}$ and the Gaussian-like spectral profile 
\begin{equation}
S_\mathrm{obs} (\nu_\mathrm{obs}) = S_0 \exp \left( - \dfrac{(\nu_\mathrm{obs} - \overline{\nu})^2}{2 \sigma_\nu^2} \right) \;.
\end{equation}
The free parameters for the peak amplitude $S_0~=~0.9017~\mbox{Jy}$, the peak frequency $\overline{\nu}~=~2.986~\mbox{GHz}$, and the width $\sigma_\nu~=~0.1984~\mbox{GHz}$ obtained from the mean values of the repeating burst FRB121102 at $z~=~0.19$, \cite{bib:Law}, are chosen as normalisation for $L_\mathrm{int}$. The second luminosity function assumes a Schechter-profile and yields higher detection rates of FRBs than the previous function, such that the detection rates of FRBs in SKA phase 1 could already distinguish between the two luminosity functions. As we consider lower bounds on the detectability, we focus on the first, constant luminosity function.

\begin{figure*}[t]
\centering
  \includegraphics[width=0.30\textwidth]{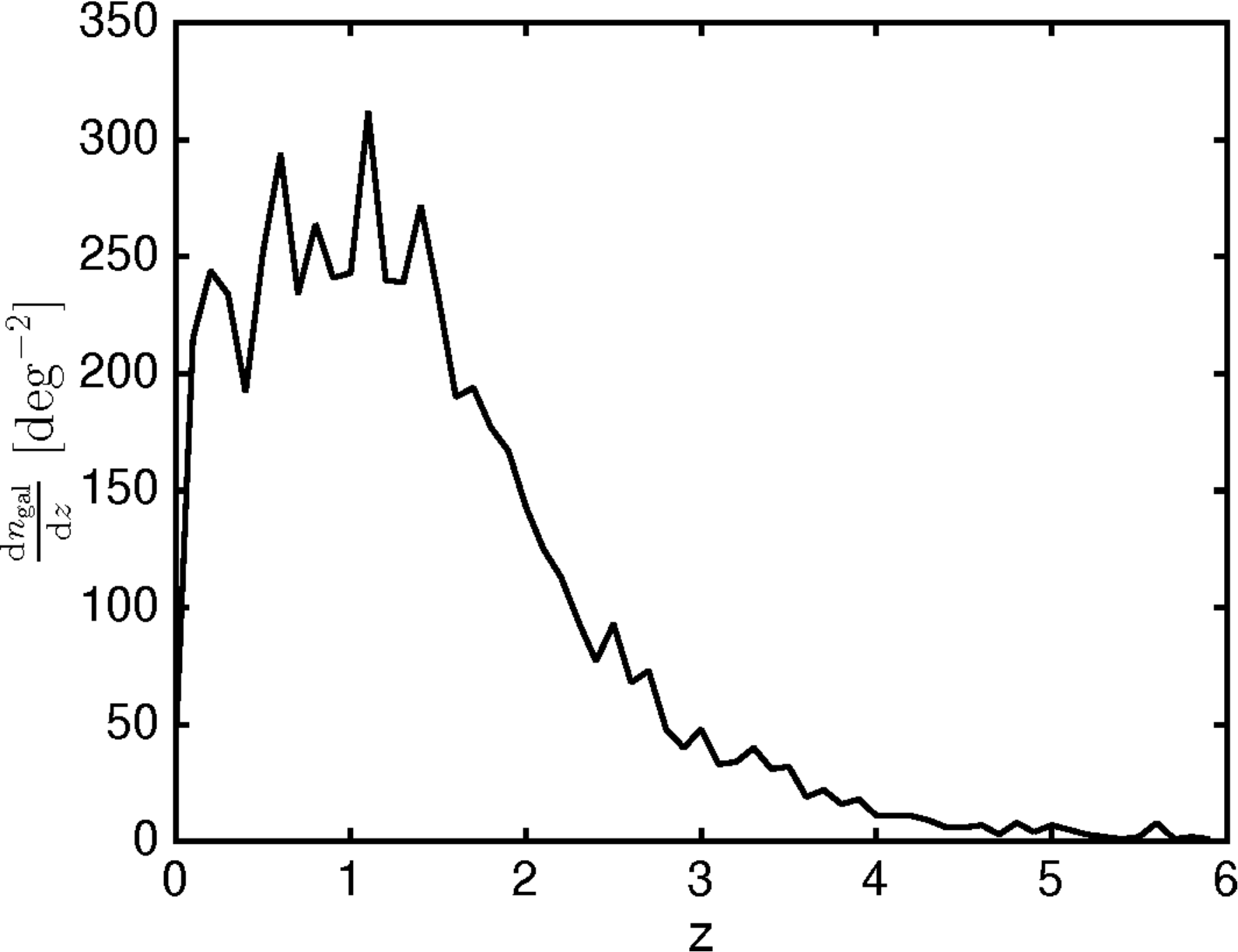} \hspace{0.01\textwidth}
  \includegraphics[width=0.30\textwidth]{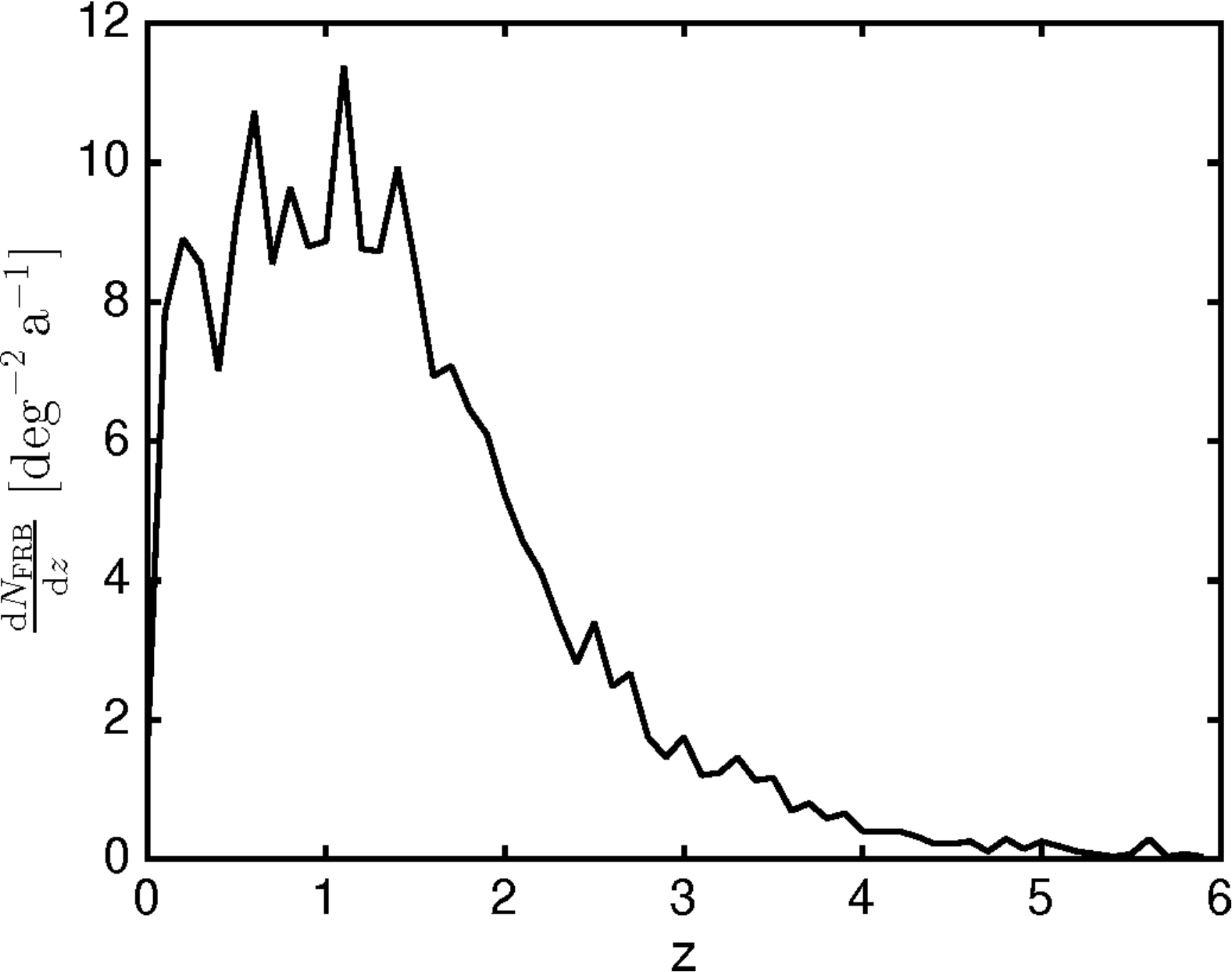} \hspace{0.01\textwidth}
  \includegraphics[width=0.30\textwidth]{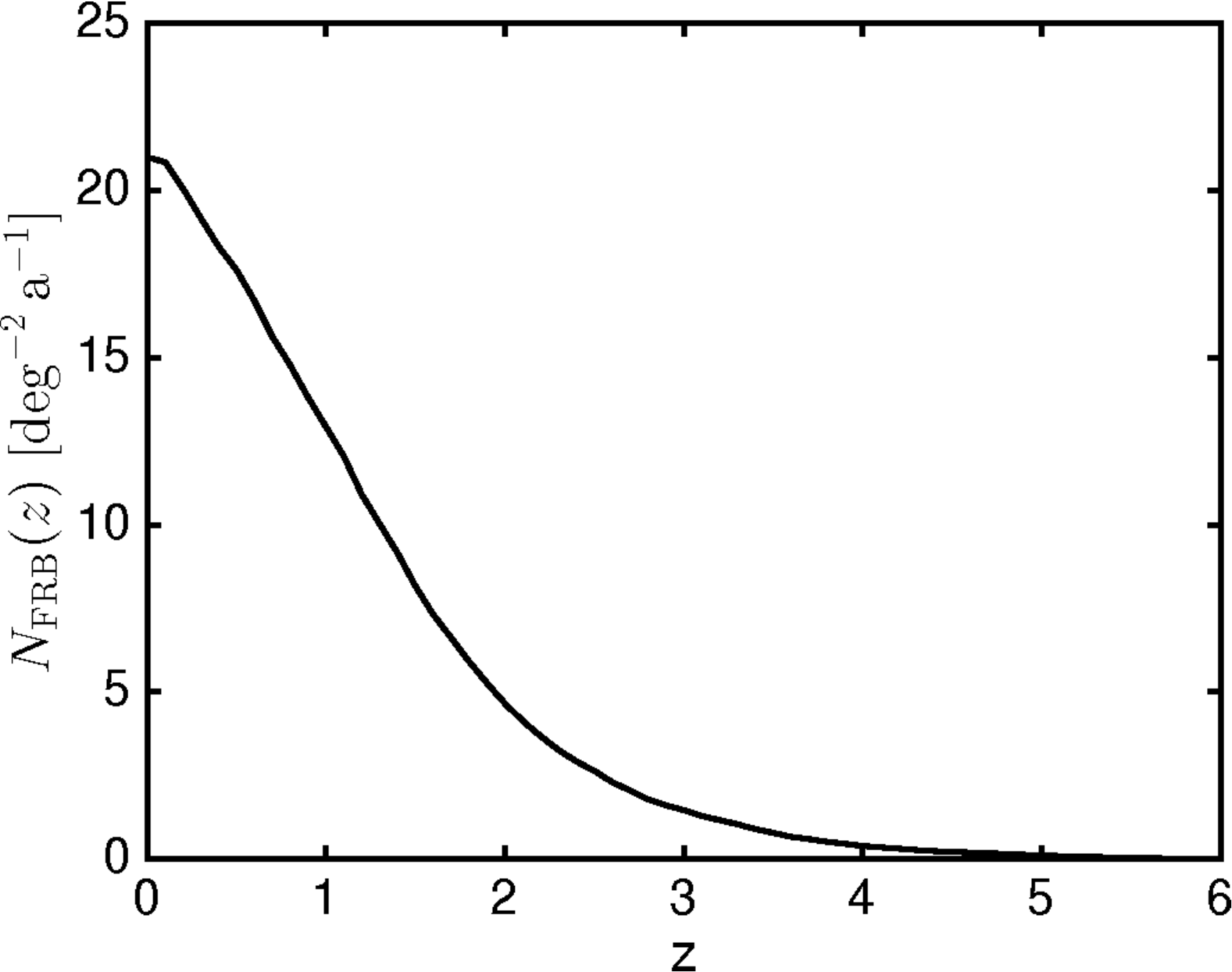}
   \caption{Left: distribution of observable FRB host galaxies with total flux above 10$\mu$Jy at 1.4~GHz per square degree, per redshift, $\mathrm{d} n_\mathrm{gal}(z)~/~\mathrm{d} z$, binned in redshift bins of 0.1 width taken from the SKA Simulated Skies. Centre: distribution of potentially observable FRBs from these host galaxies per square degree per year, assuming an internal emission rate of $R_\mathrm{int}=10^{-4}$ per galaxy per day. Right: integrated number of observable FRBs per square degree per year from $z$ to $z_\mathrm{lim}=6$.}
\label{fig:n_FRB}
\end{figure*}

For the mass range of host galaxies, \cite{bib:Fialkov1} assume $M_\mathrm{lo}~=~M_{*}/\sqrt{10}$ and $M_\mathrm{hi}~=~M_{*} \sqrt{10}$ as lower and upper bounds, respectively. $M_{*}~=~5.3~\times~10^{7}~M_\odot$ is the stellar mass of the host of FRB121102. For comparison, they also investigate $M_{*}~=~5.3~\times~10^{9}~M_\odot$, as a more common mass of a host galaxy in the low-redshift universe. 

The number of host galaxies per comoving volume is then determined in a simulation using the Sheth-Tormen halo-mass function, \cite{bib:Sheth_Tormen}, to model the distribution of dark matter halos. They are populated with stars by abundance matching and a star formation efficiency which varies with halo mass and redshift, \cite{bib:Behroozi}. 

Under the assumptions that the FRB-model described above is valid and that we observe all FRBs of 1~ms duration with peak fluxes above 1~Jy out to z=1, \cite{bib:Fialkov1} arrive at internal rates of observable FRBs per galaxy per day of $R_\mathrm{int} = 2~\times~10^{-5}$ and $R_\mathrm{int} = 3~\times~10^{-4}$ for the low and high $M_*$, respectively. 

Since we are not only interested in the rate of FRBs per square degree per year {observable by SKA phase 2, but also in resolving their host galaxies, we model the rate of observable FRBs with their hosts, $N_\mathrm{FRB}$, as
\begin{equation}
N_\mathrm{FRB} = \int \limits_0^{z_\mathrm{lim}} \mathrm{d}z \dfrac{\mathrm{d} n_\mathrm{FRB}(z)}{\mathrm{d} z} = \int \limits_0^{z_\mathrm{lim}} \mathrm{d}z \dfrac{\mathrm{d} n_\mathrm{gal}(z)}{\mathrm{d} z} R_\mathrm{int} \;,
\label{eq:n_FRB}
\end{equation}
in which $z_\mathrm{lim}$ is the survey-limiting redshift, $\mathrm{d}n_\mathrm{FRB}(z)/\mathrm{d}z$ the rate of observable FRBs with their hosts per square degree per year per redshift interval $\mathrm{d}z$, and $\mathrm{d}n_\mathrm{gal}(z)/\mathrm{d}z$ is the number density of observable FRB-emitting galaxies per square degree per redshift interval $\mathrm{d}z$. Estimates for the latter have been simulated in \cite{bib:Wilman}. This simulation takes into account the observed large-scale clustering effects of galaxies. It uses Poisson sampling of sources from luminosity functions that are based on observed luminosities and which are extrapolated over the redshift range considered in the simulation. The database of all results is publicly available\footnote{\url{http://s-cubed.physics.ox.ac.uk}} and contains about 320 million simulated radio sources of five different types in a $20 \times 20~\mbox{deg}^2$ area from redshift 0 to 20. It is designed to match the technical specifications envisioned for SKA phase 1 and 2 because the $20 \times 20~\mbox{deg}^2$ sky coverage is approximately the largest, instantaneous field of view and the lower flux density limit of 10~nJy at 1.4~GHz is on the order of the expected detection limit for a 100-h observation with the nominal SKA sensitivity.

We extract all star-forming galaxies within the central square degree of the simulation with $z \in \left[ 0, 6 \right]$ and total fluxes above 10~$\mu$Jy at 1.4~GHz to ensure that most parts of the host galaxies can be resolved above the detection limit. Figure~\ref{fig:n_FRB} (left) shows the distribution of potential FRB host galaxies per redshift, $\mathrm{d} n_\mathrm{gal}(z)~/~\mathrm{d} z$, binned in redshift bins of 0.1 width. Figure~\ref{fig:n_FRB} (right) shows the rate of observable FRBs according to Equation~\eqref{eq:n_FRB}. For this plot, we assume $R_\mathrm{int} = 10^{-4}$ per galaxy per day, which lies between the two values used in \cite{bib:Fialkov1}. The resulting 21 FRBs per square degree per year amount to approximately 2300 FRBs per sky per day, which is on the order of the estimates of \cite{bib:Fialkov1}. Hence, if \cite{bib:Keane} and the more recent estimates by \cite{bib:Law} (and references therein) of approximately 2000 FRBs per sky per day above 1~Jy~ms are the total amount of FRBs to be observable, SKA phase 2 is likely to detect the vast majority of them.

\subsection{Probability of a multiply-imaged FRB}
\label{sec:lensed_FRBs}

According to \cite{bib:SEF}, the probability of a source at redshift $z_\mathrm{s}$ being lensed by foreground masses is given by
\begin{equation}
p(z_\mathrm{s}) = \dfrac{c}{H_0 D_\mathrm{s}^2} \int \limits_{0}^{z_\mathrm{s}} \mathrm{d}z_\mathrm{l} \dfrac{\mathrm{d}D_\mathrm{l}}{\mathrm{d}z_\mathrm{l}} \int \limits_{0}^{\infty}  \sigma (M,z_\mathrm{l},z_\mathrm{s}) n(M,z_\mathrm{l}) \mathrm{d} M \;,
\label{eq:lensing_prob}
\end{equation}
in which $D_\mathrm{s}$ and $D_\mathrm{l}$ are the angular diameter distances from the observer to the source at redshift $z_\mathrm{s}$ and a lens at redshift $z_\mathrm{l}$, respectively. $\sigma(M,z_\mathrm{l},z_\mathrm{s})$ is the cross section of the lens depending on its mass $M$ and redshift, and $n(M,z_\mathrm{l}) \mathrm{d}M$ is the number density of lenses in the mass range $M$ and $M + \delta M$ at redshift $z_\mathrm{l}$.

To obtain an analytic estimate of the lensing cross section $\sigma(M,z_\mathrm{l},z_\mathrm{s})$, we approximate the mass profile of the galaxy clusters as an axisymmetric one with the Einstein radius $r_\mathrm{E}(M,z_\mathrm{l},z_\mathrm{s})$ that depends on the mass of the lens, its redshift and the redshift of the source. The mass inside a tangential critical curve of an axisymmetric lens with an Einstein radius of $r_\mathrm{E}$ can be approximated by
\begin{equation}
M(r_\mathrm{E}) \approx \pi r_\mathrm{E}^2 D_\mathrm{l}^2 \Sigma_\mathrm{cr} \approx 4.4 \times 10^{14} M_\odot \left( \dfrac{r_\mathrm{E}}{30''} \right)^2 \left( \dfrac{D_\mathrm{l} D_\mathrm{s}}{D_\mathrm{ls} \mbox{Gpc}} \right)
\end{equation}
with $\Sigma_\mathrm{cr}~=~c^2~/~(4\pi G)~D_\mathrm{s}~/~(D_\mathrm{l}~D_\mathrm{ls})$, as derived in \cite{bib:Hoekstra}.

The cross section is given by the back-projection of the area enclosed by a disk of radius $D_\mathrm{l} r_\mathrm{E}(M,z_\mathrm{l},z_\mathrm{s})$ to the source plane\footnote{For singular isothermal spheres as mass profiles, Equation~\eqref{eq:cross_section} exactly describes the region in the source plane from which multiple images arise. For other mass profiles, Equation~\eqref{eq:cross_section} is an estimate.}: 
\begin{align}
\sigma (M, z_\mathrm{l}, z_\mathrm{s}) &= \pi \left(D_\mathrm{l} r_\mathrm{E}(M,z_\mathrm{l},z_\mathrm{s}) \dfrac{D_\mathrm{s}}{D_\mathrm{l}} \right)^2 \nonumber \\
 &= 7.38 \times 10^{-5} \, \mbox{Mpc}^2 \left( \dfrac{r_\mathrm{E}}{1''} \right)^2 \left( \dfrac{D_\mathrm{s}}{1 \mbox{Gpc}} \right)^2\;.
\label{eq:cross_section}
\end{align}

We determine the Einstein radius for a given mass, assuming a Navarro-Frenk-White profile (NFW), \cite{bib:NFW}, for the galaxy-cluster scale lenses. This profile generates three multiple images of a source inside the caustic: a maximum close to the lens centre at radius $r_1$, a saddle point on the same side of the lens centre as the maximum at radius $r_2$, and a minimum on the opposite side of the lens centre at radius $r_3$, i.e.\@ $r_1 < r_2 < r_3$. The maximum and the saddle point image straddle the radial critical curve, the minimum lies outside the tangential critical curve.

\begin{figure*}[t]
\centering
  \includegraphics[width=0.32\textwidth]{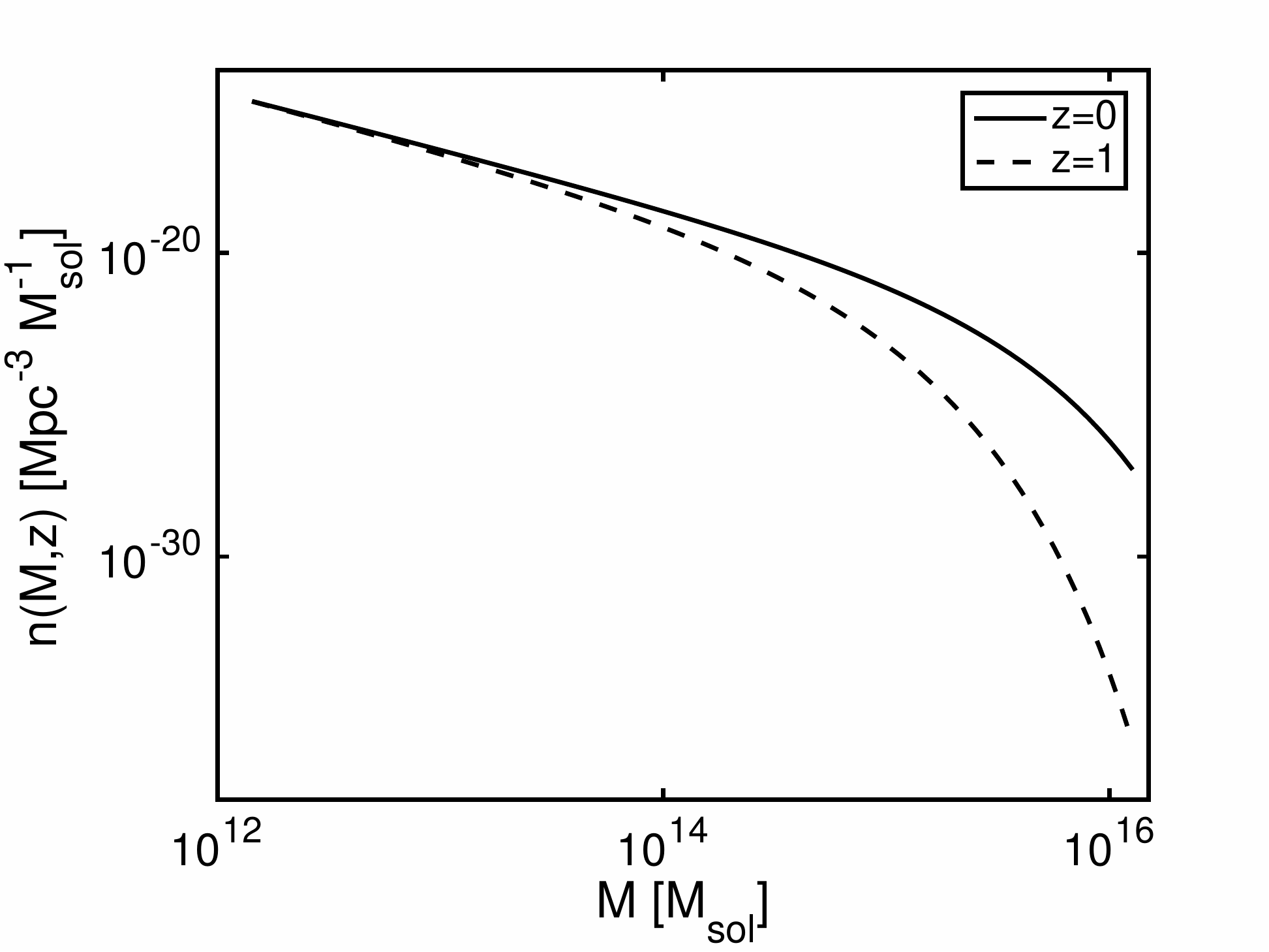} \hspace{0.01\textwidth}
  \includegraphics[width=0.32\textwidth]{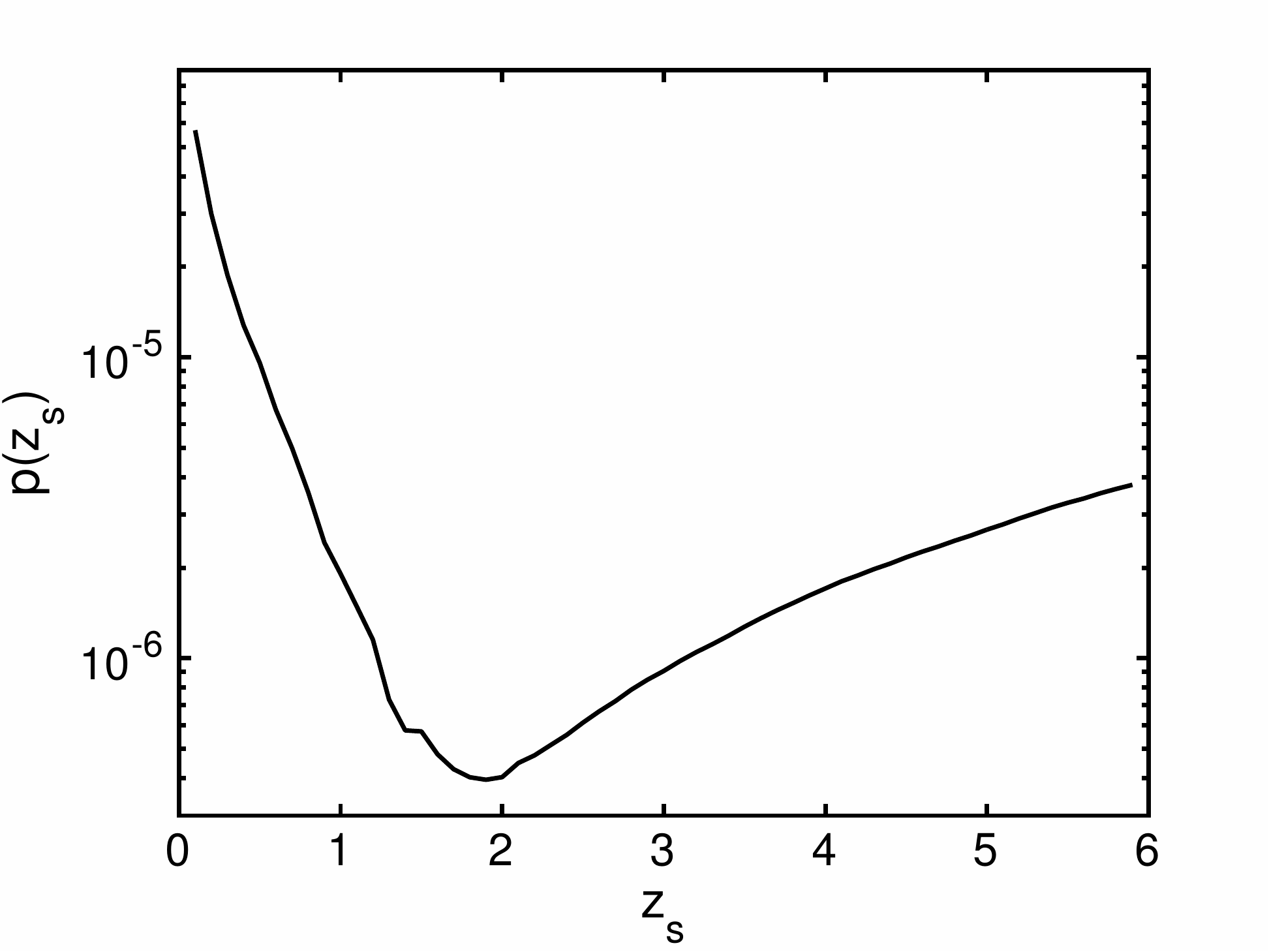} \hspace{0.01\textwidth}
  \includegraphics[width=0.32\textwidth]{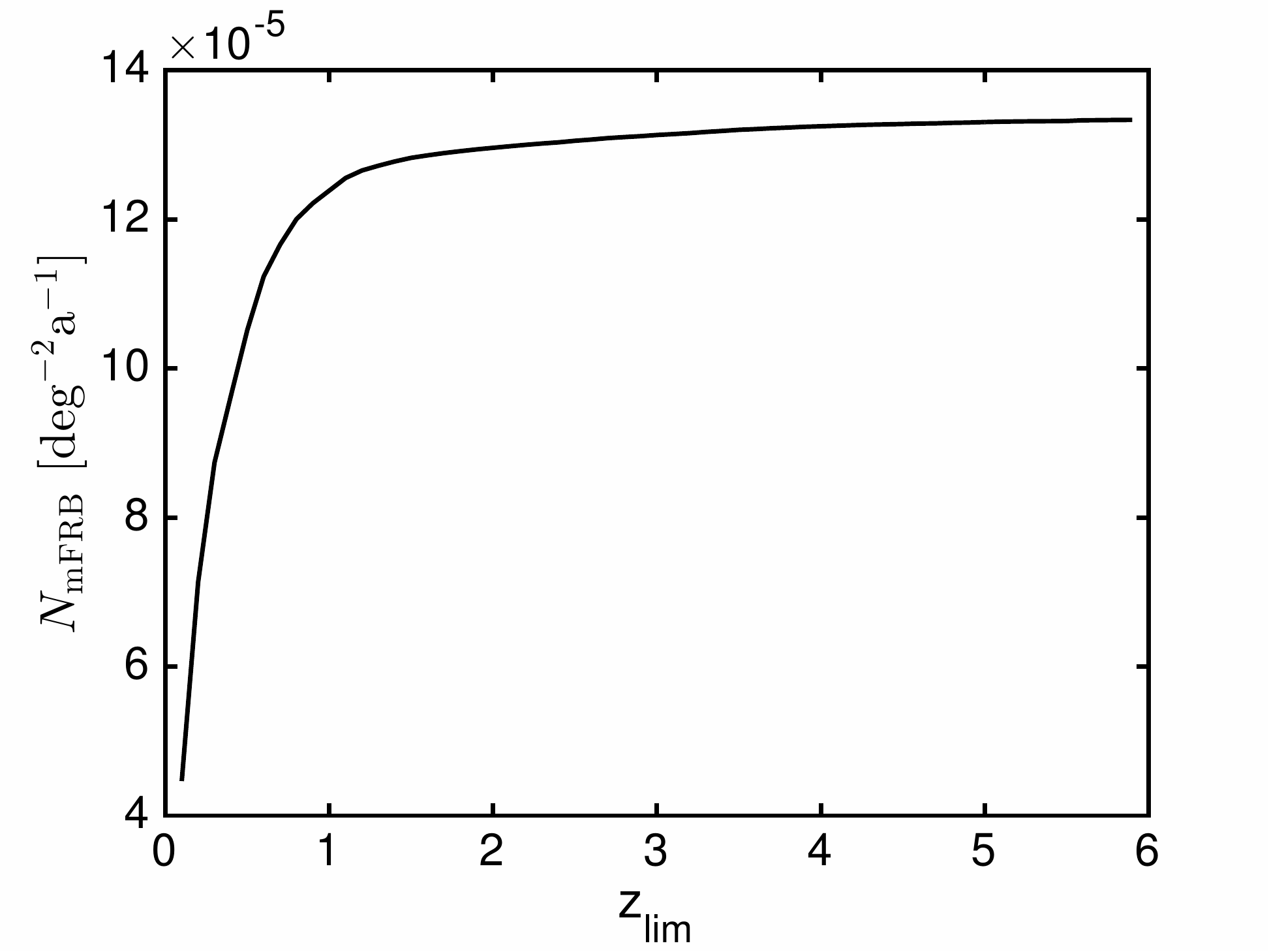}
   \caption{Left: halo-mass function according to \cite{bib:Sheth_Tormen} for $z=0$ (solid line) and $z=1$ (dashed line). Centre: lensing probability according to Equation~\eqref{eq:lensing_prob} for $z_\mathrm{s}$ from 0.1 to 6.0 and NFW halos with a minimum Einstein radius of 5 arcseconds. Right: integrated number of observable multiply-imaged FRBs per square degree per year from $z=0$ out to $z_\mathrm{lim}$ from 0.1 to 6.0 according to Equation~\eqref{eq:multiply_imaged_FRBs}.}
\label{fig:lens_probs}
\end{figure*}

The profiles we consider should generate multiple-image configurations with at least 10~arcseconds distance between the minimum and the saddle point image, so that a continuum sky survey of SKA phase 2 can resolve both images. Due to the time delays between the images, the minimum will arrive first, followed by the saddle point image, so that the detection of these two images leaves enough time (on the order of several weeks to months, as estimated in Section~\ref{sec:props_of_interest}) to observe the region of interest at subarcsecond resolution to locate the maximum. 

We approximate the angular distance between the minimum and the saddle point by the diameter of the Einstein ring when we determine the lensing probability in Equation~\eqref{eq:lensing_prob}.
In galaxy clusters, the separations of multiple images that lie on opposite sides of the lens centre usually range from several arcseconds to more than ten arcseconds, giving an estimate for $r_\mathrm{E}$. For instance, in CL0024, image separations as large as 30~arcseconds were measured, \cite{bib:Broadhurst}. 

For galaxy-scale lenses of masses less than $10^{13}~M_\odot$, a singular isothermal sphere (SIS) is usually employed as lens model. The combination of the SIS for the galaxies and the NFW profiles for masses of more than $10^{13}~M_\odot$ has been shown to match observations well, \cite{bib:Li_Ostriker}. 

To determine the number density of lenses $n(M,z_\mathrm{l})$ in a mass interval d$M$, various analytical and numerical approaches have been established, \cite{bib:Jenkins, bib:Linke, bib:Press_Schechter, bib:Sheth_Tormen, bib:Tinker}. Although the halo mass function by \cite{bib:Press_Schechter} is often employed, it is known to overestimate the number density of galaxy-size halos and underestimate the number density of galaxy-cluster scale halos. A comparison of the most common approaches can be found in \cite{bib:Linke}. They develop an analytical approach to obtain the number density of galaxy clusters taking into account correlated, non-linear structure growth and show that the standard halo-mass functions based on numerical simulations come close to their results. 
Therefore, we resort to the halo-mass function of \cite{bib:Sheth_Tormen} which is a commonly used standard halo-mass function. Figure~\ref{fig:lens_probs} (left) shows the halo-mass function for $z=0$ (solid line) and $z=1$ (dashed line), as obtained by the software described in \cite{bib:Murray}. 

We restrict our calculations to the dark matter halos and neglect the luminous part of the galaxy clusters. In this way, we obtain a lower bound on the probability of multiply-imaged FRBs because \cite{bib:Hilbert} showed that the probability of lensing is only increased when luminous matter is taken into account.

% singular isothermal sphere, for which the lensing cross section is given by
%\begin{equation}
%\sigma(r_\mu) = 8 \pi^2 G M \dfrac{D_\mathrm{l} D_\mathrm{s}}{D_\mathrm{ls}} \left( \dfrac{r_\mu - 1}{r_\mu + 1} \right)^2 \;,
%\end{equation}
%in which $r_\mu$ denotes the brightness ratio of the two multiple images, as derived in \cite{bib:Li_Ostriker}. 

Combining Equations~\eqref{eq:n_FRB} and \eqref{eq:lensing_prob}, the overall expected number of multiply-imaged FRBs per square degree per year is
\begin{equation}
N_\mathrm{mFRB} = \int \limits_0^{z_\mathrm{lim}} \mathrm{d}z p(z) \dfrac{\mathrm{d} n_\mathrm{FRB}(z)}{\mathrm{d} z} \;.
\label{eq:multiply_imaged_FRBs}
\end{equation}
The lensing probabilities for sources at $z_\mathrm{s}$ from 0.1 to 6.0 by NFW profiles with an Einstein radius of at least 5 arcseconds are shown in Figure~\ref{fig:lens_probs} (centre). The estimated detection rate of multiply-imaged FRBs out to $z_\mathrm{lim}$ from 0.1 to 6.0 based on the FRB model as outlined in Section~\ref{sec:FRBs} is shown in Figure~\ref{fig:lens_probs} (right). 

As a rough estimate for the number of repeating FRBs, we could assume that at least 1/60 of all FRBs are repeating, given that FRB121102 is the only repeating FRB out of a sample of 60 detected so far. Consequently, the rate of multiply-imaged repeating FRBs could be estimated to be at least 1/60 $N_\mathrm{mFRB}$ because some of the remaining 59 FRBs could turn into repeating FRBs. %\com{yet, as investigated in \cite{bib:Shannon}, it currently seems unlikely because the observed spectral properties of the non-repeating FRBs rather hint at two different kinds of FRBs, one that is repeating and the other one that is non-repeating}. 

%%%%%%
\subsection{Estimated time delays and image separations}
\label{sec:props_of_interest}

For cluster-scale lenses, we determine estimates for the expected time delays from an NFW-profile, assuming that the background source is located close to the caustic curve, so that we observe the maximum and the saddle point images close to the radial critical curve. Then, as derived in detail in Appendix~\ref{app:NFW_config}, we can approximate the expected time delay between these two images close to the radial critical curve
\begin{align}
\tau_\mathrm{f} =& \dfrac{r_\mathrm{s}^2 D_\mathrm{s}}{c D_\mathrm{l} D_\mathrm{ls}} \left(1+z_\mathrm{l} \right) \left[ \left(r_1 - r_2\right)\left(r_\mathrm{r} + y_\mathrm{r} \right) + \psi(r_2) - \psi(r_1)  \right] \nonumber \\
 =& 1.19 \times 10^9~\mbox{d} \dfrac{D_\mathrm{s}(1+z_\mathrm{l})}{D_\mathrm{ls}} \left[ \dfrac{\pi^2}{4.20 \times 10^{11}}  \left( \dfrac{\theta_1 - \theta_2}{1''}\right)\left( \dfrac{\theta_\mathrm{r} + \beta_\mathrm{r}}{1''}\right)\right. \nonumber \\
& \left. + \dfrac{r_\mathrm{s}}{D_\mathrm{l}} \left( \psi\left( \theta_2\dfrac{D_\mathrm{l}}{r_\mathrm{s}}\right) - \psi\left( \theta_1\dfrac{D_\mathrm{l}}{r_\mathrm{s}} \right)  \right) \right]\;,
\label{eq:tau_fold}
\end{align}
in which the deflection potential $\psi(r)$ is given by
\begin{equation}
\psi(r) = 4 \kappa_\mathrm{s} \left[ \dfrac12 \ln^2 \left( \dfrac{r}{2} \right) + 2 \, \mbox{atanh}^2 \sqrt{\dfrac{1-r}{1+r}}\right] \;,
\label{eq:NFW_potential}
\end{equation}
and the parameters of the NFW-profile are the scaling convergence $\kappa_\mathrm{s}$ and the scale radius $r_\mathrm{s}$.  The radial distance $r_i~=~\theta_i~D_\mathrm{l}/r_\mathrm{s}$ is the observed angular distance $\theta_i$ from the lens centre to the position of the image $i$ multiplied by the ratio between the angular diameter distance from the observer to the lens plane and the scale radius. The radius of the radial critical curve is denoted by $r_\mathrm{r}$ and $y_\mathrm{r}~=~\beta_\mathrm{r}~D_\mathrm{l}/r_\mathrm{s}$ is its back-projection into the source plane. 
Expected time delays are thus on the order of 100 days with image separations of a few arcseconds for typical values of $\kappa_s = 0.2$ and $r_\mathrm{s} = 500~\mbox{kpc}$, $z_\mathrm{l}=0.5$, and $z_\mathrm{s}=1.5$ (see e.g. \cite{bib:Merten2} for typical values of $\kappa_\mathrm{s}, r_\mathrm{s}, z_\mathrm{l}$, and $z_\mathrm{s}$). 

Analogously, we derive an estimate for the time delay between the saddle point image and the minimum image, which is located close to the tangential critical curve with radius $r_\mathrm{t}$
\begin{align}
\tau_\mathrm{o} &= \dfrac{r_\mathrm{s}^2 D_\mathrm{s}}{c D_\mathrm{l} D_\mathrm{ls}} \left(1+z_\mathrm{l} \right) \left[ \dfrac{\left(r_2 - r_3\right)\left(r_\mathrm{t} + r_\mathrm{r} \right)}{2} + \psi(r_3) - \psi(r_2)  \right] \;.
\label{eq:tau_opp}
\end{align}
The transformation from $r$ to observable angular distances $\theta$ is the same as for Equation~\eqref{eq:tau_fold} and the detailed calculation can also be found in Appendix~\ref{app:NFW_config}. Hence, typical time delays $\tau_\mathrm{o}$ are on the order of years for image separations in the range of 10 arcseconds for the same values of $\kappa_\mathrm{s}$ and $r_\mathrm{s}$ as used to estimate $\tau_\mathrm{f}$. Measured values for quasars, as shown in Section~\ref{sec:comparison_quasars}, are in agreement with estimates for $\tau_\mathrm{f}$ and $\tau_\mathrm{o}$.

%%%%%%
\subsection{Lensing effects due to smaller-scale structures}
\label{sec:microlensing}

For galaxy-scale lenses, we can approximate the lensing mass profile by a SIS and estimate the time delay $\tau$ between the two images to be
\begin{align}
\tau &= - \dfrac{4 \pi}{c} \left( \dfrac{\sigma_\mathrm{v}}{c}\right)^2 D_\mathrm{l} (1+z_\mathrm{l}) \left( r_1 - r_2 \right) \nonumber \\
 &= - 8.0594 \, \mbox{d} \, \left( 1+z_\mathrm{l} \right) \left( \dfrac{\sigma_\mathrm{v}}{100 \, \mbox{km}/\mbox{s}}\right)^2 \left( \dfrac{D_\mathrm{l}}{1 \mbox{Gpc}}\right) \left( \dfrac{r_1 - r_2}{1''}\right) \;,
\end{align}
in which $\sigma_\mathrm{v}$ is the velocity dispersion along the line of sight, $r_1$ is the observed angular distance from the centre of the lens to the minimum image located outside of the critical curve and $r_2$ the observed angular distance from the centre to the saddle point image inside the critical curve. 
Measured time delays for quasars, as listed, for instance, in \cite{bib:Oguri}, are usually in the range of days to weeks with image separations on the order of an arcsecond.  

As discussed in detail in \cite{bib:Zheng}, the lensing effects due to intervening massive compact halo objects (MACHOs) will not lead to observable separations between the two images and lead to time delays on the order of 0.1 ms for impact parameters that are approximately equal to the Einstein radius of the MACHO. 

Hence, image separations and typical time delays for the different scales of lensing can be distinguished from each other. This is corroborated by observations from quasar lensing. For instance, \cite{bib:Oguri} compares observed time delays due to galaxy-scale lenses that are isolated in the field and due to those that are embedded in a cluster environment. 

%%%%%%
\subsection{Discrimination from repeating and independent FRBs}
\label{sec:confusion}

If the luminosity function of the FRBs has a steep slope at the faint end, the main contribution of observed FRBs could come from member galaxies of clusters, as investigated in \cite{bib:Fialkov2}. Hence, precise redshift measurements for the FRBs are necessary in order to distinguish the multiply-imaged FRBs from background host galaxies from the FRBs of cluster member galaxies.

If the spatial resolution of the telescope is not sufficient to resolve the angular positions of two or three adjacent multiple images in a fold or cusp configuration\footnote{See \cite{bib:SEF, bib:Wagner1} for a detailed characterisation of both configurations.}, a repeating, temporally resolved signal will be observed, twice for a fold, thrice for a cusp. 

Repeatingly detected FRBs due to spatially unresolved multiple images have to be distinguished from repeating signals coming from the same source that is not multiply-imaged, as, for instance, the repeating FRB121102. These two cases can be distinguished by observing the signal from the multiple image on the opposite side of the lens centre after time delays on the order of days to weeks for galaxy-scale lenses and on the order of months to years for cluster-scale lenses. 

As FRBs are events of very short duration, resolving the host galaxy before or after the occurrence of an FRB is easier than in the presence of a bright quasar. Thus, a spectroscopic follow-up analysis of the multiply-imaged host galaxy at an increased resolution (also in different bands) corroborates the lensing hypothesis for the observed repeating FRB-signal. 
The same spectroscopic test of the host distinguishes multiple images of an FRB from a single source from several FRBs from different hosts. 

%%%%%%%%%%%%%%%%%
\section{Usage of multiply-imaged FRBs}
\label{sec:usage}

In Section~\ref{sec:mi_information}, we discuss the information multiply-imaged FRBs yield about the galaxy cluster at the positions of the multiple images without assuming a specific lens model. Subsequently, in Section~\ref{sec:mb_information}, we show benefits of multiply-imaged FRBs for global reconstructions of the cluster mass or potential. 
%Finally, in Section~\ref{sec:n_e}, we derive an equation to estimate differences in the cluster gas density between the positions of the multiple images from their dispersion measures. 
While the main focus lies on the galaxy-cluster scale lenses, we also comment on the applicability to galaxy-scale lenses where it is possible. While Section~\ref{sec:occurrence} was concerned about the detection rates of multiply-imaged FRBs and therefore strongly relied on our choice of the underlying FRB-model, the findings of Section~\ref{sec:usage} are based on general principles of the gravitational lensing formalism and can be applied to any transient multiply-imaged source.

%%%%
\subsection{Model-independent, local information gain}
\label{sec:mi_information}

The time delay $\tau_{ij}$ between two multiple images $i$ and $j$ of an FRB located at angular position $\vec{y}$ in the source plane is given by
\begin{equation}
\tau_{ij} = \dfrac{D_\mathrm{l} D_\mathrm{s}}{c D_\mathrm{ls}} (1+z_\mathrm{l}) \Delta \phi \left(\vec{y},\vec{x}_i, \vec{x}_j \right) \equiv \dfrac{D}{c}  \Delta \phi \;,
\label{eq:mi_time_delay}
\end{equation}
with measured angular positions $\vec{x}_i$ of the two images in the lens plane and the lensing potential $\phi(\boldsymbol{y},\boldsymbol{x})$, as defined in Appendix~\ref{app:NFW_config}. Using Equation~\eqref{eq:mi_time_delay}, we can infer the difference of the lensing potentials between the two positions, $ \Delta \phi \left(\vec{y},\vec{x}_i, \vec{x}_j \right)$ without inserting a specific lens model, for a fixed cosmological model that determines the distances $D_\mathrm{l}, D_\mathrm{s}, D_\mathrm{ls}$ along the line of sight. Vice versa, if $\Delta \phi$ and its dependence on the cosmological model is known, we can determine properties of this cosmological model, e.g. the Hubble constant, $H_0$, or the matter density parameter $\Omega_{m0}$ in a Friedmann model, as for instance detailed in \cite{bib:Grillo}. 

Neither knowing the underlying cosmology nor $\Delta \phi$, Equation~\eqref{eq:mi_time_delay} is subject to a degeneracy: we can scale $D$ by an arbitrary factor $\lambda \in \mathbb{R}$. At the same time $\Delta \phi$ is subject to the mass sheet degeneracy (see \cite{bib:Liesenborgs2, bib:Schneider} and references therein) stating that $\Delta \phi$ can also be scaled by an arbitrary factor $\tilde{\lambda} \in \mathbb{R}$, while leaving the observables invariant. For $\lambda = \tilde{\lambda}^{-1}$, the measured time delay remains invariant. 
 
This degeneracy is broken, if the distances $D_\mathrm{l}$ and $D_\mathrm{s}$ can be measured. Then, $\Delta \phi$ can be determined from Equation~\eqref{eq:mi_time_delay}, which becomes feasible, if FRBs turn out to be standardisable candles. Observing a multiply-imaged FRB and an FRB from a cluster member galaxy, the positions of the lens and the source along the line of sight with respect to the observer can be fixed. This is true for any combination of observed standardisable candles, for instance for a multiply-imaged supernova coinciding with an FRB from the image plane. However, the latter case seems less frequent (see Section~\ref{sec:comparison_supernovae}) and it remains to be shown that FRBs can be standardised.

As shown in \cite{bib:Wagner1}, we can approximate Equation~\eqref{eq:mi_time_delay} for two images that straddle a critical curve by
\begin{equation}
\tau_\approx = \dfrac{D}{12 c} (\delta x)^3 \phi_{222}^{(0)} \;,
\label{eq:tau_approx}
\end{equation}
with $\delta x = \delta \theta D_\mathrm{l}/r_\mathrm{s}$ being the rescaled measured angular separation between the two images. The third order derivative in $x_2$-direction $\phi_{222}^{(0)}$ is determined at the position $\vec{x}_0 = 1/2(\vec{x}_i + \vec{x}_j)$ in the coordinate system in which the multiple images are stretched along the $x_2$-direction. Figure~\ref{fig:tau_approx} shows the ratio of $\tau_\approx$ to $\tau$ for the two multiple images at the radial critical curve of an NFW profile, i.e.\@ $\delta x = r_1 - r_2$, with typical values of $\kappa_\mathrm{s}$ for $r_\mathrm{s} = 0.5$~Mpc.

Instead of $\Delta \phi$, i.e. an information between two points in the image plane, we can determine $\phi_{222}^{(0)}$ at a single point from the measured time delays and image separation by Equation~\eqref{eq:tau_approx}. This may be more convenient for some global lens reconstruction algorithms, as discussed in Section~\ref{sec:mb_information}. 

\begin{figure}[t]
\centering
  \includegraphics[width=0.45\textwidth]{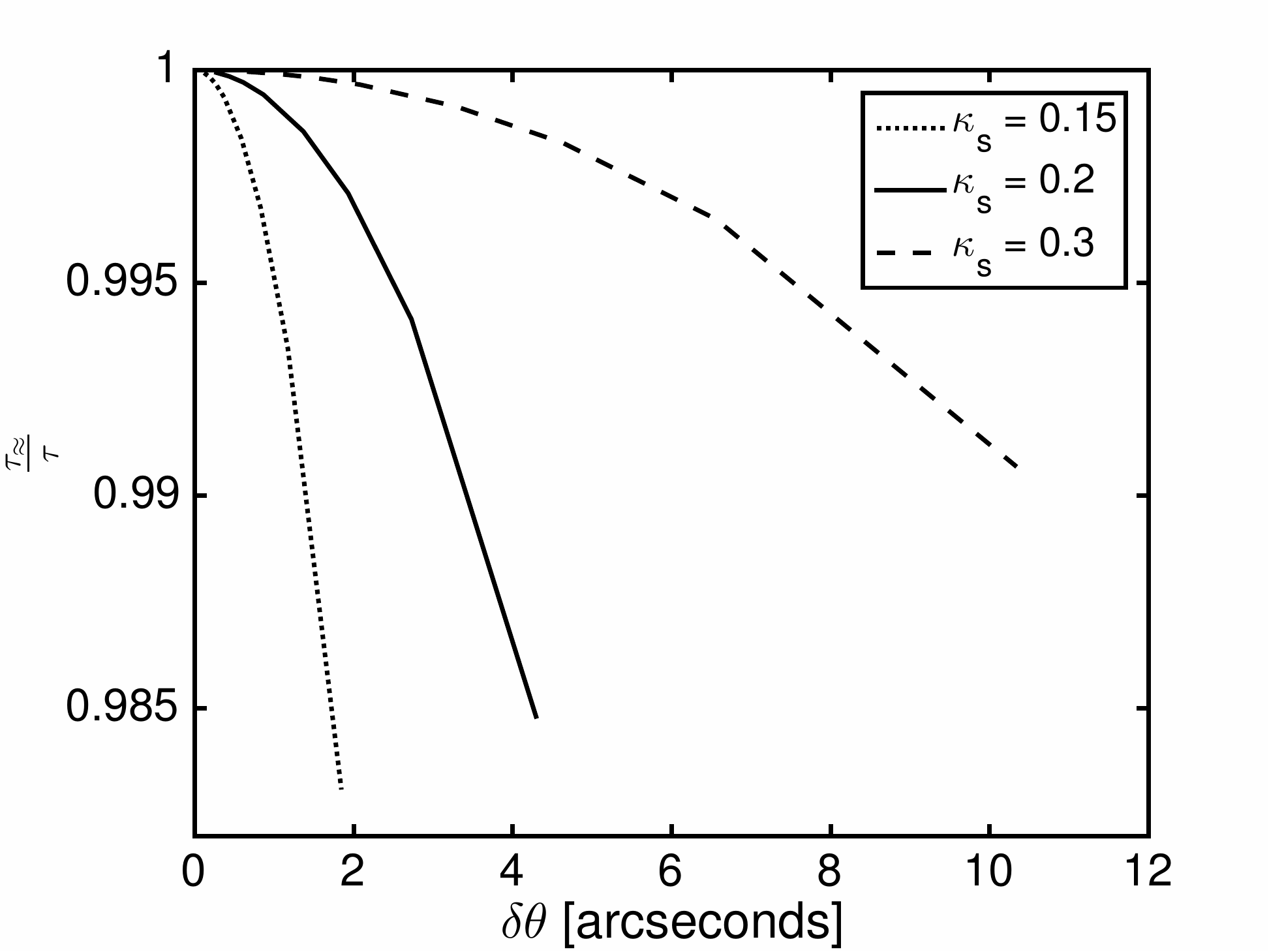}
   \caption{Accuracy of the approximation of the time delay: ratio between $\tau_\approx$, Equation~\eqref{eq:tau_approx}, and $\tau$, Equation~\eqref{eq:mi_time_delay}, for the two images straddling the radial critical curve in NFW profiles with $r_\mathrm{s}=0.5$~Mpc for $\kappa_\mathrm{s}=0.15$ (dotted curve), $\kappa_\mathrm{s}=0.2$ (solid curve), and $\kappa_\mathrm{s}=0.3$ (dashed curve) versus the observed angular separation of the two images $\delta \theta = (r_1 - r_2) r_\mathrm{s}/D_\mathrm{l}$.}
\label{fig:tau_approx}
\end{figure}

All results obtained in this section can be applied to multiple images of any time-varying source and generated by a galaxy-scale and cluster-scale lenses.
For both scales, the precision to determine $\Delta \phi$ or $\phi_{222}^{(0)}$ as determined by Equations~\eqref{eq:mi_time_delay} and \eqref{eq:tau_approx} is dominated by the uncertainties in the distances $D$ and for Equation~\eqref{eq:tau_approx} also by the angular separation $\delta x$. Assuming an FRB as source with uncertainties of 1~ms for time delays of 100~days, the relative uncertainty is on the order of $10^{-10}$. Assuming typical lens and source redshifts of $z_\mathrm{l}=0.5\pm0.01$ and $z_\mathrm{s}=1.5\pm0.01$, the relative uncertainty of $D$ -- only considering the imprecision on $z$ and neglecting the uncertainties in the cosmological parameters -- is of the order of 2-3\% and the relative uncertainty of $\delta x$ is usually just below 1\%, so that the time delay is the most precise quantity, which is not the case for the other time-varying sources (see Table~\ref{tab:summary} for a comparison).
%Comparing this estimate to the precision of time delays measured from multiply-imaged quasars (see Section~\ref{sec:comparison_quasars}), the gain is several orders of magnitude. 

\begin{figure}[t]
\centering
  \includegraphics[width=0.47\textwidth]{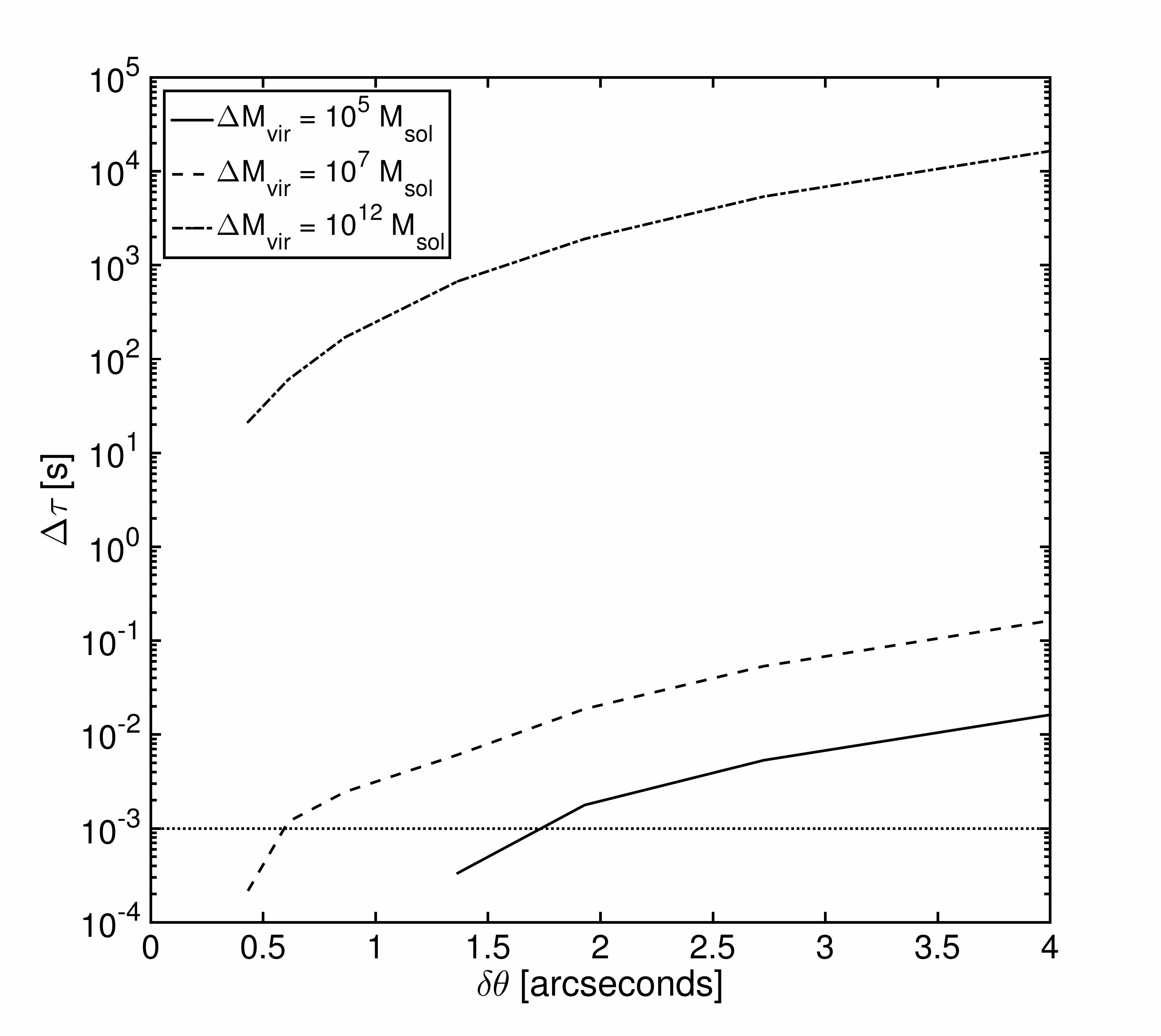}
   \caption{Difference between time delays of repeating multiply-imaged FRBs for increasing angular distances at the radial critical curve of the NFW profile of Section~\ref{sec:props_of_interest} when the virial mass of the NFW profile is increased by $10^5$ solar masses (solid line), by $10^{7}$ solar masses (dashed line), and by $10^{12}$ solar masses (dash-dotted line). The measurement precision of $\tau_{ij}$ for one measurement, approximately 1 ms, is shown as dotted line.}
\label{fig:substructure}
\end{figure}

For repeating bursts from the same source, we can measure the time delays between the multiple images several times. Significant deviations between the results indicate a change in the potential difference, $\delta(\Delta \phi)$. Given that the absorption properties of the intracluster gas and other extinction effects do not change 
%(see Section~\ref{sec:n_e} for a method to test this hypothesis) 
and the emission properties of the source remain the same, this difference could be attributed to small-scale structure, as detailed in Section~\ref{sec:microlensing}, that enters and leaves the light paths of the multiple images between the repetitions. 

To study the change in time delays caused by small-scale structure moving in and out the galaxy cluster potential, $\phi$, we consider a change in the virial mass $\Delta M_\mathrm{vir}$ of the NFW profile of Section~\ref{sec:props_of_interest} from $10^5$ to $10^{12}$ solar masses. Figure~\ref{fig:substructure} shows the change in the time delay, $\Delta \tau$, between the repeating multiply-imaged FRBs close to the radial critical curve for different angular distances between the multiple images. The figure shows that depending on the distance between the images, a change in the virial mass of $10^{5} M_\odot$ between the repetitions of the FRBs can be detected. Determining the change in the angular distances for increasing changes in mass, we find them to be on the order of $10^{-9}$, $10^{-8}$, and $10^{-3}$ arcseconds. Thus, they are too small to be observed. Altogether, we find that changes in the time delay due to a change in the cluster mass between FRB repetitions cannot be used to detect MACHOs of a few solar masses and repetition and observation times are way too short to track the motion of larger objects, as galaxies.

%%%%
\subsection{Information gain in global lens reconstructions}
\label{sec:mb_information}

\subsubsection{Time-delay constraints in different lens reconstruction algorithms}

Reconstructions of the lensing mass distribution from several sets of multiple images in a galaxy cluster are obtained by parametric methods as Lenstool, \cite{bib:Kneib, bib:Jullo}, the approach by \cite{bib:Grillo15}, non-parametric approaches on a grid as Grale, \cite{bib:Liesenborgs}, or model-free and mesh-free approaches as SaWLens, \cite{bib:Merten}. As time delays in galaxy clusters are rarely available, Lenstool does not offer to use time delays as constraints for the reconstructions of the lens.

In SaWLens, constraints from time delays are also not implemented yet. This mesh-free algorithm sets constraints to the lensing potential and its derivatives in form of a $\chi^2$-minimisation term at positions where observables are available. As Equation~\eqref{eq:mi_time_delay} cannot be assigned to a single point, it will require some extension of the algorithm to include this constraint. But it is easy to use Equation~\eqref{eq:tau_approx} for all time delays of multiple images $i$, $j$ that straddle a critical curve by adding
\begin{equation}
\chi^2_\tau = \sum \limits_{i,j, j>i} \dfrac{\left( \tfrac{12c \tau_{\approx,ij}}{D(1+z_\mathrm{l})(\delta x)^3 }- \phi_{222}^{(0)}\right)^2}{\sigma_\tau^2}
\end{equation}
to the already existing $\chi^2$-terms. 

\cite{bib:Grillo} includes time delays in the approach of \cite{bib:Grillo15} to determine the posterior probability density function of cosmological parameters. The Bayesian parameter estimation employs a Gaussian likelihood for a given $\chi_\tau^2$  
\begin{equation}
\chi^2_\tau = \sum \limits_{(i,j)} \left( \dfrac{\tau_{ij} - \tau_{ij}^{(m)}}{\sigma_{\tau_{ij}}} \right)^2 \;,
\label{eq:Jori_time_delays}
\end{equation}
in which the sum is taken over all image pairs $(i,j)$ with measured time delays $\tau_{ij}$. The measurement uncertainty for each $\tau_{ij}$ is denoted by $\sigma_{\tau_{ij}}$ and $\tau_{ij}^{(m)}$ is the model-predicted time delay between the images. The latter is determined by Equation~\eqref{eq:mi_time_delay} after inserting a cosmological model to determine $D$ and a lens model to determine $\Delta \phi$. A joint reconstruction from all multiple images of the source, as detailed in \cite{bib:Suyu4}, is employed to determine the source position that is inserted into $\Delta \phi$. Instead of constraining cosmological parameters with this approach, the time delays could be used to improve the reconstruction of the lensing potential as described in Section~\ref{sec:mi_information}.

\begin{table}[t]
 \caption{Grale models and their constraints (2nd to 4th column) used for the lens reconstructions as shown in Figures~\ref{fig:Jori_critical_curves}, \ref{fig:Jori_radial_profiles}, and \ref{fig:Jori_reconstruction} (from top left to bottom right).}
\label{tab:Jori_models}
\begin{center}
\begin{tabular}{ccccc}
\hline
\noalign{\smallskip}
 model & mass sheet & $\tau_{12}$ & $\tau_{13}$ & $\tau_{23}$ \\
\noalign{\smallskip}
\hline
\noalign{\smallskip}
1 & -- & -- & -- & -- \\
\noalign{\smallskip}
\hline
\noalign{\smallskip}
2 & \checkmark & -- & -- & -- \\
\noalign{\smallskip}
\hline
\noalign{\smallskip}
3 & -- & \checkmark & \checkmark & \checkmark \\
\noalign{\smallskip}
\hline
\noalign{\smallskip}
4 & \checkmark & \checkmark & \checkmark & \checkmark \\
\noalign{\smallskip}
\hline
\noalign{\smallskip}
5 & -- & \checkmark & -- & -- \\
\noalign{\smallskip}
\hline
\noalign{\smallskip}
6 & \checkmark & \checkmark & -- & -- \\
\noalign{\smallskip}
\hline
\noalign{\smallskip}
7 & -- & -- & -- & \checkmark \\
\noalign{\smallskip}
\hline
\noalign{\smallskip}
8 & \checkmark & -- & -- & \checkmark \\
\noalign{\smallskip}
\hline
\end{tabular}
\end{center}
\end{table}

\cite{bib:Liesenborgs3} and \cite{bib:Liesenborgs2} investigated how time delays improve the reconstruction of the lensing mass distribution. They implemented a term in the optimisation routine of Grale that minimises the difference between the time delay predicted by the lens model and the observed one as follows:
\begin{equation}
\chi_\tau^2 = \sum \limits_{i \in T} \sum \limits_{j \in T} \sum \limits_{k=1}^N \sum \limits_{h=1}^N \left( \dfrac{t(\boldsymbol{x}_i, \boldsymbol{y}_k) - t(\boldsymbol{x}_j, \boldsymbol{y}_h) - \tau_{ij} }{\tau_{ij}} \right)^2 \;,
\label{eq:Jori_time_delays}
\end{equation}
in which the first and second sum are taken over all pairs of images $(i,j)$ of a source for which a time-delay measurement, $\tau_{ij}$, is available. The last two sums are taken over all $N$ multiple images. As the source position is unknown, the back-projected image positions, $\boldsymbol{y}_k$ and $\boldsymbol{y}_h$, are employed. Then, $\chi_\tau^2$ searches for the minimum difference between the arrival times $t$ for all image pairs $(i,j)$ and the observed time delay $\tau_{ij}$. Due to the weighting of all time delays by the duration of the measured time delay, time delays of all lengths are treated equally. As $\boldsymbol{y}_k$ and $\boldsymbol{y}_h$ are not necessarily equal, a bias is introduced that may affect the reconstructed time delay. It will be subject of a more detailed investigation of the general form of $\chi_\tau^2$ and a comparison to the other approaches in a later work.

\subsubsection{Reconstructing a simulated cluster-scale lens with Grale including time delays}

For a simulated example of a generalised NFW profile, \cite{bib:Liesenborgs2} used time delays of one set of three multiple images to demonstrate that time delay information greatly alleviates the mass sheet degeneracy in their lens reconstruction. They compared lens models based on the constraints of positions of 26 multiple images from 8 sources with $z_\mathrm{s} \in \left[2.7, 3.4\right]$ with and without an additional mass-sheet basis function, with and without the three time-delay constraints between the three multiple images. We extend their analysis and investigate the reconstruction accuracy for the mass density distribution when only one time-delay constraint is available, with and without adding an additional mass sheet. Instead of 20 models, we average over 30 individual reconstructions from the genetic algorithm. Yet, a systematic investigation performed in \cite{bib:Wagner3} shows that the effect on the standard deviations is negligible, which can be corroborated comparing the plots in the top row of Figure~\ref{fig:Jori_radial_profiles} with the respective ones in \cite{bib:Liesenborgs2}. 

We summarise all Grale models with the different constraints in Table~\ref{tab:Jori_models}. Figures~\ref{fig:Jori_critical_curves}, \ref{fig:Jori_radial_profiles}, and \ref{fig:Jori_reconstruction} show the reconstruction results of these models starting with model~1 from top left and ending with model~8 at the bottom right. 

Comparing the results for the reconstructed critical curves between the different models, Figure~\ref{fig:Jori_critical_curves} shows that the inclusion of an additional mass sheet improves the reconstruction accuracy, has a smoothing effect on the critical curves, and leads to less complex and more accurate caustics. These results are in agreement with the findings of \cite{bib:Liesenborgs2} and occur because the Plummer basis functions alone can hardly account for an overall offset in the mass density distribution; a mass-sheet basis function alleviates this. Comparing the reconstructed critical curves for models~4 and 6, Figure~\ref{fig:Jori_critical_curves} shows that both are of comparable quality, while the reconstructed critical curves for model~8 are of minor quality with additional critical curves in the central part and larger deviations between the true and the reconstructed inner critical curve. These results can be explained by the fact that the multiple images with the time-delay constraint in model~8 lie very closely together and yield a more local constraint than the time-delay constraints in model~6 or model~4 do. For the same reason, the reconstruction of the inner critical curve in model~8 is accurate only in the vicinity of the two multiple images with the time-delay constraint. 

\begin{figure*}[p]
\centering
  \includegraphics[width=0.3\textwidth]{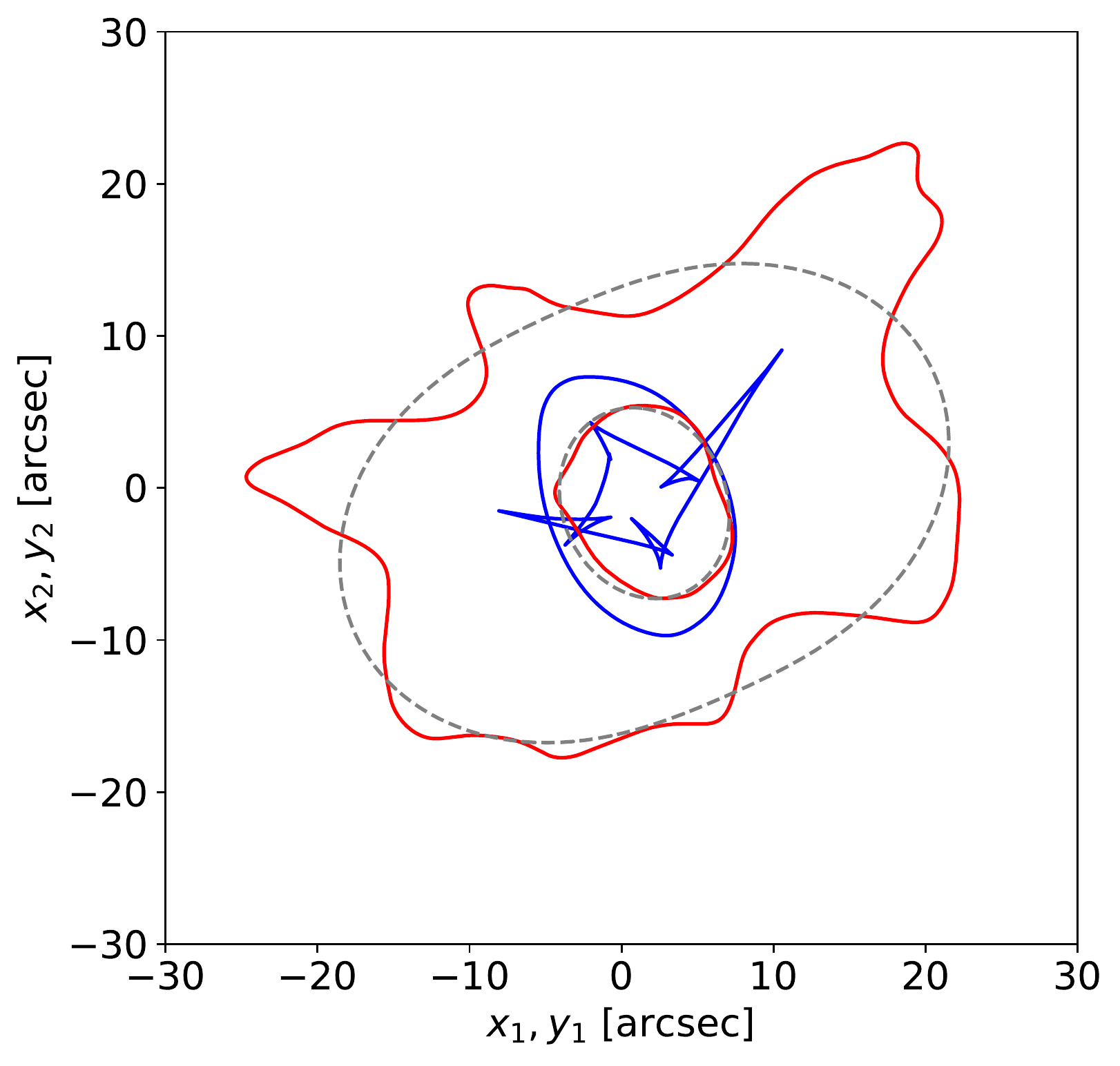} \hspace{3ex}
  \includegraphics[width=0.3\textwidth]{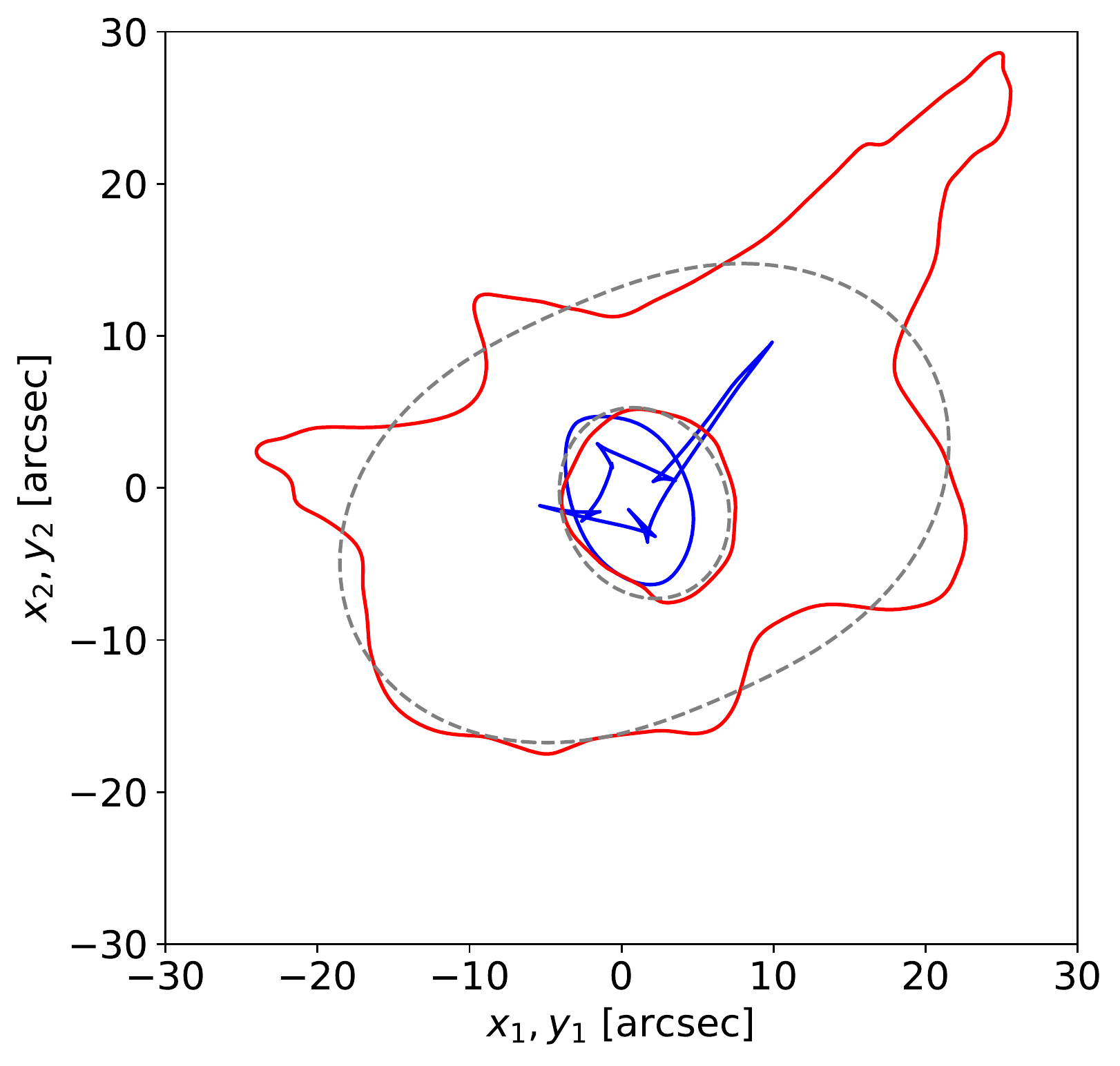} \\
  \includegraphics[width=0.3\textwidth]{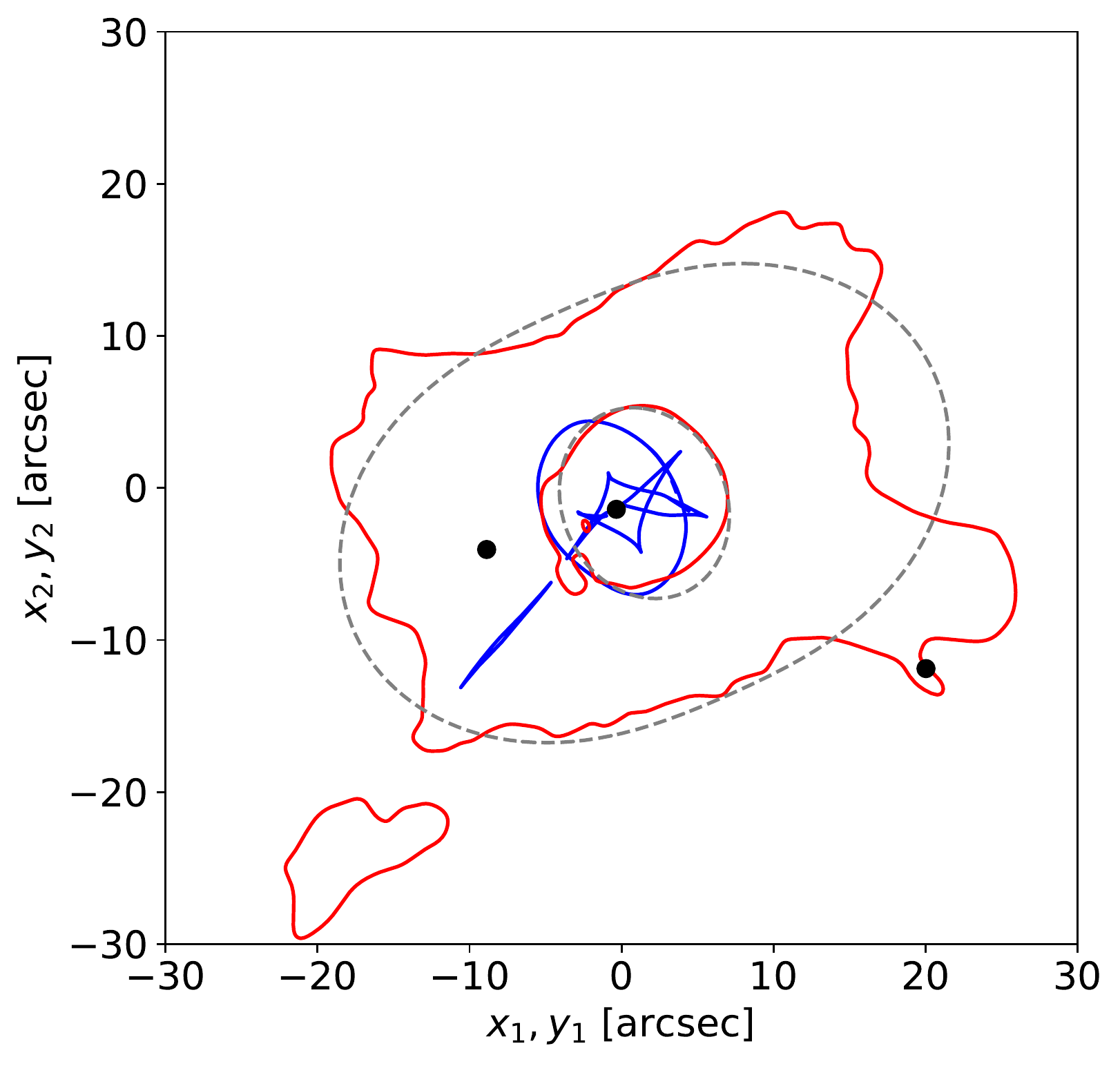} \hspace{3ex}
  \includegraphics[width=0.3\textwidth]{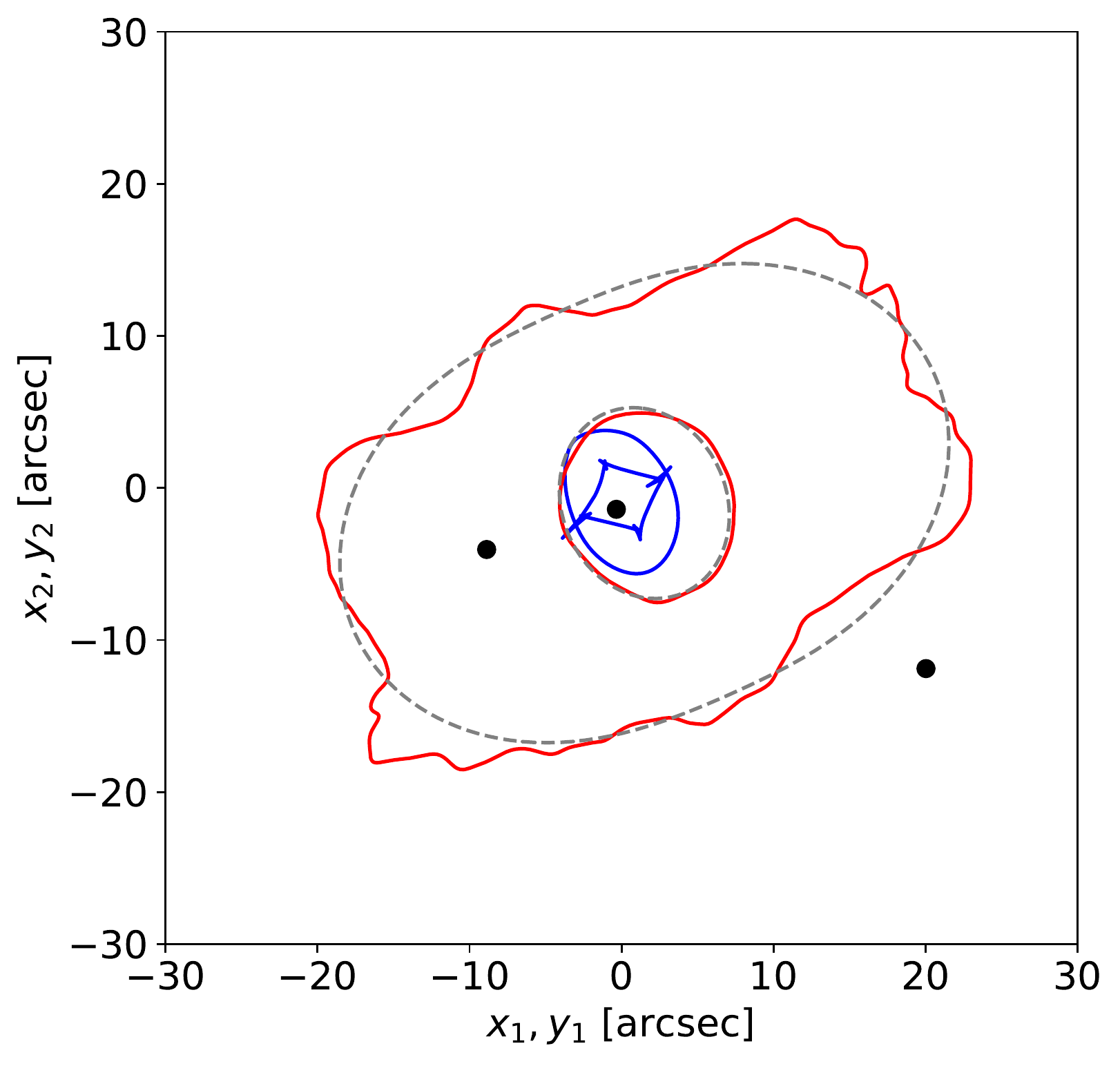} \\
  \includegraphics[width=0.3\textwidth]{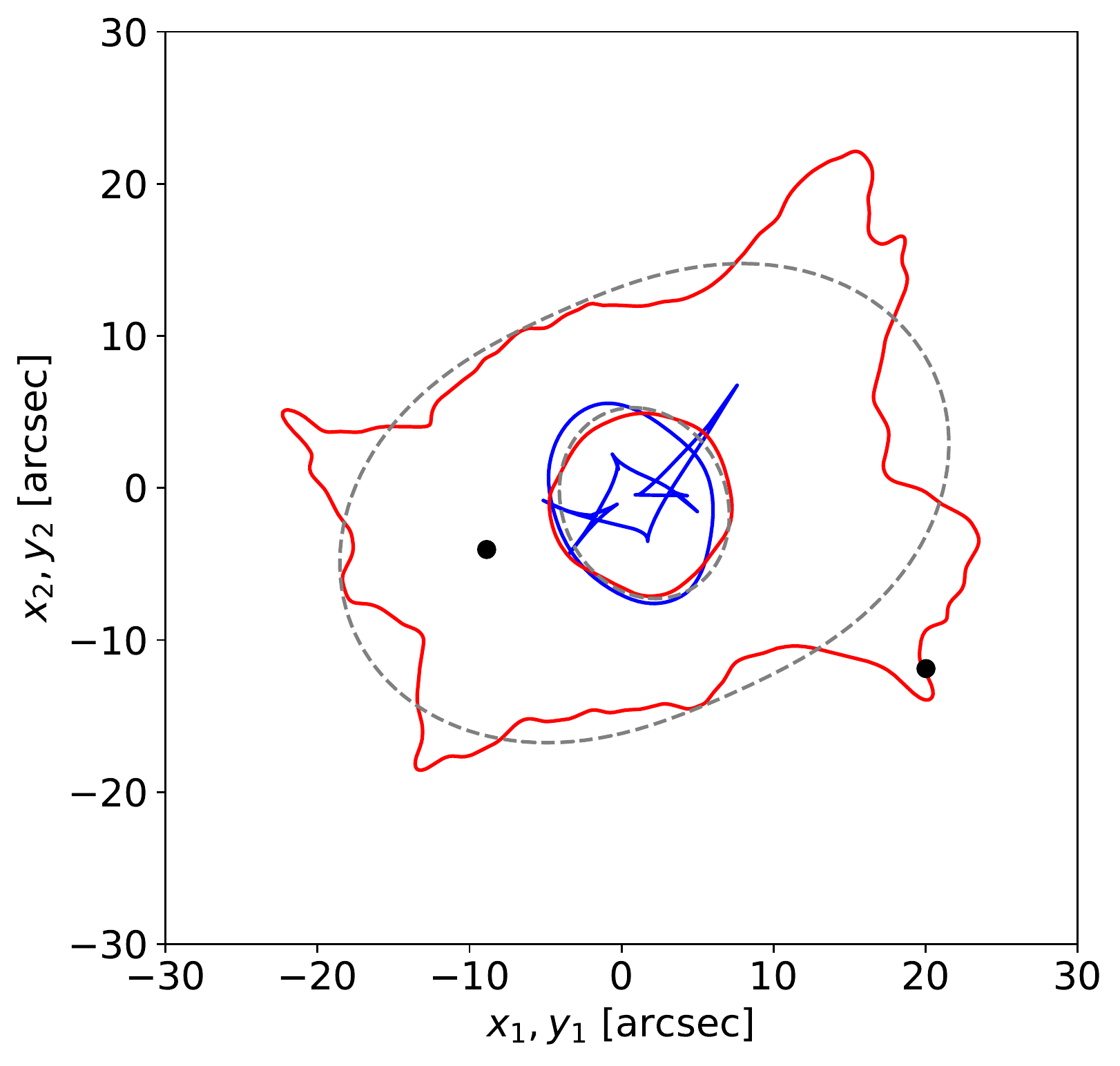} \hspace{3ex}
  \includegraphics[width=0.3\textwidth]{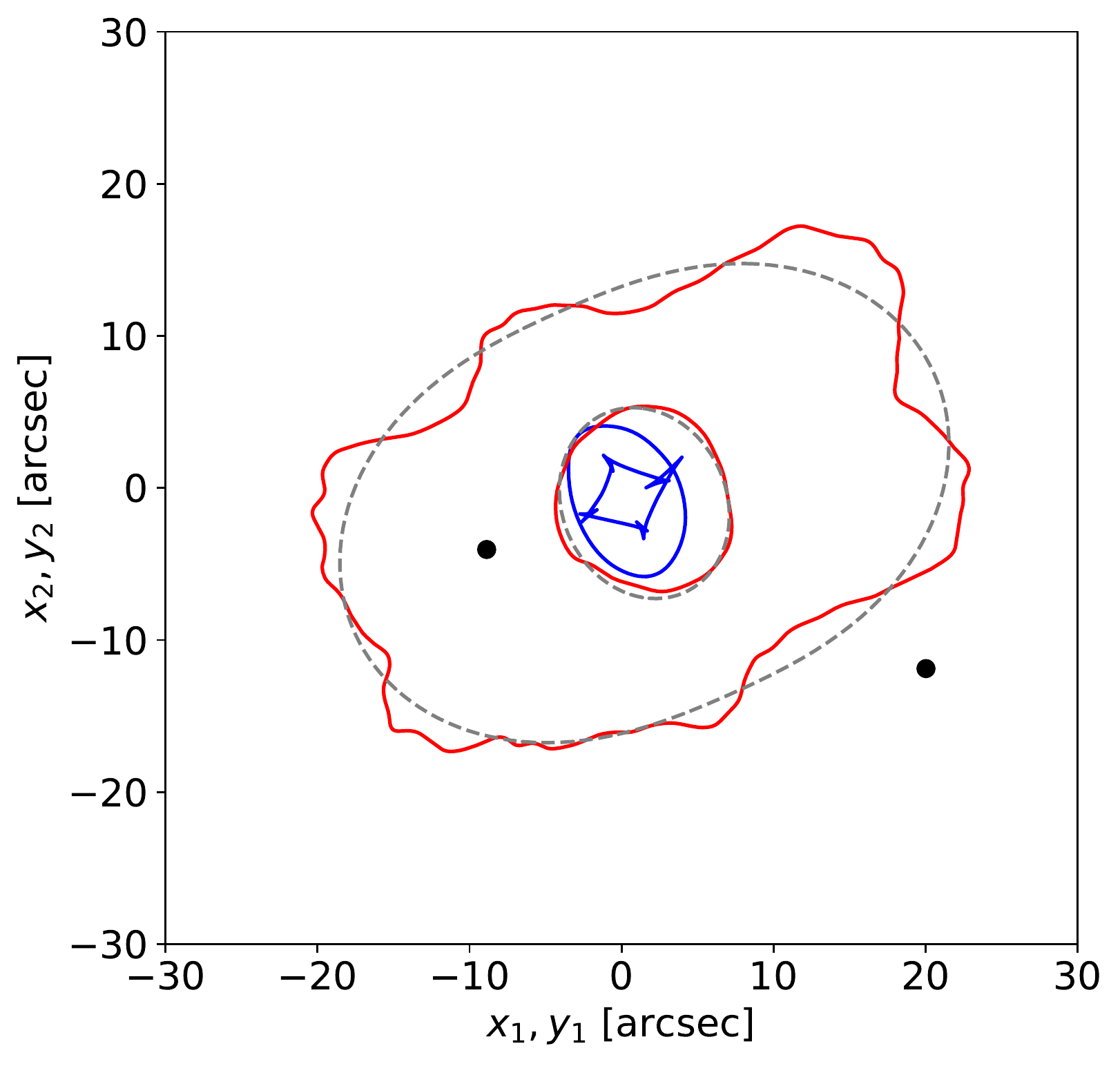} \\
  \includegraphics[width=0.3\textwidth]{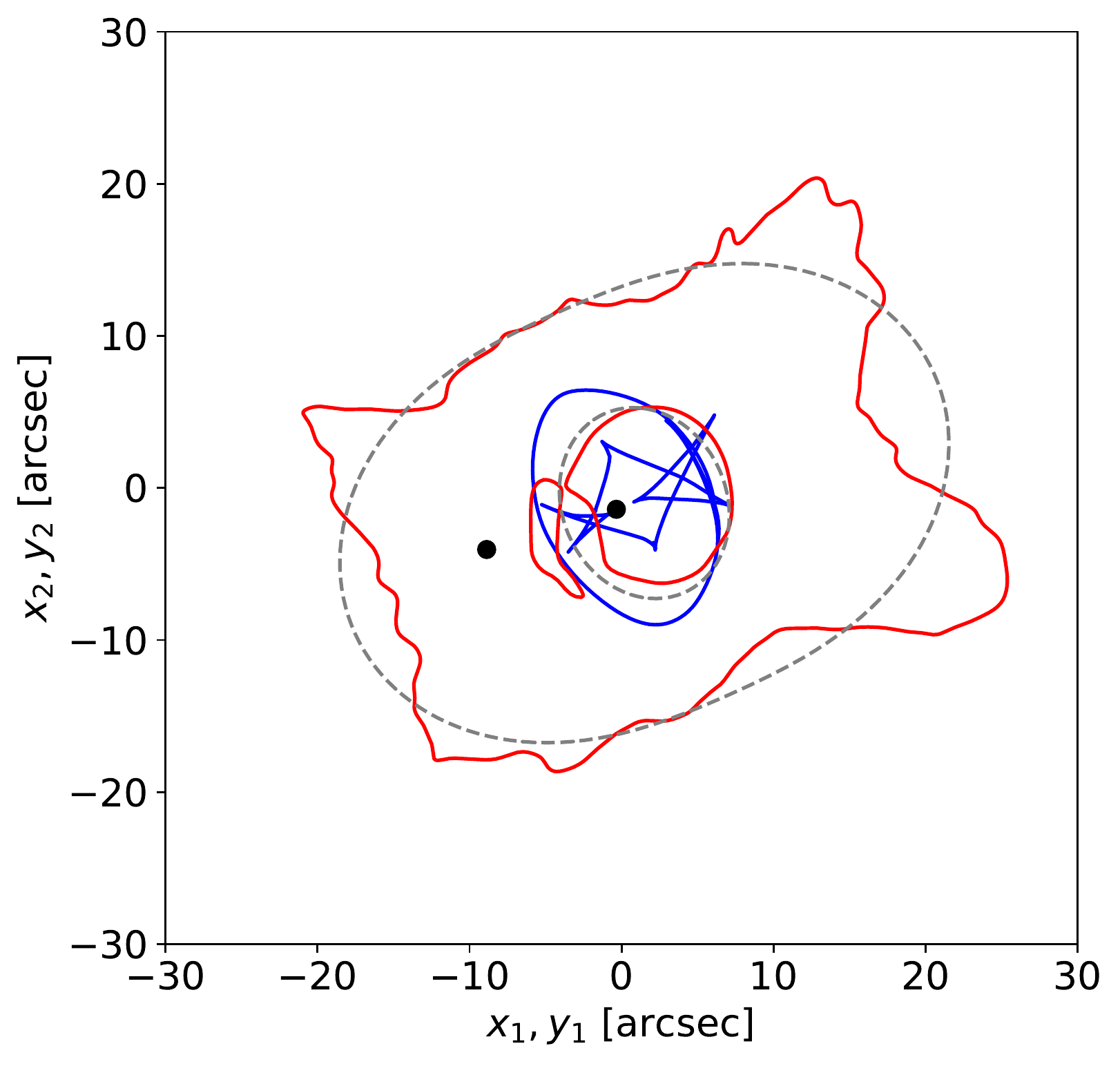} \hspace{3ex}
  \includegraphics[width=0.3\textwidth]{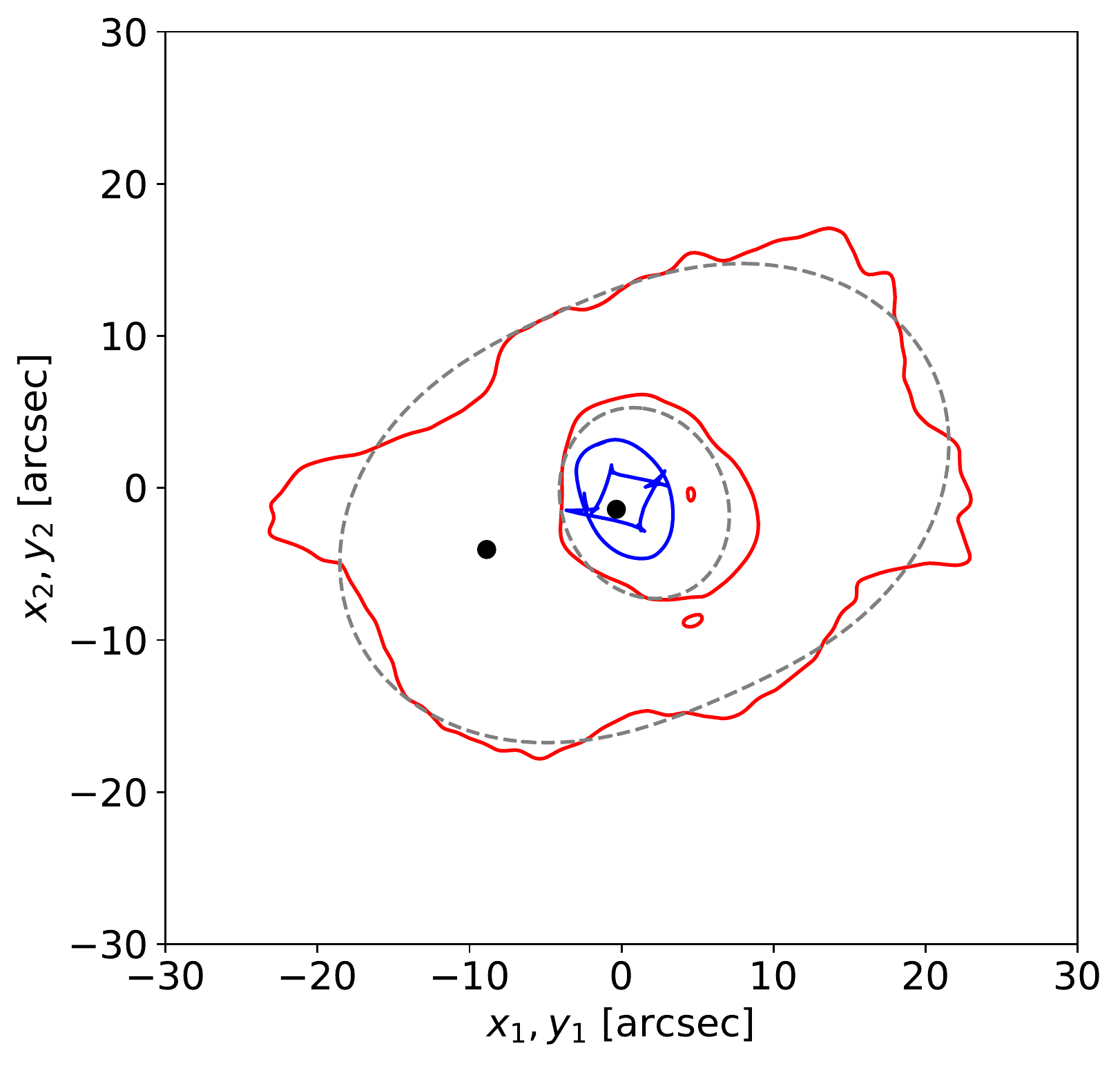}
   \caption{Comparison between Grale lens models for the simulated, generalised NFW profile that is detailed in \cite{bib:Liesenborgs2}. Each map shows the reconstructed critical curves (red, solid lines) and the reconstructed caustics (blue, solid lines) together with the true critical curves (black, dashed lines) for the Grale models with the same constraints detailed in Table~\ref{tab:Jori_models} and Fig.~\ref{fig:Jori_reconstruction} (i.e.\ the left and right columns show models without and including an additional mass-sheet basis function, respectively). Top left: the model includes no additional mass sheet and no time-delay constraints; top right: using an additional mass sheet; second row left: including time-delay constraints between the three multiple images marked by the black dots; second row right: same as the left-hand side using an additional mass sheet; third row left: including one time-delay constraint between the multiple images marked by the black dots; third row right: same as for the left-hand side with an additional mass sheet; fourth row left: same as above but for a different pair of multiple images; fourth row right: same as for the left-hand side with an additional mass sheet.}
\label{fig:Jori_critical_curves}
\end{figure*}

\begin{figure*}[p]
\centering
  \includegraphics[width=0.315\textwidth]{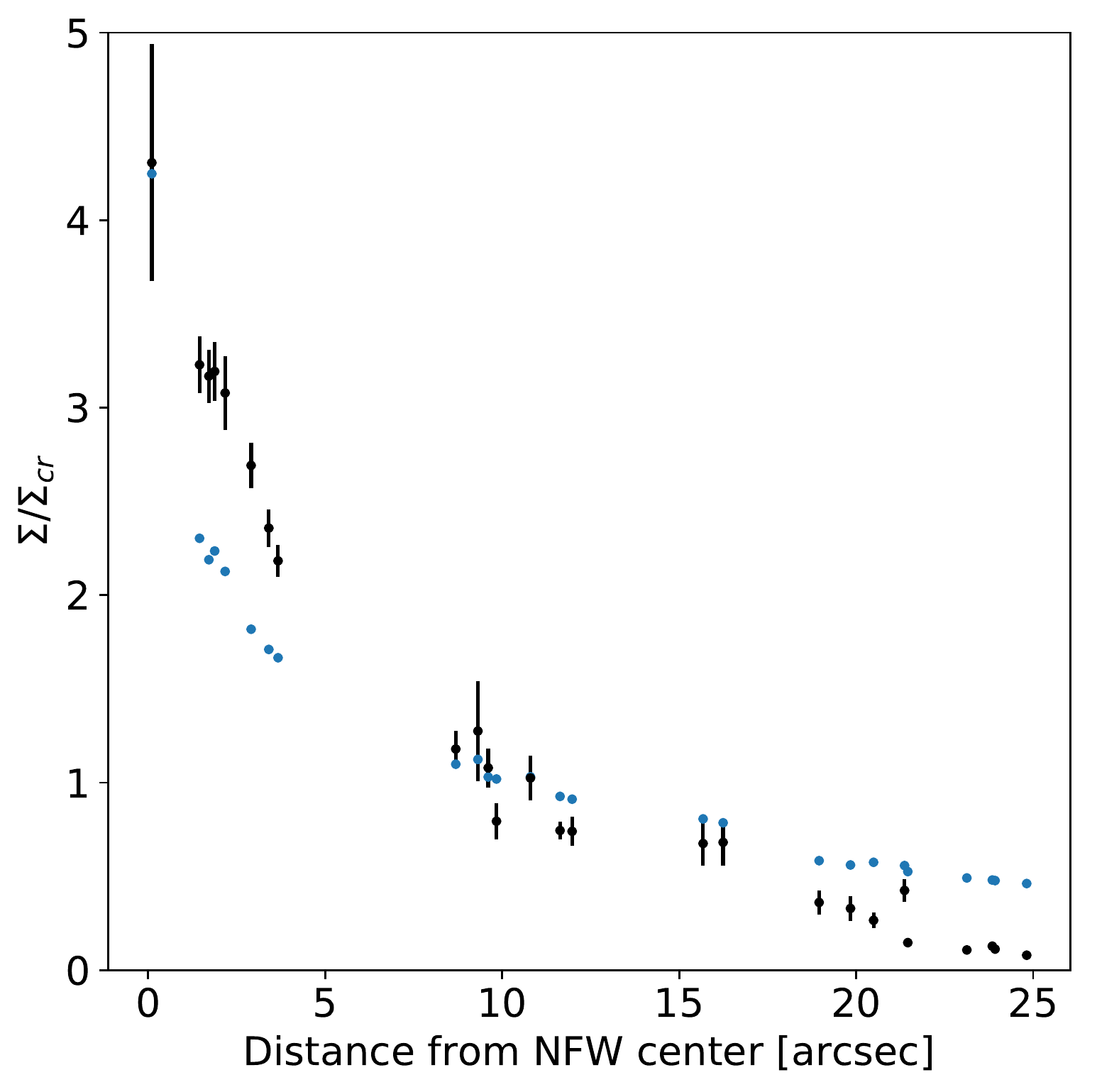} \hspace{3ex}
  \includegraphics[width=0.315\textwidth]{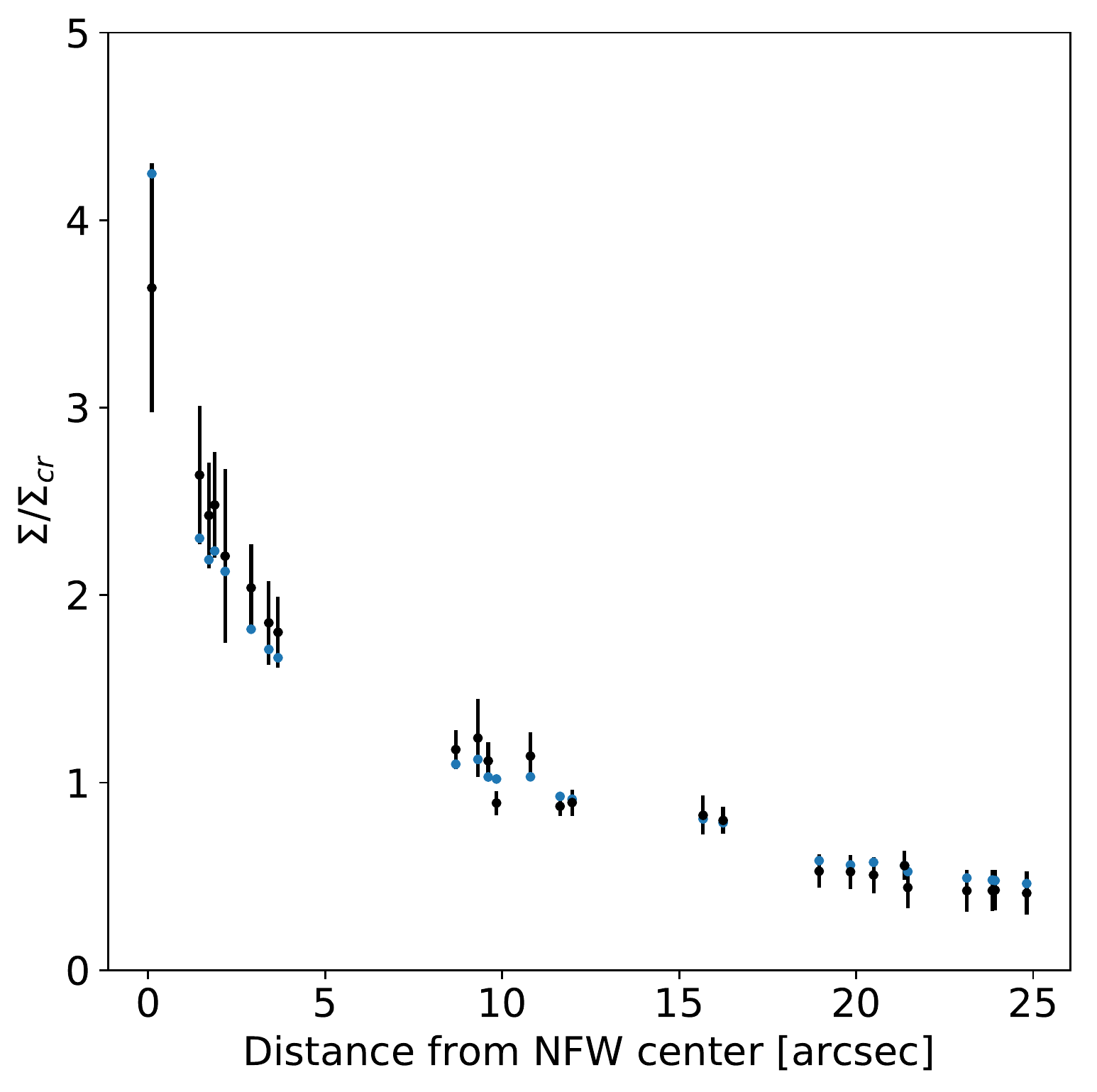} \\
  \includegraphics[width=0.315\textwidth]{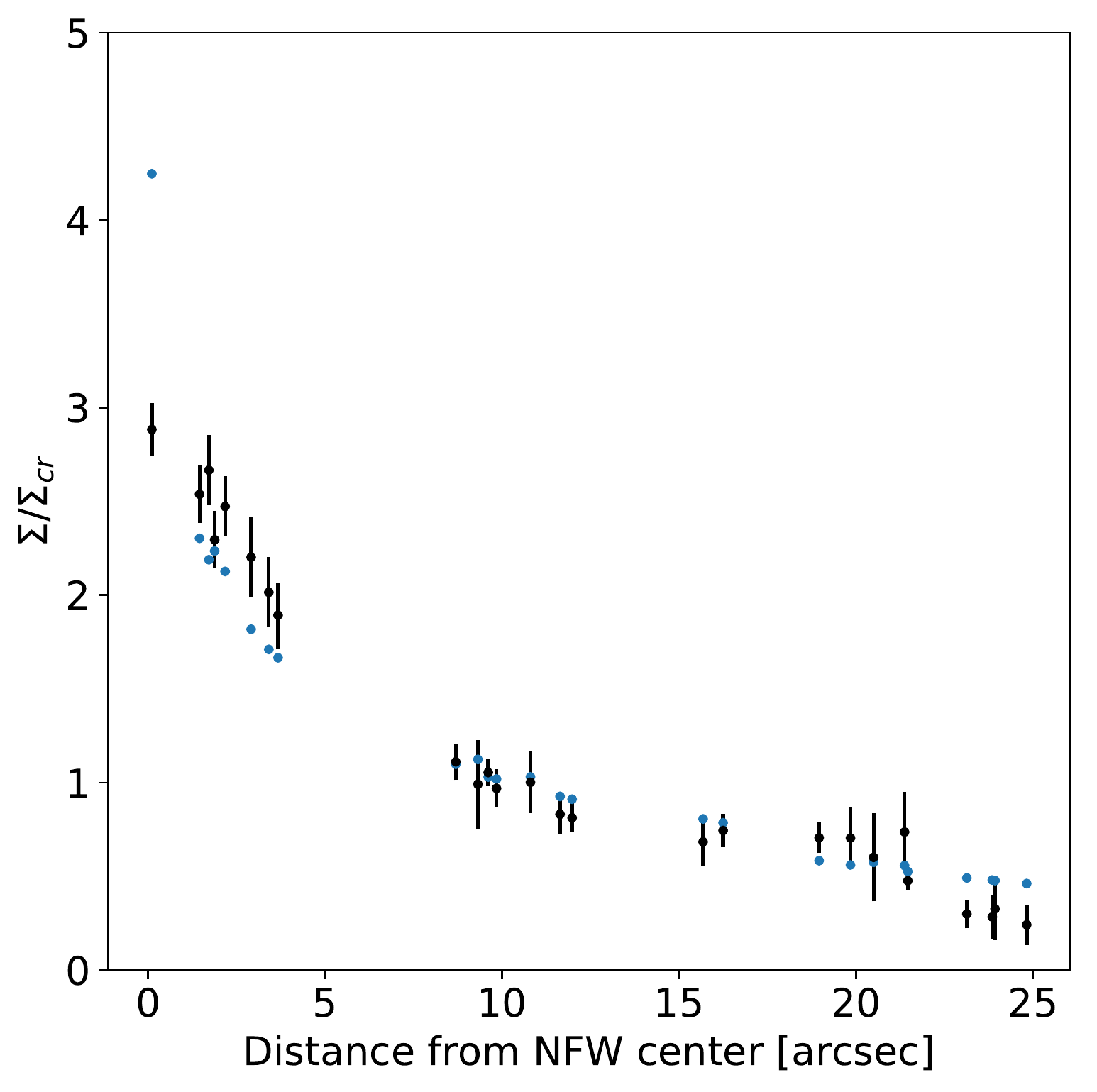} \hspace{3ex}
  \includegraphics[width=0.315\textwidth]{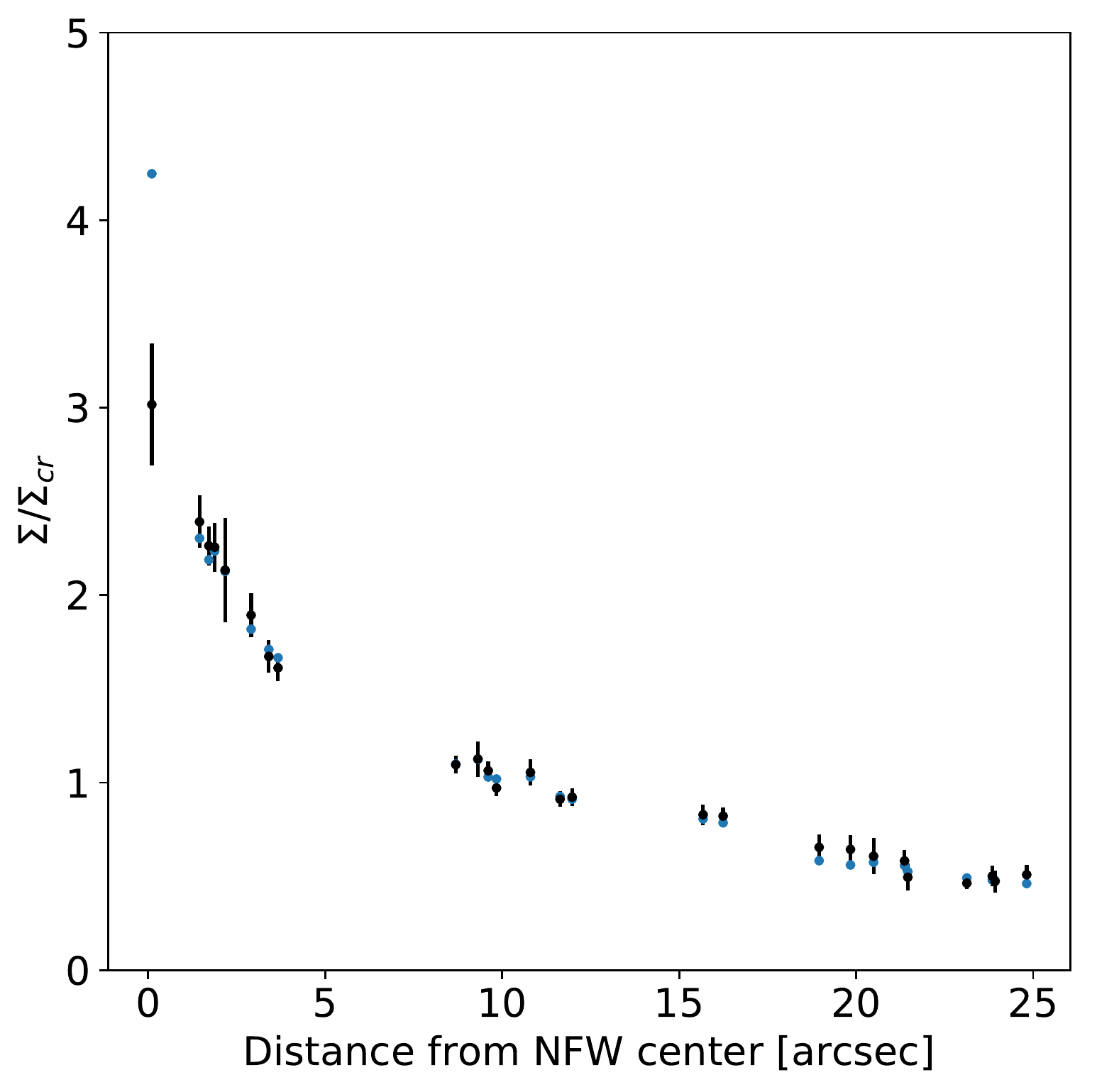} \\
  \includegraphics[width=0.315\textwidth]{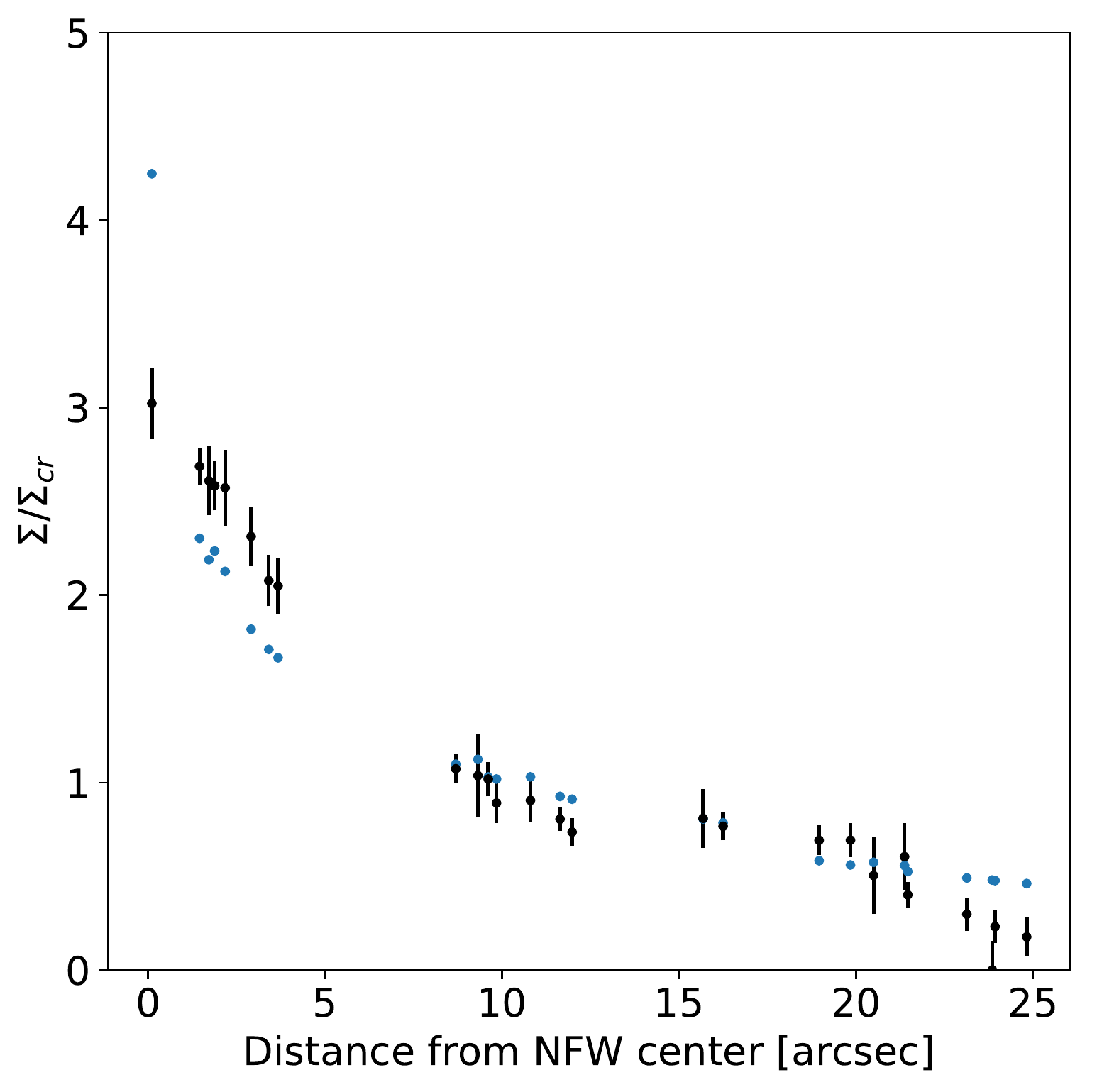} \hspace{3ex}
  \includegraphics[width=0.315\textwidth]{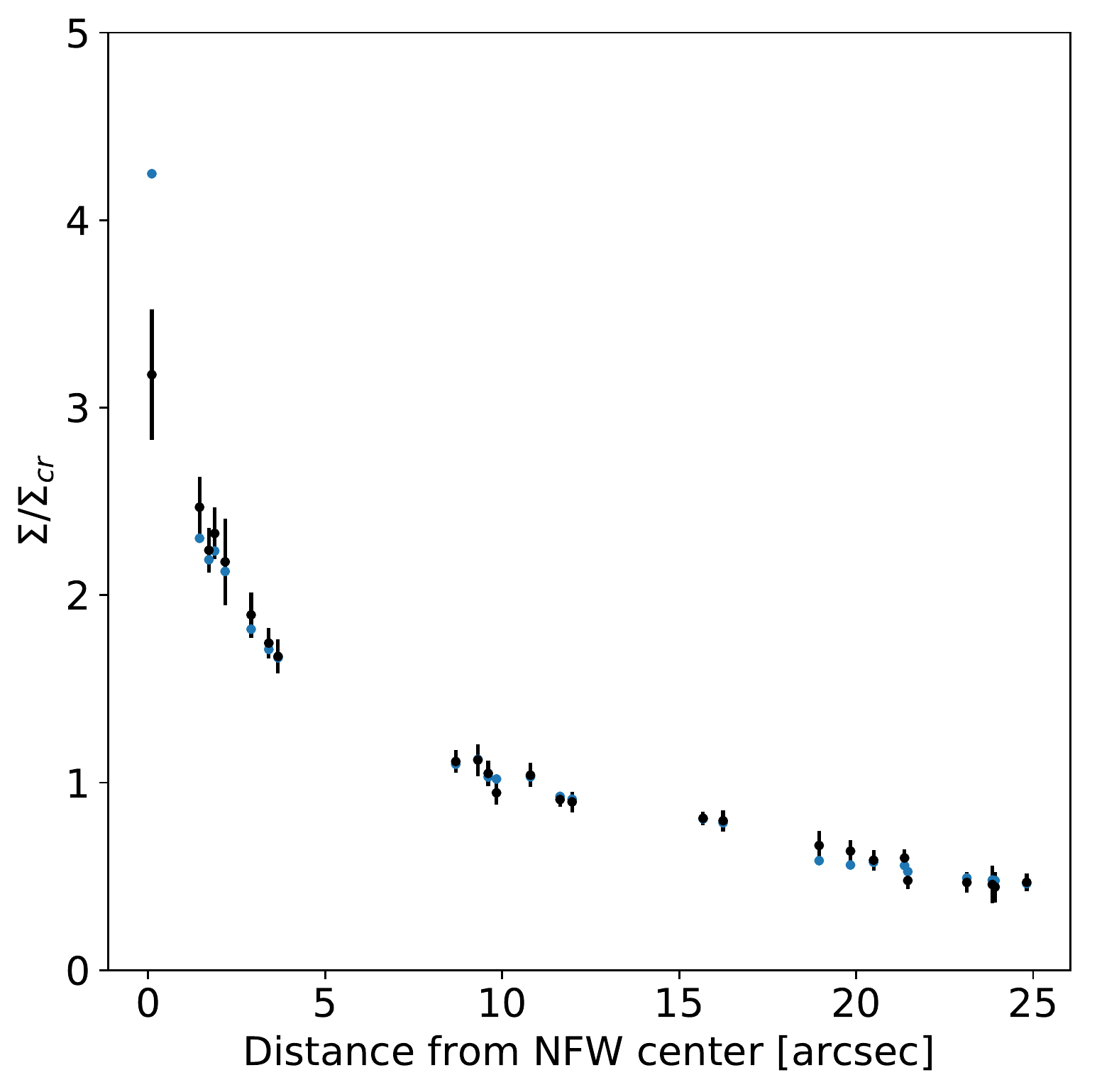} \\
  \includegraphics[width=0.315\textwidth]{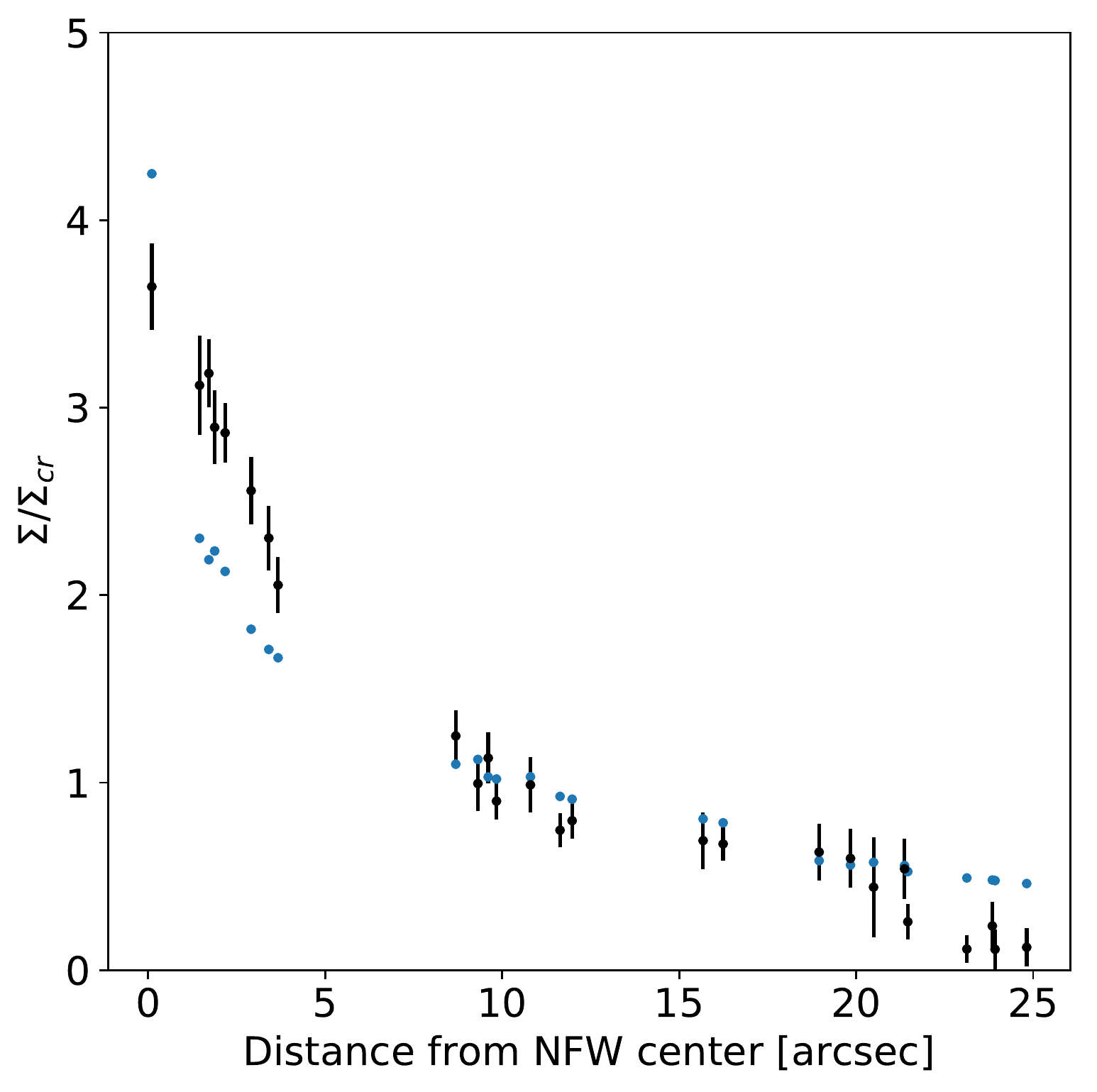} \hspace{3ex}
  \includegraphics[width=0.315\textwidth]{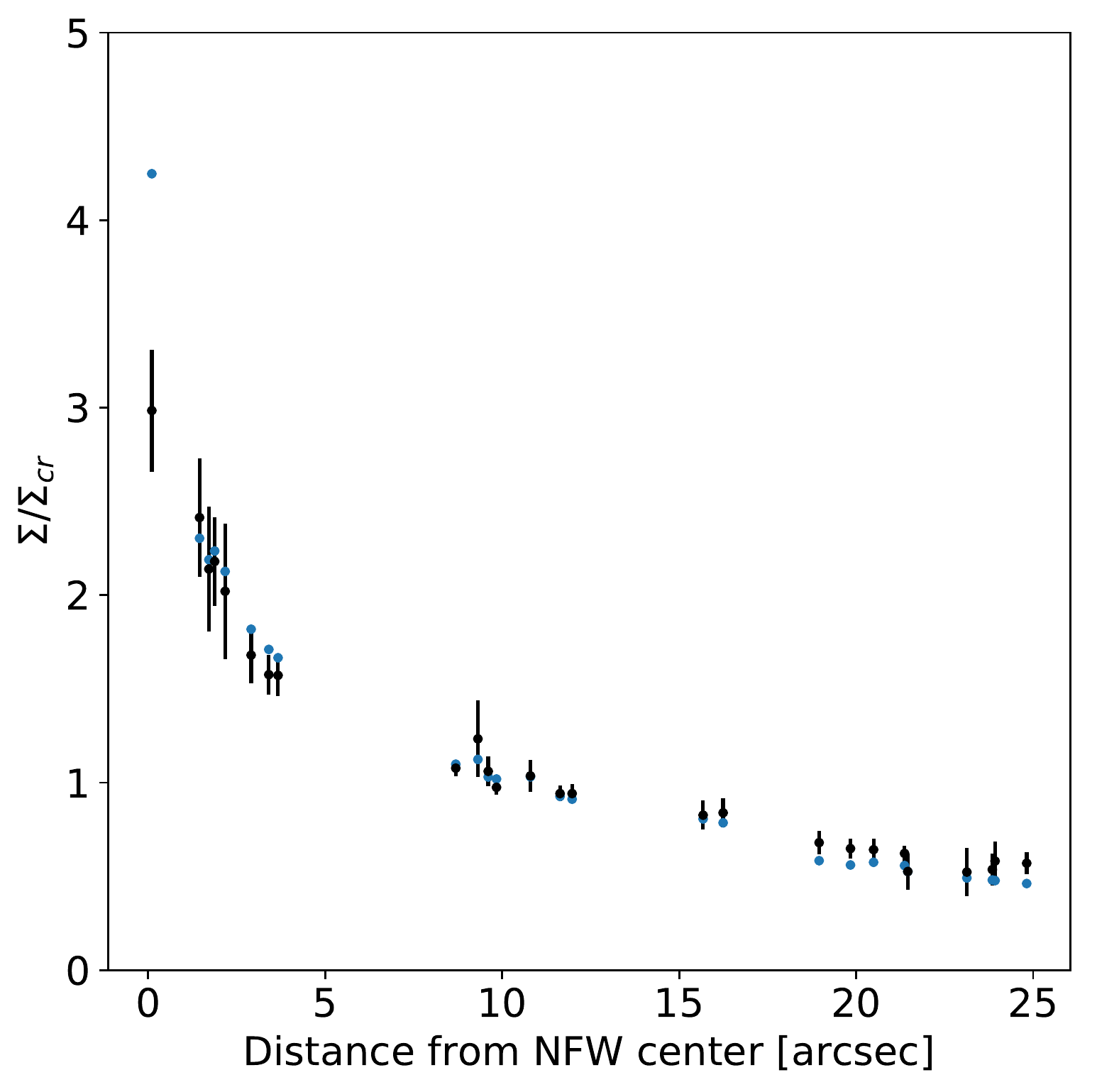}
   \caption{Comparison between Grale lens models for the simulated, generalised NFW profile that is detailed in \cite{bib:Liesenborgs2}. Each plot shows the reconstructed convergence values and their uncertainties (black points with error bars) and the true convergence values (blue points) at the positions of the multiple images measured as radial distance from the centre of the NFW profile for the Grale models with the same constraints as detailed in Table~\ref{tab:Jori_models} and Fig.~\ref{fig:Jori_critical_curves}.}
\label{fig:Jori_radial_profiles}
\end{figure*}

\begin{figure*}[p]
\centering
  \includegraphics[width=0.38\textwidth]{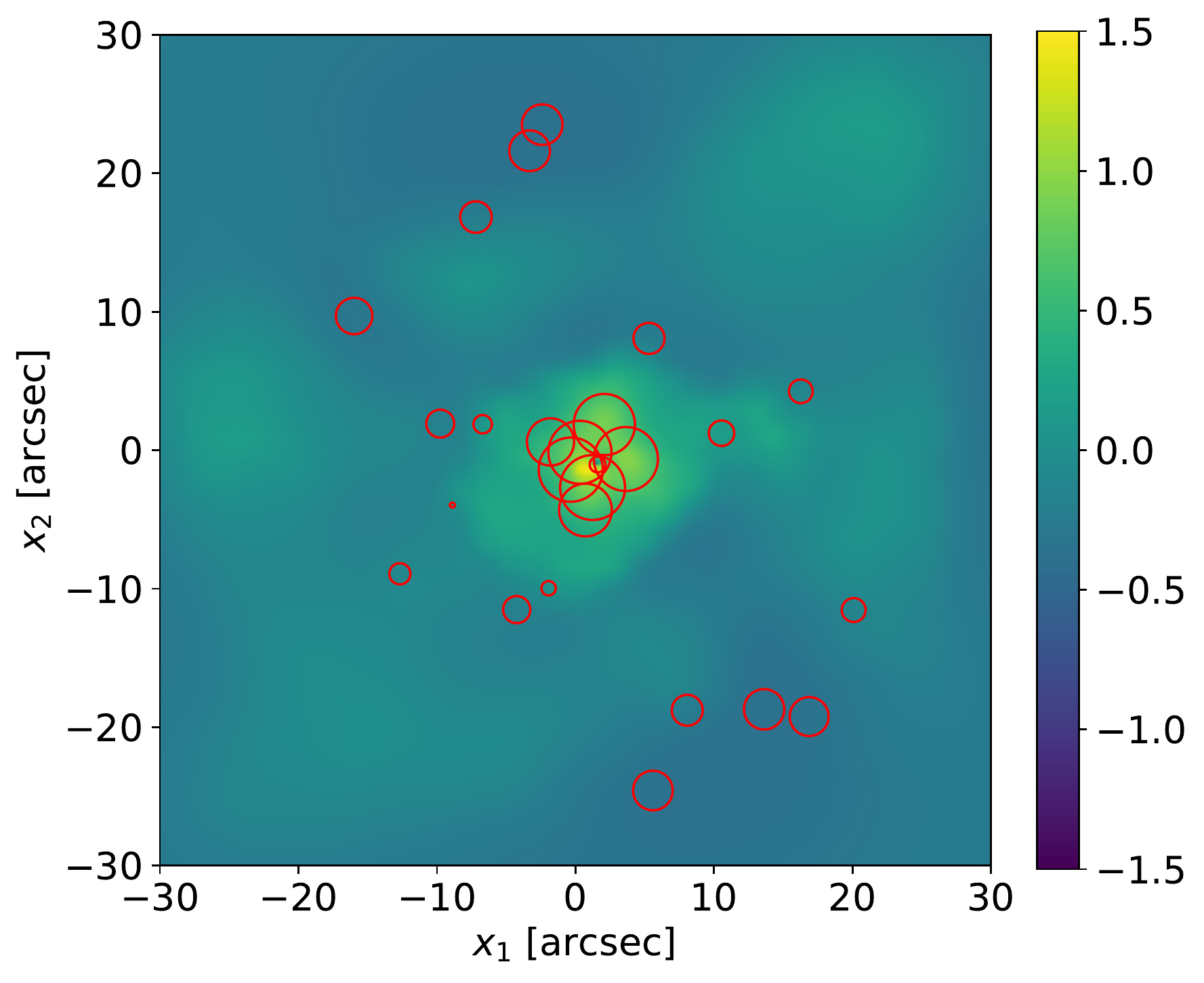} \hspace{3ex}
  \includegraphics[width=0.38\textwidth]{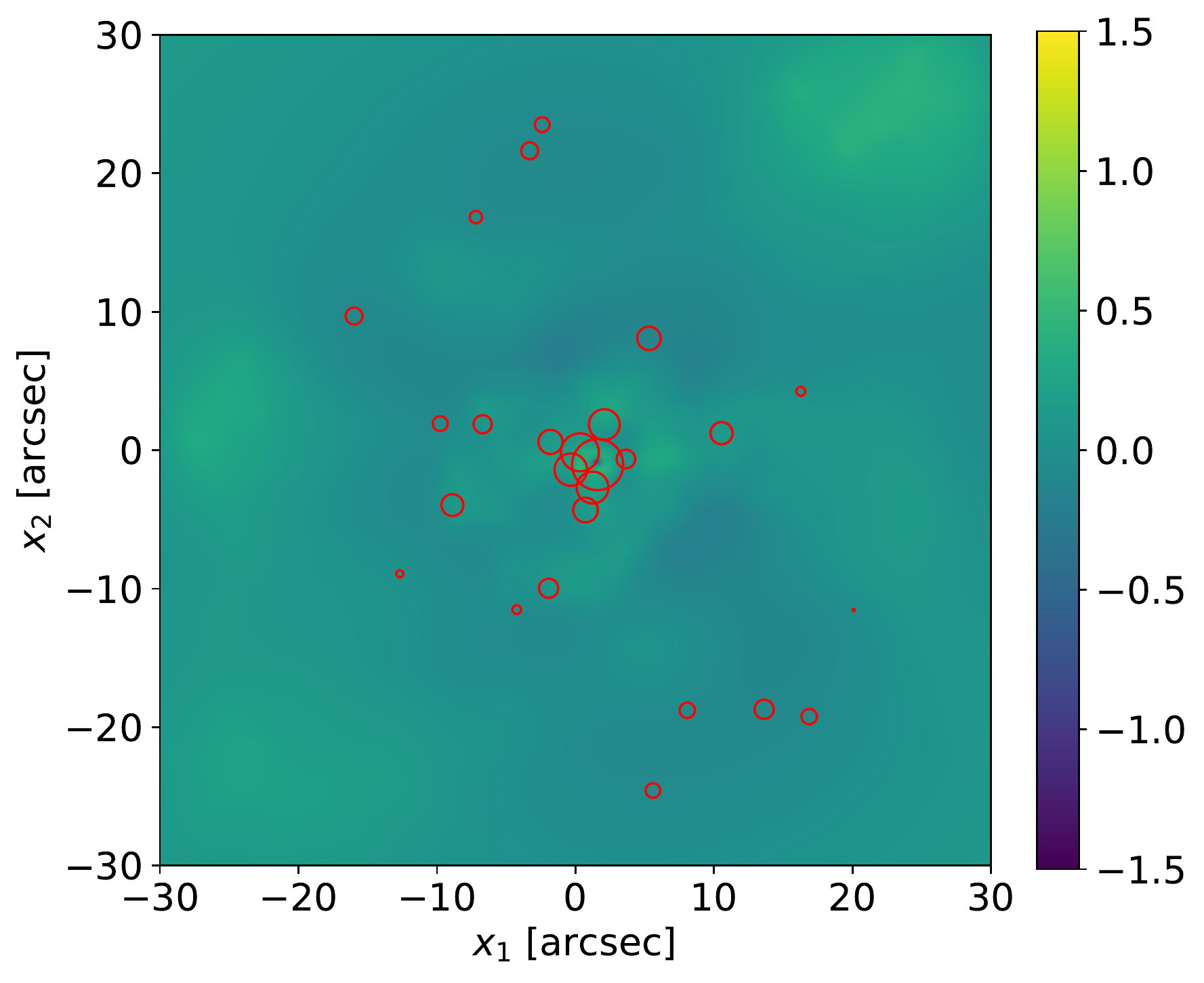}
  \includegraphics[width=0.38\textwidth]{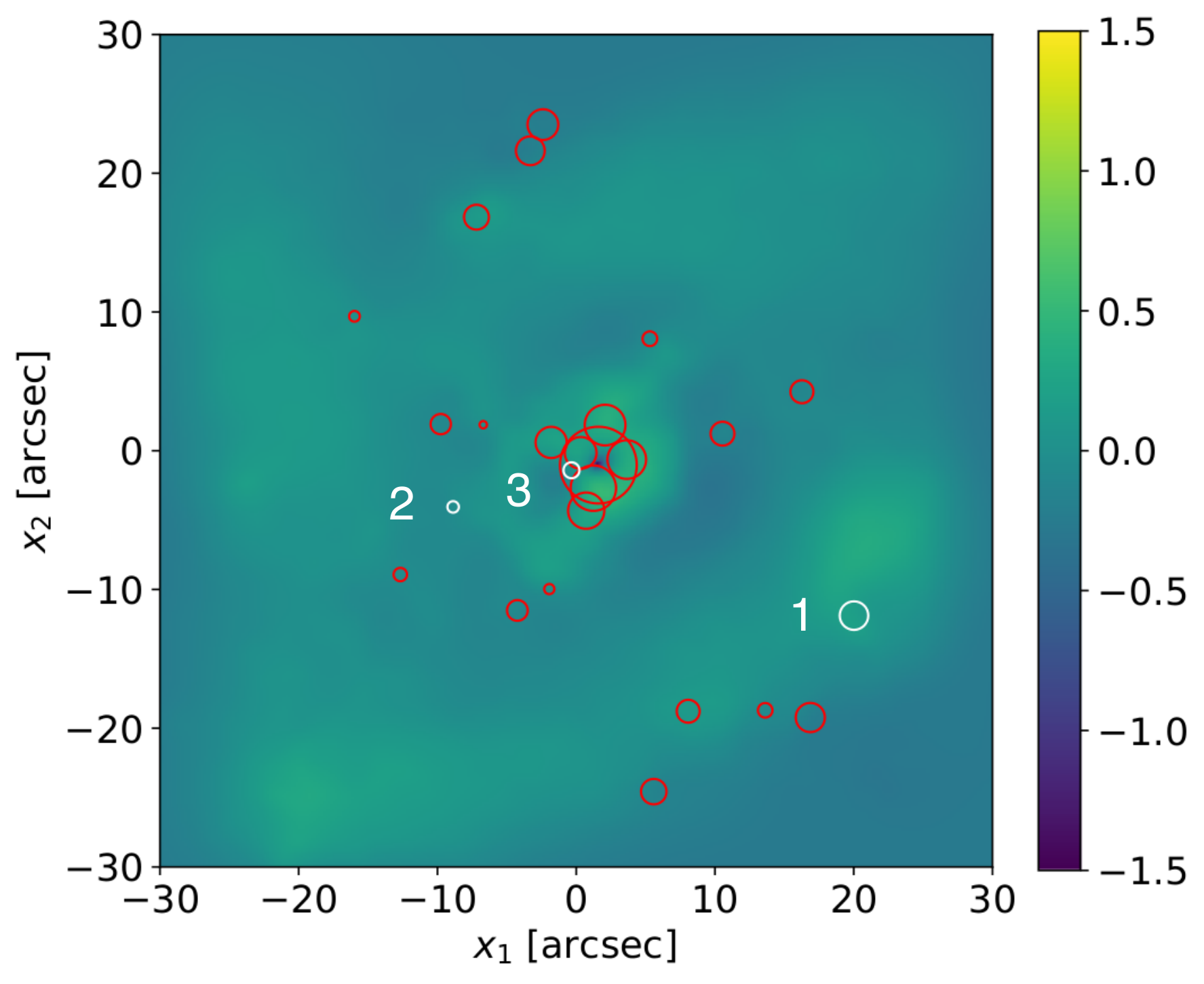} \hspace{3ex}
  \includegraphics[width=0.38\textwidth]{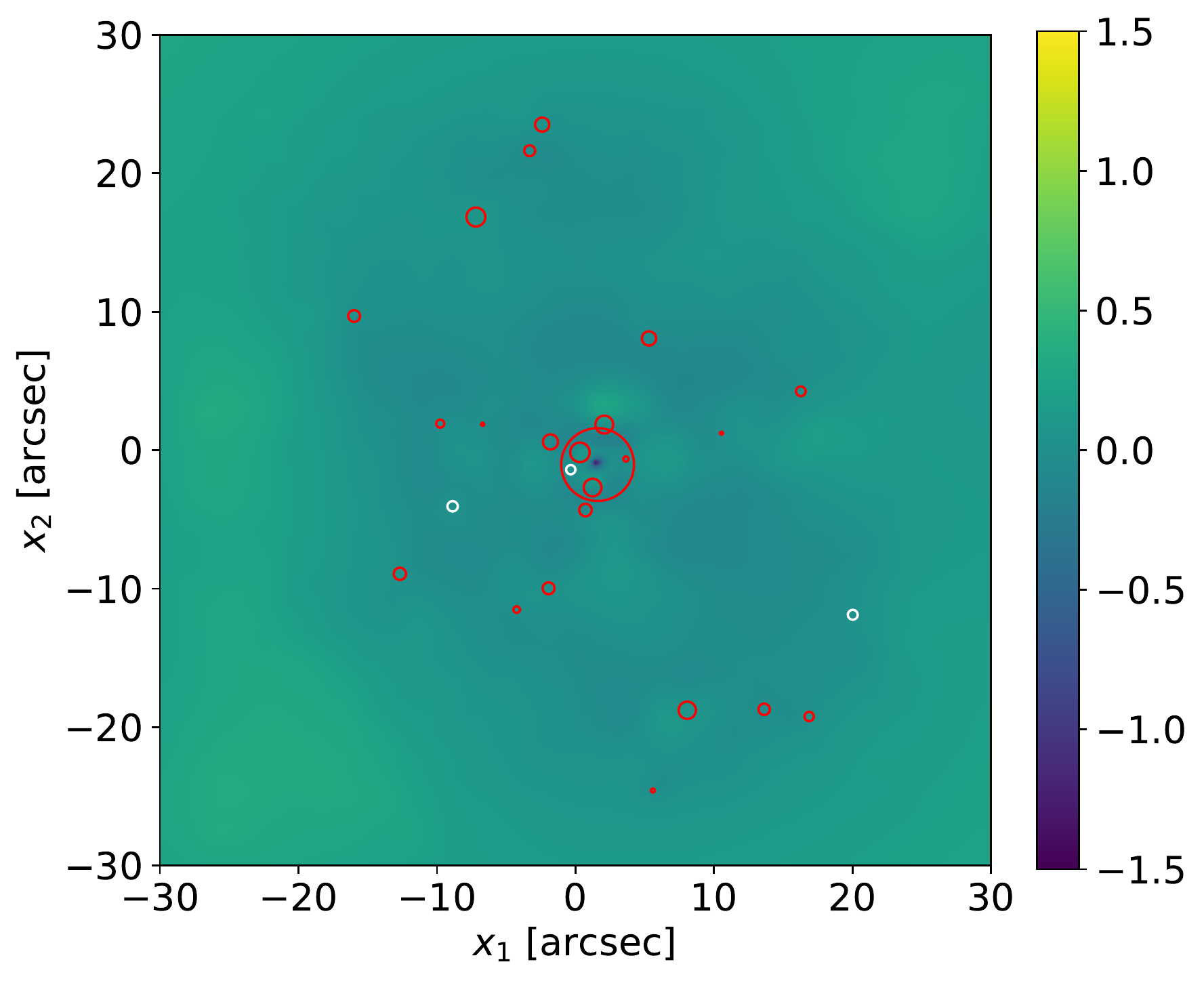}
  \includegraphics[width=0.38\textwidth]{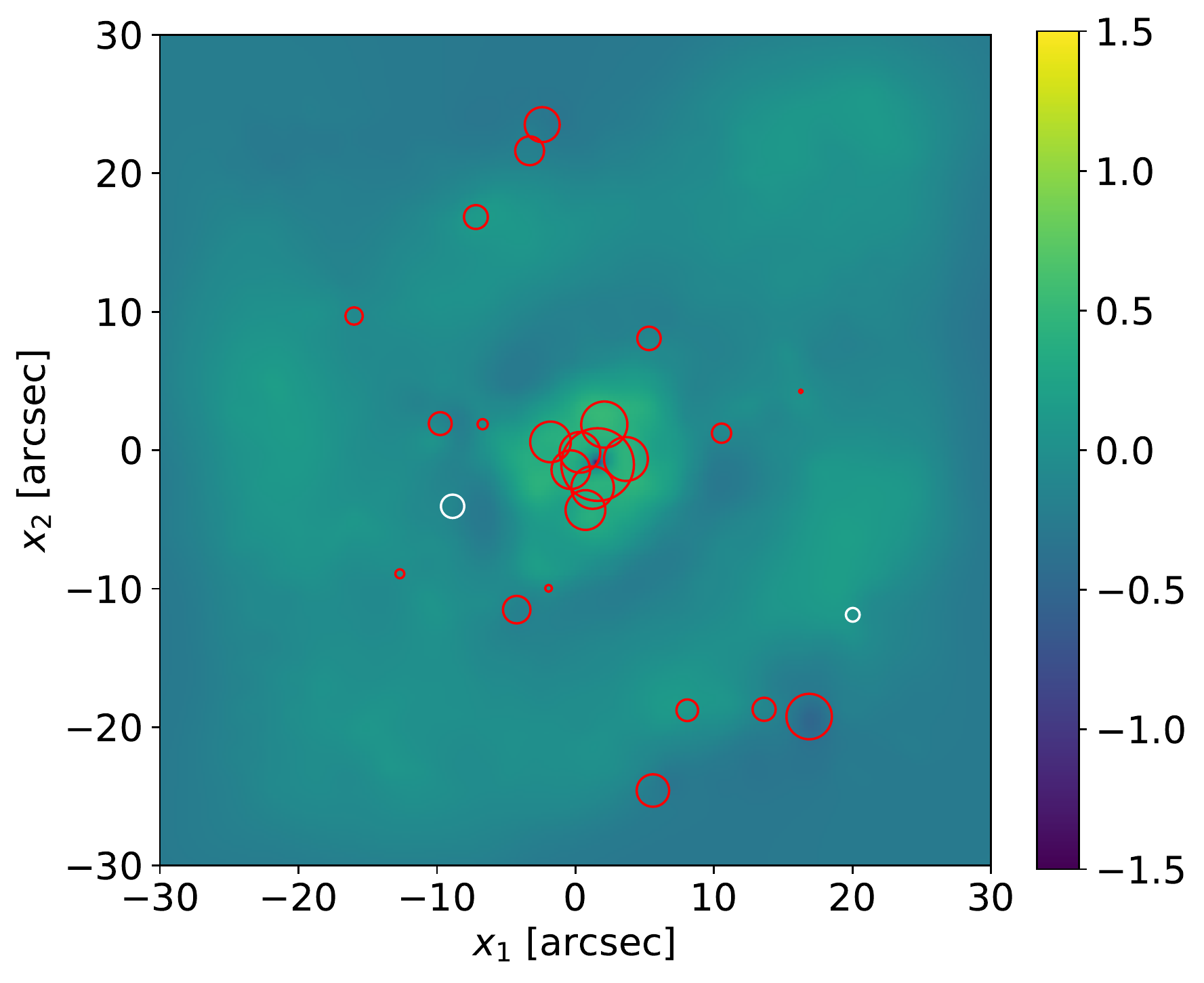} \hspace{3ex}
  \includegraphics[width=0.38\textwidth]{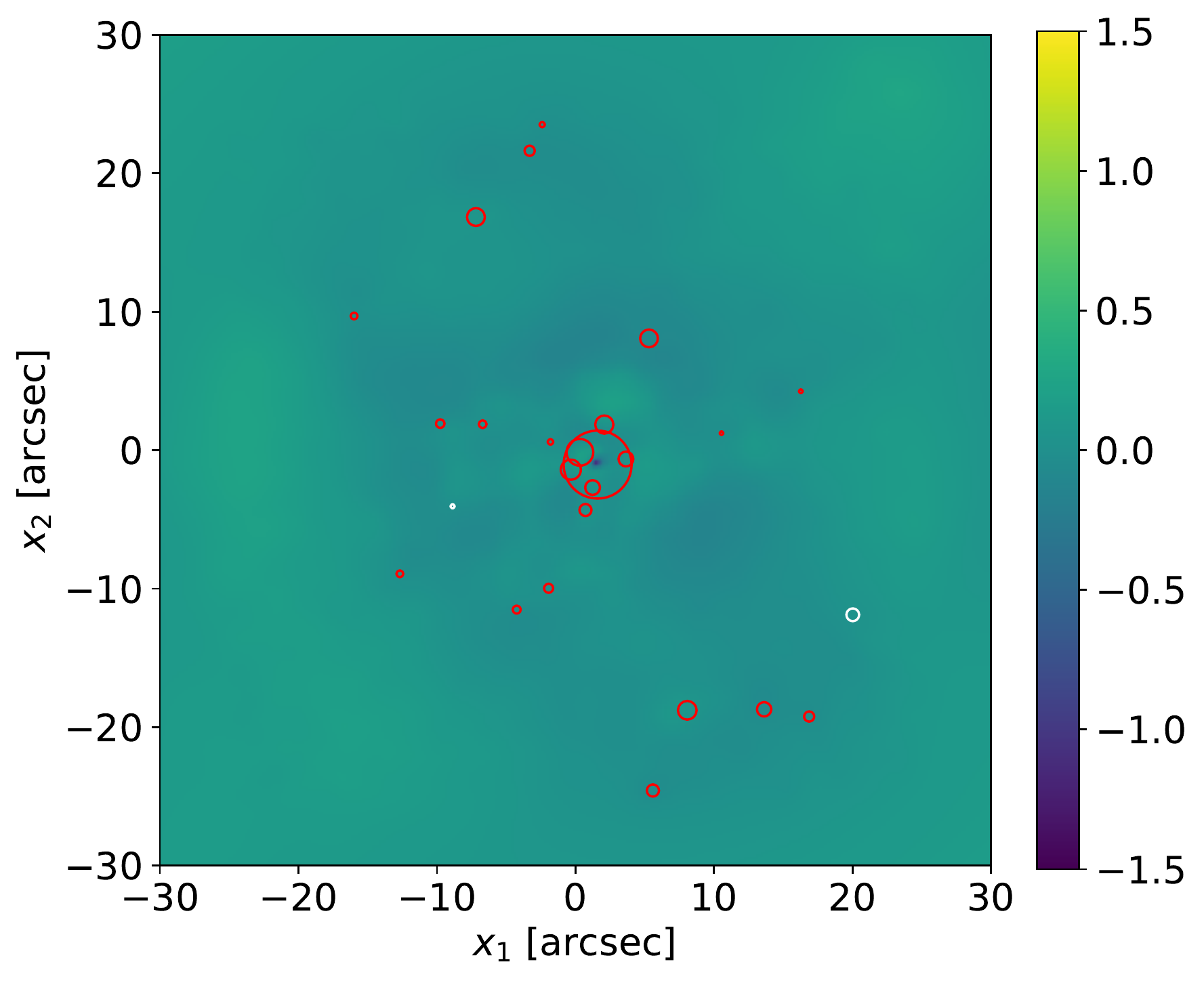}
  \includegraphics[width=0.38\textwidth]{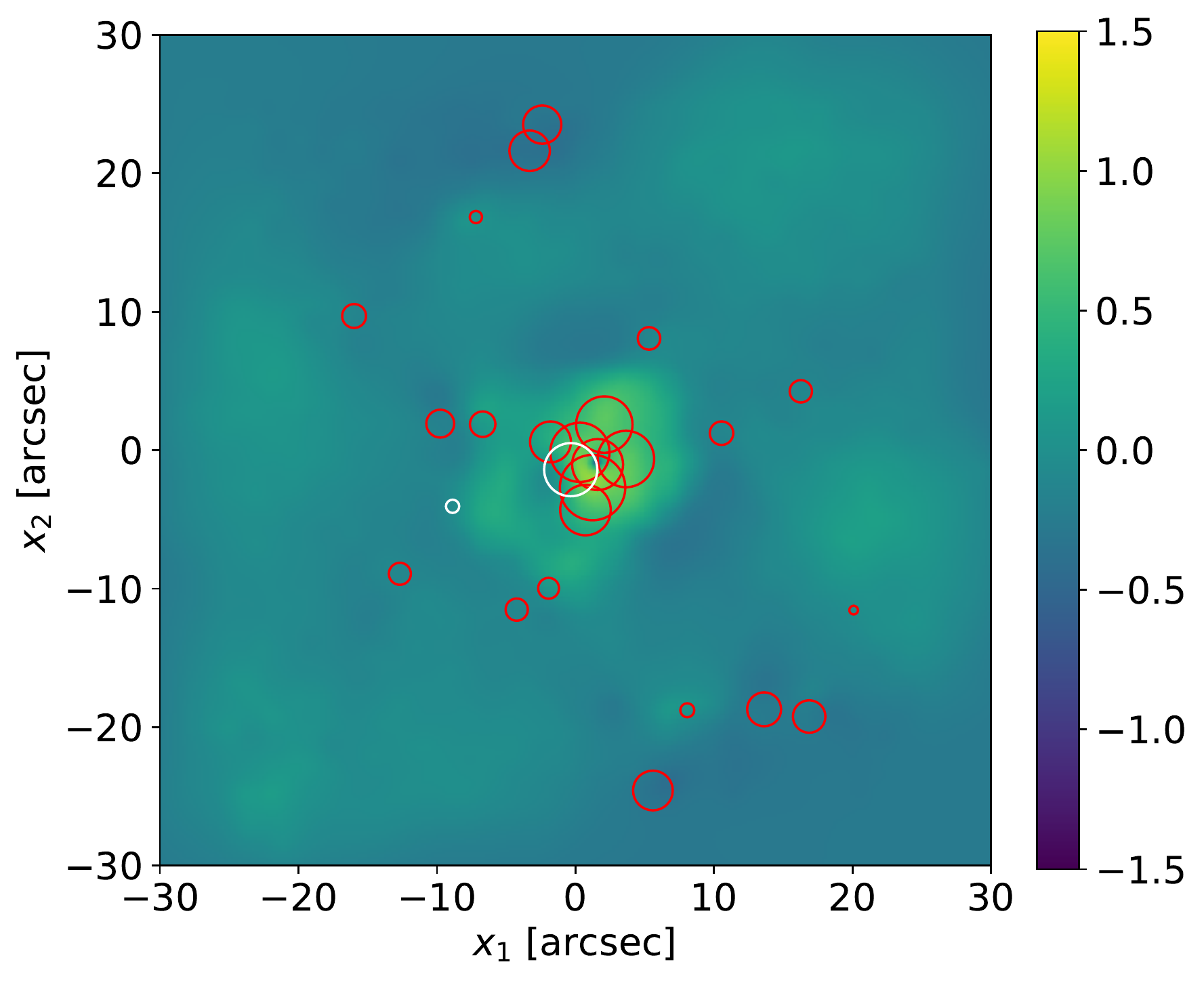} \hspace{3ex}
  \includegraphics[width=0.38\textwidth]{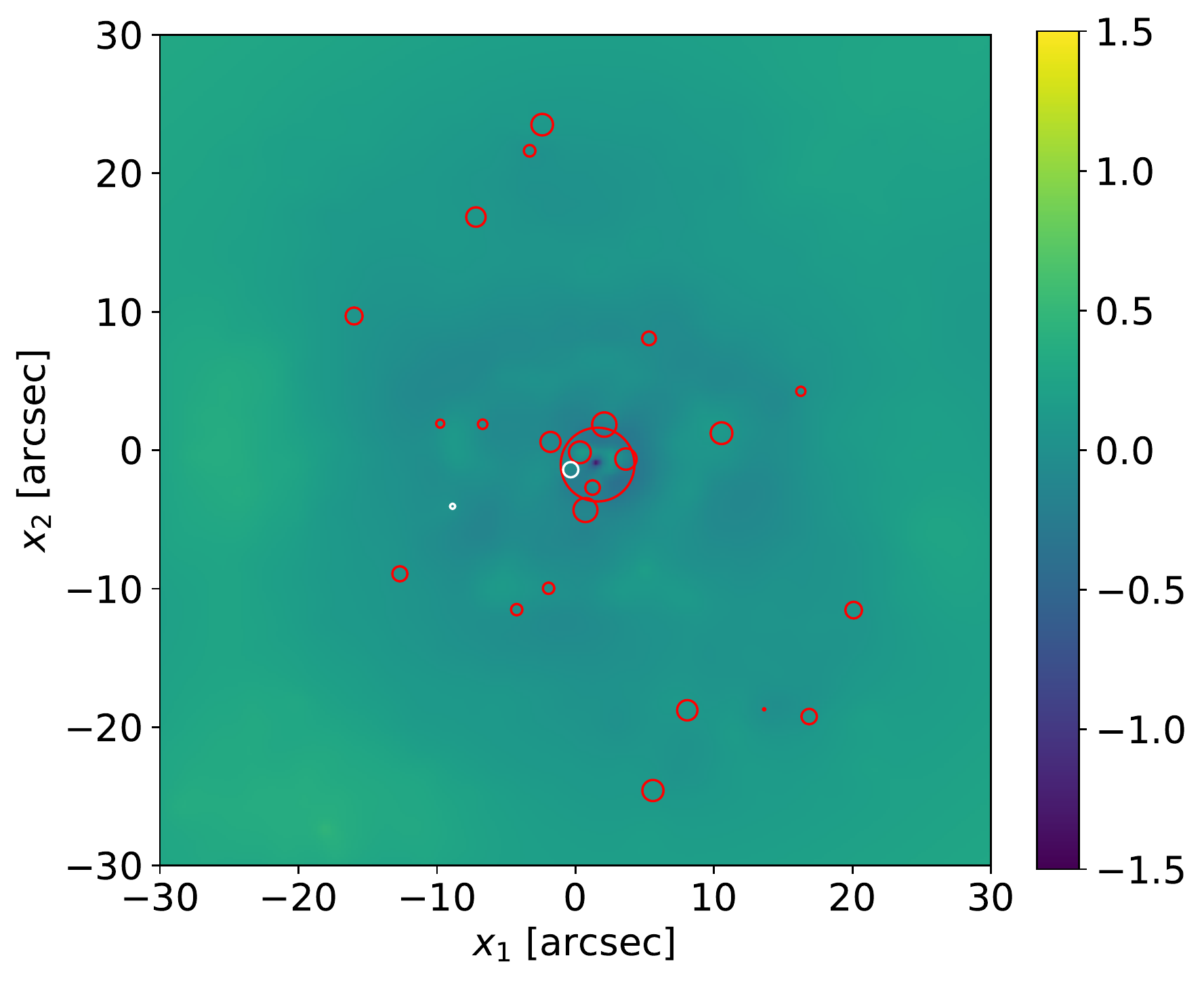}
   \caption{Comparison between Grale lens models for the simulated, generalised NFW profile that is detailed in \cite{bib:Liesenborgs2}. Each map shows the difference between the reconstructed convergence map and the simulation result for the models with  the same constraints as detailed in Table~\ref{tab:Jori_models} and Fig.~\ref{fig:Jori_critical_curves}. The circles mark the positions of the multiple images and their size is proportional to the difference between the reconstructed and real convergence at these positions. The white circles indicate the multiple images with the time-delay constraints. For the multiple images marked by red circles, no time-delay constraint is given.}
\label{fig:Jori_reconstruction}
\end{figure*}

Consistent results for the reconstruction of the convergence are found in Figure~\ref{fig:Jori_radial_profiles}. The plots show the true and the reconstructed convergences at the radial distances of the multiple images from the centre of the NFW profile for all models. Adding a mass sheet to the Plummer basis increases the reconstruction accuracy, as can be observed when comparing models with odd numbers (left-hand side of Figure~\ref{fig:Jori_radial_profiles}) and models with even numbers (right-hand side of Figure~\ref{fig:Jori_radial_profiles}). Taking into account time-delay constraints without adding a mass sheet increases the overall accuracy but the reconstruction is biased towards a less cuspy profile in the centre. Three time-delay constraints yield the highest accuracy (model~3), followed by the time-delay constraint of the two multiple images with farthest distance (model~5), and then followed by the time-delay constraint of the two multiple images that lie closely together (model~7). Adding a mass sheet the accuracy is further increased, except for model~8, but it does not alleviate the bias. This bias might originate from the sparse density of constraints in favour of a cuspy centre because the distribution of basis functions over the cluster region is similar to that of the first two models which are able to reconstruct the steep slope in the centre of the cluster. 

Comparing the standard deviations between the models, we observe that including the time-delay constraints, and thus reducing the degrees of freedom, reduces the differences between the individual reconstructions obtained from the genetic algorithm. Due to the locality of the time-delay constraint in model~7 and model~8, their standard deviations are increased compared to the other models with time-delay constraints. For the latter models, the standard deviations are comparable with each other. Comparing the standard deviation between the reconstructions obtained from the genetic algorithm (i.e. the reconstruction precision) with the differences between the true convergence value and the reconstructed value (i.e. the reconstruction accuracy) in the entire field, we find that both are on the same order of magnitude. 

Figure~\ref{fig:Jori_reconstruction} shows the map of differences between the reconstructed convergence values and the true values over the entire region of the simulation. As already observed in Figures~\ref{fig:Jori_critical_curves} and \ref{fig:Jori_radial_profiles}, including time-delay constraints increases the reconstruction accuracy of the lensing mass distribution and adding a mass sheet reduces the deviations further. 

To investigate the accuracy to which the time delays between the images are recovered for models~1 to 8, we proceed as follows: for each individual model, the source position is estimated as the average of the back-projected image positions. For this source position, updated image positions are calculated, in order to circumvent the bias that Equation~\eqref{eq:Jori_time_delays} is subject to. Then, time delays are determined between these image positions. For all 30 time delays per image pair of the individual reconstructions, the average and its standard deviation are calculated as summarised in Table~\ref{tab:Jori_time_delays} for the three image pairs in all models~1 to 8. Comparing these time delays to our estimates for time delays of NFW profiles in Section~\ref{sec:props_of_interest}, we find that our time delays are shorter because we use parameter values that are comparable to observed ones, while the simulated lens of \cite{bib:Liesenborgs2} has a higher concentration parameter. The scale radius is of comparable size.

\begin{table}[t]
 \caption{Time delays between the three images marked as $1,2,3$ in Figure~\ref{fig:Jori_reconstruction} for the models~1--8 (Table~\ref{tab:Jori_models}). True values used as constraints are $\tau_{12} = 29.247~\mbox{a}$, $\tau_{13} = 31.247~\mbox{a}$, and $\tau_{23} = 2.001~\mbox{a}$.}
\label{tab:Jori_time_delays}
\begin{center}
\begin{tabular}{lllllll}
\hline
\noalign{\smallskip}
  model & $\tau_{12}$ & $\tau_{13}$ & $\tau_{23}$ & $\Delta \tau_{12}$ & $\Delta \tau_{13}$ & $\Delta \tau_{23}$ \\
  & $\left[\mathrm{a}\right]$ & $\left[\mathrm{a}\right]$ & $\left[\mathrm{a}\right]$ & $\left[\mathrm{a}\right]$ & $\left[\mathrm{a}\right]$ & $\left[\mathrm{a}\right]$ \\
\noalign{\smallskip}
\hline
\noalign{\smallskip}
1 & 49.20 & 52.32 & 3.12 & 2.53 & 2.07 & 0.74 \\
2 & 33.82 & 35.55 & 1.74 & 6.21 & 6.45 & 0.75 \\
3 & 29.27 & 31.28 & 2.00 & 0.11 & 0.11 & 0.03 \\
4 & 29.26 & 31.26 & 2.00 & 0.05 & 0.05 & 0.02 \\
5 & 29.45 & 33.99 & 4.54 & 0.16 & 0.59 & 0.57 \\
6 & 29.30 & 31.34 & 2.04 & 0.11 & 0.70 & 0.70 \\
7 & 44.57 & 46.49 & 1.92 & 2.95 & 2.94 & 0.14 \\
8 & 24.26 & 26.26 & 2.00 & 3.37 & 3.38 & 0.04 \\
\noalign{\smallskip}
\hline
\end{tabular}
\end{center}
\end{table}

Comparing the reconstructed time delays for the different models listed in Table~\ref{tab:Jori_time_delays}, we conclude that, as expected, the time delays are retrieved very accurately when using them as constraints. In addition, we find that adding a mass sheet is necessary to reconstruct most of the time delays correctly because time delays break the mass-sheet degeneracy. Only the time delays $\tau_{12}$ and $\tau_{13}$ cannot be correctly reconstructed within the range of their standard deviations in model~8. 
Leaving these time delays aside, we observe that the standard deviation of the reconstructed time delay due to the variation in the individual reconstructions by the genetic algorithm is reduced by about one order of magnitude when time-delay constraints from the remaining pair of multiple images for that source are included in the lens reconstruction, except for $\Delta \tau_{23}$ in model~6.
For another visualisation, three dimensional plots of the maps in Figure~\ref{fig:Jori_reconstruction} are available online\footnote{\url{http://research.edm.uhasselt.be/~jori/tdmodels/}}. Equation~\eqref{eq:Jori_time_delays} assumes that measurement uncertainties of the time delays are negligible, so that the results shown in Figure~\ref{fig:Jori_reconstruction} resemble those that are obtainable with FRBs. 

On the whole, this example shows that even a single time delay measurement improves the accuracy of the lens reconstruction. If the pair of images consists of a minimum and a saddle point lying on opposite sides of the lens centre, the reconstruction accuracy for the convergence in the entire region around the galaxy cluster core is of comparable quality as if time-delay constraints between three multiple images had been employed. 

\subsubsection{Using time delays to determine the Hubble constant}

Galaxies can be phenomenologically described by power-law volume mass density profiles, $\rho(R) \propto R^{-\gamma}$, which is supported by observations of galaxy dynamics, X-ray emissions, and strong lensing measurements. By assuming the power-law profile, the mass sheet degeneracy is broken. This may introduce a bias, e.g. in the determination of $H_0$ from Equation~\eqref{eq:mi_time_delay}, \cite{bib:Sonnenfeld, bib:Xu}. However, selecting galaxies that have an approximately isothermal density profile (i.e. with a power-law index $\gamma \approx 2$), simulations analysed in \cite{bib:Sonnenfeld, bib:Xu} showed that the bias can be reduced and $H_0$ determined with less than 5\% inaccuracy. Hence, assuming a power-law density profile as lens model, time-delay measurements between multiple images caused by a galaxy-scale lens can be used to probe cosmology by means of Equation~\eqref{eq:mi_time_delay}. 

For galaxy clusters, the deflecting mass density profile is much more complex, so that the bias caused by breaking the mass sheet degeneracy with a lens model can be larger than for galaxies (see \cite{bib:Meneghetti2} for the precision and accuracy of local lens properties obtained by the most common lens reconstruction approaches). In addition, the cosmological model need not enter as a simple scale factor of the lens potential as is the case for some galaxy-scale lens models. Therefore, galaxy clusters seem to be less suitable to determine $H_0$ compared to galaxy-scale lenses. Nevertheless, \cite{bib:Vega} and \cite{bib:Grillo} determined $H_0$ from the supernova Refsdal behind the galaxy cluster MACS1149 with 17\% and 6\% precision, respectively. In the same way, $H_0$ could be determined from multiply-imaged FRBs as well.

\section{Comparison to other time-varying sources}
\label{sec:comparison}

In this section, we compare the expected abundances of multiply-imaged quasars, supernovae, and GRBs behind galaxy clusters under their most favourable observation conditions in different bands. In addition, we compare the precision and accuracy of the time delays obtained for these time-varying sources with the precision and accuracy achievable by FRBs. For these comparisons, we rely on the FRB-model assumptions as stated in Section~\ref{sec:FRBs}, so that these estimates should be treated with caution given the limited knowledge of the FRB-properties we currently have. Yet, our estimates are conservative in the sense that we choose an FRB model with low detection rates, requiring SKA phase 2 for the observation.

\subsection{Quasars}
\label{sec:comparison_quasars}

Estimated abundances of multiply-imaged quasars with image splittings of at least 10~arcseconds are 8 in 8000~deg$^2$ when considering lensing by dark matter only for quasars up to $z_\mathrm{s}=5$ using the specifications of the SDSS photometric survey, \cite{bib:Hennawi}. Including the influence of the brightest cluster galaxy, the rate increases to 12 in 8000~deg$^2$. Assuming a data collection time on the order of 10 years, the estimated detection rate is $1.5 \times 10^{-4}$ multiply-imaged quasars per square degree per year. This result is of the same order as the $1.3 \times 10^{-4}$ multiply-imaged FRBs per square degree per year up to $z_\mathrm{s} = 5$.

The photometrical variability of quasars is on the order of a few days, \cite{bib:Bonvin}. Apart from the time scale of the intrinsic variability, the precision of a time-delay measurement also depends on the cadence of the observations as investigated by \cite{bib:Liao} for simulated time delays caused by galaxy-scale lenses. In \cite{bib:Liao}, cadences on the scale of 3~days are considered, so imprecisions of the time-delay measurements are on the order of 1--2~days at best. 
% while the time delays are several weeks to months (see Sections~\ref{sec:props_of_interest} and \ref{sec:microlensing}). 

\cite{bib:Liao} find that the best algorithms obtained time delays with sub-percent inaccuracy and 3\% imprecision. Yet, the rate of light curves from which time delays could be retrieved was only around 50\% of all simulated ones. For current observational data, as collected in \cite{bib:Oguri}, the precision is on the order of 90\% for galaxy-scale lenses. Table~\ref{tab:quasars} lists the observed multiply-imaged quasars in galaxy clusters together with the largest image separation and measured time delays. The table shows that there is a high variability of measurement uncertainties ranging from sub-percent level to over 10\% with absolute uncertainties on the order of the cadence mentioned above.

\begin{table}[t]
 \caption{Synopsis of observed multiply-imaged quasars with their largest image separation $\delta\theta$ and measured time delays $\tau$.}
\label{tab:quasars}
\begin{center}
\begin{tabular}{lllll}
\hline
\noalign{\smallskip}
  Name & $z_\mathrm{l}$ & $z_\mathrm{s}$ & $\delta\theta$ & $\tau$  \\
 (Ref.) & & &$\left[\text{''}\right]$ & $\left[\text{d}\right]$ \\
\noalign{\smallskip}
\hline
\noalign{\smallskip}
Q0957+561 & 0.36 & 1.413 & 6.0 & $(417.0 \pm 1.5)_{AB}$ \\
(1) & & & & \\
\noalign{\smallskip}
\hline
\noalign{\smallskip}
RX J0911 & 0.77 & 2.800 & 3.1 & $(143.0 \pm 6.0)_{A1B}$ \\
(1) & & & & $(149.0 \pm 8.0)_{A2B}$ \\
 & & & & $(154.0 \pm 16.0)_{A3B}$ \\
\noalign{\smallskip}
\hline
\noalign{\smallskip}
SDSS J1004 & 0.68 &1.734 & 14.6 & $(40.6 \pm 1.8)_{BA}$ \\
 (2) & & & & $(821.6 \pm 2.1)_{CA}$ \\
\noalign{\smallskip}
\hline
\noalign{\smallskip}
SDSS J1029 & 0.58 & 2.197 & 22.6 & $(744 \pm10)_{AB}$ \\
(3) & & & & \\
\noalign{\smallskip}
\hline
\noalign{\smallskip}
SDSS J2222 &  0.49 & 2.82 & 15.1 & $(47.7 \pm 6.0)_{AB}$ \\
(4) & & & & $(722 \pm 24)_{CA}$ \\
\noalign{\smallskip}
\hline
\end{tabular}
\end{center}
\tablebib{
(1)~\citet{bib:Oguri}; (2)~\citet{bib:Fohlmeister1}; (3)~\citet{bib:Fohlmeister2}; (4)~\citet{bib:Dahle2}; 
}
\end{table}

Hence, assuming an uncertainty of 1.5~days for measured time delays of 100~days of quasars, the relative uncertainty amounts to 0.015, which is several orders of magnitudes larger than for FRBs (see Section~\ref{sec:props_of_interest}) and not negligible compared to the uncertainties of the other observables.

%\begin{itemize}
%\item \textbf{Microlensing vs. Macrolensing disentanglement in spectrum} \\ \url{http://cdsads.u-strasbg.fr/abs/2012A%26A...544A..62S}
%\item \textbf{Follow-up on the previous paper} \\ \url{http://cdsads.u-strasbg.fr/abs/2017A%26A...607A..32B}
%\end{itemize}

%%%%
\subsection{Supernovae}
\label{sec:comparison_supernovae}

The abundance of multiply-imaged supernovae observable behind massive galaxy clusters is estimated in \cite{bib:Gunnarsson}. They consider an NFW-profile of virial mass $2~\times~10^{15}~M_\odot$ located at $z_\mathrm{l}=0.2$ and sources in the redshift range $z_\mathrm{s} \in \left[1.5, 5 \right]$. These configurations cover minimum image splittings in the range of 6--21~arcseconds depending on $z_\mathrm{s}$. For observations in the $J$-band, they expect 7--10 multiply-imaged supernovae Type Ia per cluster per year and 21--24 multiply-imaged supernovae of Type II per cluster per year. These numbers are rough estimates because the supernova rates, especially at high redshifts, are not known, analogous to the FRB rates (see Section~\ref{sec:FRBs}). Improved searching criteria, as discussed in \cite{bib:Goldstein} may better constrain the abundance estimates in the future. 

As further detailed in Appendix~\ref{app:cluster_abundance}, the total number of about 30 supernovae per cluster per year translates to an estimated detection rate of $2.34 \times 10^{-5}$ multiply-imaged supernovae per square degree per year.
For comparison, we determine the detection rate of multiply-imaged FRBs with source redshifts $z_\mathrm{s} \in \left[ 1.5, 5 \right]$ that are caused by clusters of $2 \times 10^{15} M_\odot$ at lens redshift $z_\mathrm{d}$, which amounts to $5.51 \times 10^{-9}$ multiply-imaged FRBs per square degree per year. Hence, we expect less FRBs than supernovae given the limited range of redshifts of \cite{bib:Gunnarsson}. 

The distribution of observable, FRB-emitting galaxies peaks around $z=1$ and decreases quickly for $z>1.5$. Thus, a much more significant comparison would extend the calculations for the supernovae to smaller redshifts and a broader range of cluster masses. Yet, to our knowledge, these calculations are still in preparation for large surveys and not available yet. Furthermore, the numbers given in \cite{bib:Gunnarsson} do not take into account that supernovae are only visible for about 100~days. 
From the estimated numbers of supernovae and quasars behind galaxy-scale lenses, as discussed in \cite{bib:Oguri2}, we find, for example, that the predicted detection rates of multiply-imaged quasars per year per square degree are more than one order of magnitude higher than those for supernovae for the Large Synoptic Survey Telescope (LSST). Extrapolating from these results to the galaxy-cluster lensing regime, we can assume that both rates will be scaled-down by the same lower probability of the sources lying behind a galaxy cluster. Hence, we conclude that the detectable rate of multiply-imaged supernovae for a broader range of lens redshifts and masses could be lower than the rate of detectable multiply-imaged quasars and FRBs.

The sole, serendipitously found, multiply-imaged supernova with image splittings of 2~arcseconds and measured time delays, \cite{bib:Kelly,bib:Rodney}, indicates that uncertainties in time-delay measurements are on the order of days amounting to more than 28\% of the measured time delays\footnote{The time delay between the multiple images of the supernova on galaxy-cluster scale is not yet measured but only estimated to 345 days in \cite{bib:Grillo}.}. Among the effects causing this high uncertainty, there can be microlensing as discussed in \cite{bib:Goldstein2}, which is a disadvantage of using supernovae compared to FRBs. We also expect the measurements of time delays between multiple images of a supernova to be less precise than those between multiple images of a quasar due to the limited life time of supernovae. 

\begin{figure*}[t]
\centering
  \includegraphics[width=0.32\textwidth]{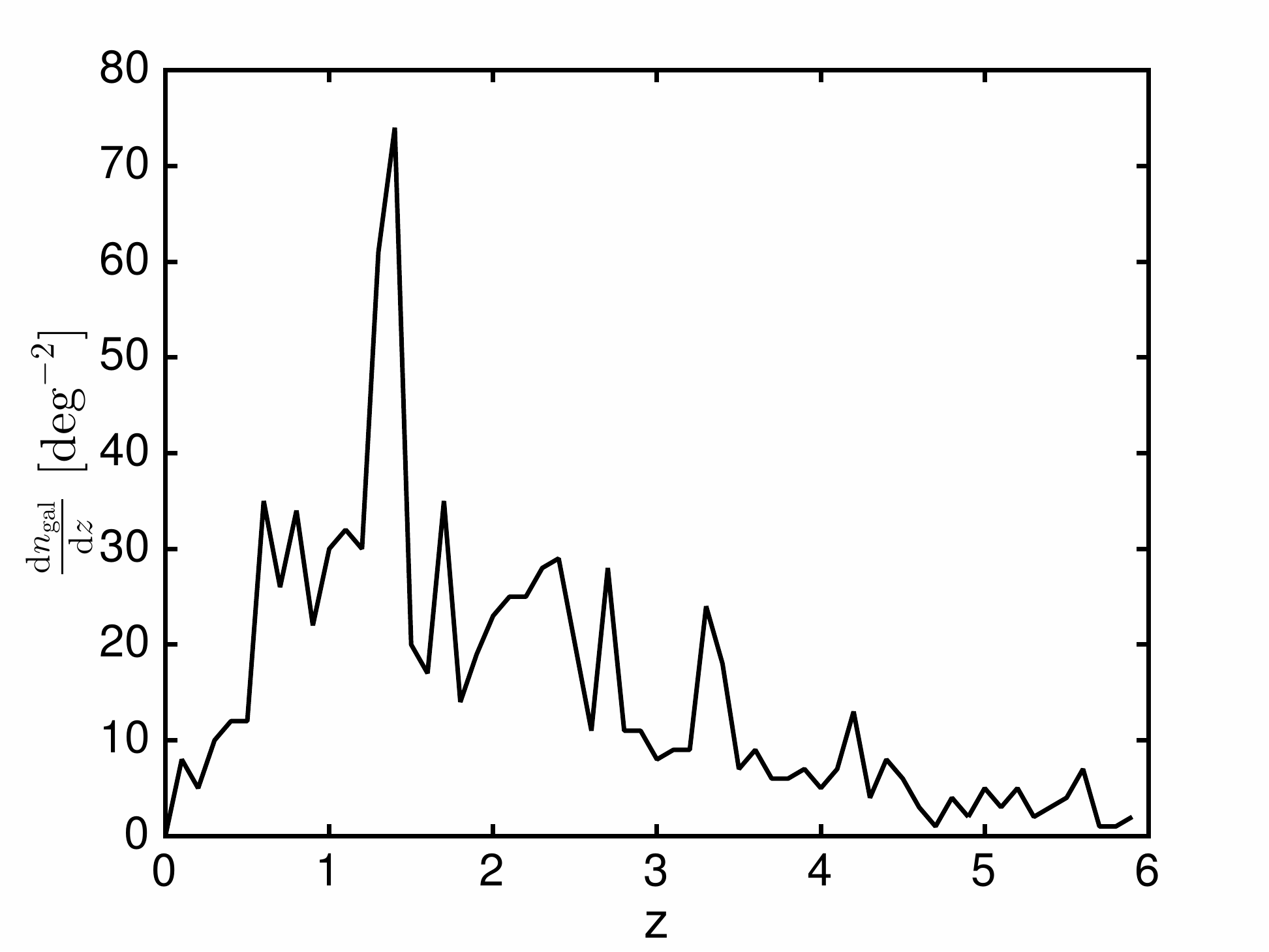} \hspace{0.01\textwidth}
  \includegraphics[width=0.32\textwidth]{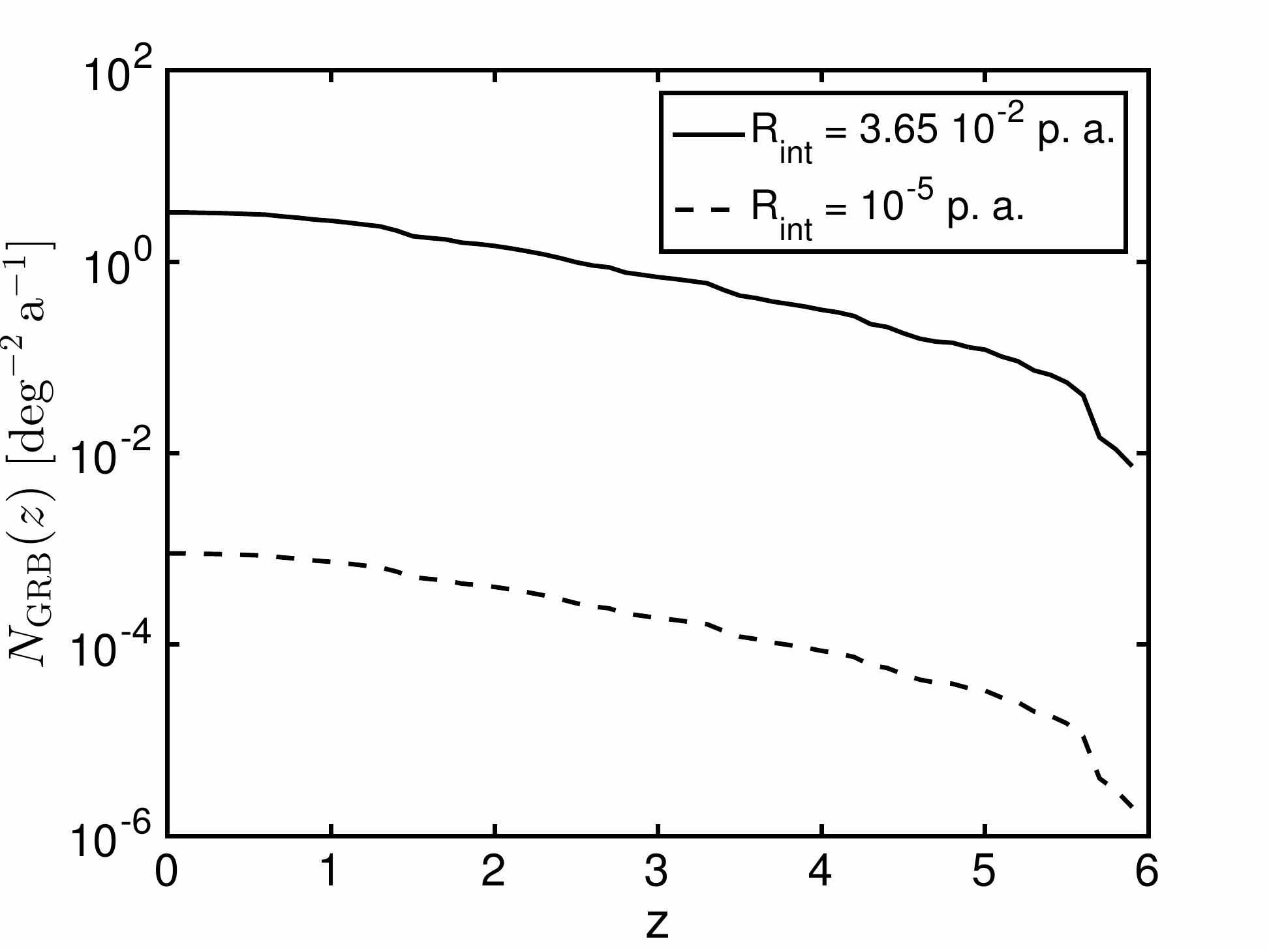} \hspace{0.01\textwidth}
  \includegraphics[width=0.32\textwidth]{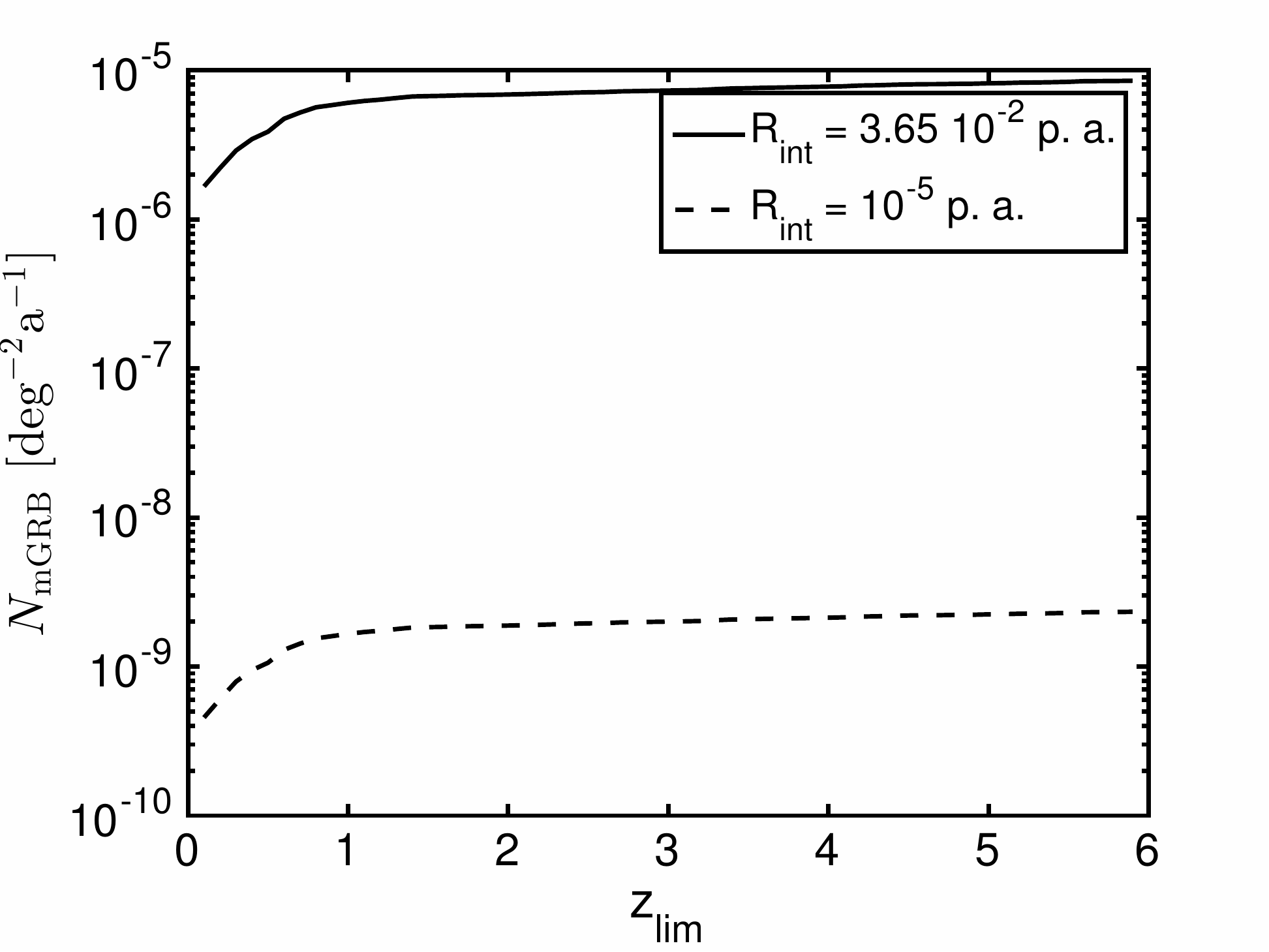}
   \caption{Left: distribution of observable GRB starburst host galaxies with total flux above 10$\mu$Jy at 1.4~GHz per square degree, per redshift, $\mathrm{d} n_\mathrm{gal}(z)~/~\mathrm{d} z$, in redshift bins of 0.1 width taken from the SKA Simulated Skies. Centre: integrated number of observable GRBs per square degree per year from $z$ to $z_\mathrm{lim}=6$ assuming an internal emission rate of $R_\mathrm{int}=10^{-4}$ per galaxy per day (solid line) or $R_\mathrm{int}=10^{-5}$ per galaxy per year (dashed line). Right: integrated number of observable multiply-imaged GRBs per square degree per year out to $z_\mathrm{lim}$ from 0.1 to 6.0 according to Equation~\eqref{eq:multiply_imaged_FRBs}.}
\label{fig:n_GRB}
\end{figure*}

%%%%
\subsection{GRBs}
\label{sec:comparison_grbs}

GRBs are divided into two classes, short-duration GRBs lasting less than 2~seconds and long-duration GRBs lasting more than 2~seconds. Only 30\% of all GRBs belong to the first group and the observed cases have a mean redshift around $z=0.5$, while the mean redshift of all observed long-duration GRBs is around $z = 2.0$, \cite{bib:Berger}. Therefore, we focus on the latter class and, unless stated otherwise, GRB refers to a long-duration GRB.  

\cite{bib:Chary} find that starburst galaxies are likely to host GRBs. In order to estimate the number of multiply-imaged GRBs by galaxy-cluster lenses which also have observable hosts in the radio band, we extract all starburst galaxies from the central square degree simulated in \cite{bib:Wilman} with the same specifications as in Section~\ref{sec:FRBs}. Their distribution over the redshift range $z \in \left[0, 6 \right]$ is shown in Figure~\ref{fig:n_GRB} (left). Processing them analogously to the star-forming galaxies (see Sections~\ref{sec:FRBs} and \ref{sec:lensed_FRBs}), we obtain the distribution of GRBs (see Equation~\eqref{eq:n_FRB}) and multiply-imaged GRBs (see Equation~\eqref{eq:multiply_imaged_FRBs}), as shown in Figure~\ref{fig:n_GRB} (centre) and (right), respectively. As internal emission rate, we employ $R_\mathrm{int}=10^{-4}$ per galaxy per day\footnote{Similar to the FRB internal emission rate, we assume that beaming effects are incorporated in this rate.} as for the FRBs and $R_\mathrm{int}=10^{-5}$ per galaxy per year, as is the estimated rate for our Galaxy. A comparison of both values shows that, even for an internal emission rate as high as for the FRBs, the number of observable multiply-imaged GRBs with their hosts is one order of magnitude smaller than the one for FRBs. For the more realistic $R_\mathrm{int} = 10^{-5}$ per galaxy per year, the expected number of multiply-imaged GRBs is even five orders of magnitude smaller. 

Hence, compared to all other time-varying sources, the abundance of multiply-imaged GRBs is the smallest. But observing a multiply-imaged GRB, the probability that we will also observe a core-collapse supernova in its vicinity is increased, \cite{bib:Hjorth}.

As the spatial resolution of gamma-ray telescopes is limited to approximately 0.1~degrees by their detection mechanism, the localisation precision is too low to identify multiply-imaged GRBs as such. Instead, they are detected by light-curve matching as further detailed in \cite{bib:Barnacka1, bib:Davidson}. Being able to observe an afterglow in X-ray or optical and radio bands, the localisation precision is increased to 10~arcseconds and subarcseconds, respectively, \cite{bib:Prochaska}.
Due to these difficult observation and detection conditions, only two multiply-imaged GRBs caused by galaxy-scale lenses have been detected so far, \cite{bib:Barnacka1, bib:Barnacka2}. 

Given uncertainties of 0.5~days for these two observed cases, the expected imprecision of time delays in the galaxy cluster regime with time delays on the order of 100~days is $0.5\%$. Hence, the precision is higher than those of quasars (see Table~\ref{tab:summary}). 
%The abundance of multiply-imaged GRBs has been investigated in \cite{bib:Davidson}. 
%
%While a systematic search for GRBs multiply-imaged by galaxy-scale lenses was performed without yielding candidates, \cite{bib:Davidson}, two multiply-imaged GRBs were detected later, \cite{bib:Barnacka, bib:}. These are the only ones so far with time delay precisions of a few hours, with a minimum of 2.4~hours, \cite{bib:Barnacka}. Spatial resolutions of X-ray observations are on the order of 0.1~degrees, so that the multiply-imaged GRBs can only be resolved in the time domain and their hosts by follow-up higher-resolution observations in other bands. Furthermore, the sensitivity of current instruments limits observations of GRBs in the field out to redshifts of 1, so that observations of magnified, multiply-imaged GRBs at high redshifts does not seem possible. 

%%%%%%%%%%%%%%%%%
\section{Redshifts for FRBs based on gravitational lens models}
\label{sec:redshifts}

If the FRB and its host are positioned at such a high redshift that the peak flux of the FRB is close to the sensitivity limit of the detector and the host galaxy is below it, the latter cannot be observed anymore. Consequently, the redshift of the FRB cannot be determined from the spectrum of the host. 
%Knowing the volume density of electrons, $n_\mathrm{e}$, we can employ Equation~\eqref{eq:DM} to determine the redshift of the FRB for a given cosmology ($H_0, E(z)$). Yet, this ansatz will lead to imprecise redshifts if the FRB is located at $z_\mathrm{s} \gg z_\mathrm{l}$. 
For this case, a gravitational lens model can be employed to determine the (comoving) cosmic distance to the FRB or its redshift:

Assume that the time delay between two adjacent multiple images in a fold configuration, $\tau$, the distance between these two images $\delta x$, and the redshift of the lens were measured. Then, we rewrite Equation~\eqref{eq:tau_approx}
\begin{equation}
D_\mathrm{r} \equiv \dfrac{D_\mathrm{s}}{D_\mathrm{ls}} = \dfrac{12 c \tau}{(1+z_\mathrm{l})D_\mathrm{l} (\delta x)^3 \phi_{222}^{(0)}} \;,
\label{eq:distance_ratio}
\end{equation}
in order to estimate the uncertainty up to which we can determine the distance to the emitted FRB. 
$\phi_{222}^{(0)}$, is obtained from a global, model-based reconstruction of the gravitational lensing potential or its second-order derivatives, e.g. by \cite{bib:Grillo15, bib:Jullo, bib:Liesenborgs, bib:Merten}. 
A Gaussian propagation of the uncertainties of the observables in Equation~\eqref{eq:distance_ratio} leads to a relative uncertainty of $D_\mathrm{r}$ given by
\begin{align}
\dfrac{\Delta D_\mathrm{r}}{D_\mathrm{r}} = \sqrt{\left( \dfrac{\Delta \tau}{\tau} \right)^2 + \left( \dfrac{\Delta \left( (1+z_\mathrm{l}) D_\mathrm{l}\right)}{(1+z_\mathrm{l}) D_\mathrm{l}} \right)^2 + \left( \dfrac{3\Delta \delta x}{\delta x} \right)^2 + \left( \dfrac{\Delta \phi_{222}^{(0)}}{\phi_{222}^{(0)}} \right)^2}  \;.
\end{align}
In addition, the derivation of Equation~(2) of \cite{bib:Wagner1} involves a Taylor expansion around $x_0$ which adds a systematic bias of less than 2\% to the measurement uncertainty (see Figure~\ref{fig:tau_approx}). 

Typical values for the observables of two images of an FRB in a galaxy cluster are $\tau~=~100~\mbox{d}$, $\delta x~=~5~''$, $z_\mathrm{l}~=~0.5$ with uncertainties $\Delta \tau~=~1~\mbox{ms}$, $\Delta \delta x~=~0.01~''$, $\Delta z_\mathrm{l}~=~0.01$ which yield
\begin{align}
\dfrac{\Delta D_\mathrm{r}}{D_\mathrm{r}} = \sqrt{\left( 10^{-10} \right)^2 + \left( 0.0067 \right)^2 + \left( 0.006 \right)^2 + \left( 0.1 \right)^2} \;.
\end{align}
Hence, the dominant source of uncertainty is the lens model, if the uncertainty in $\phi_{222}^{(0)}$ is on the order of 10\%.
We can neglect the uncertainty in $\tau$, $z_\mathrm{l}$, and $D_\mathrm{l}$ and solve $D_\mathrm{r}$ for the comoving distance to the FRB
\begin{align}
C_\mathrm{s}~=~\int \limits_0^{z_\mathrm{s}} \dfrac{\mathrm{d}z}{E(z)} = \dfrac{C_\mathrm{l}}{1-D_\mathrm{r}} \;,
\end{align}
in which $E(z)$ is the expansion function of the universe\footnote{In a flat Friedmann model, it is usually approximated by $E(z) \approx \sqrt{ \Omega_\mathrm{m} (1+z)^3 + \Omega_\Lambda}$ for $z \in \left[0, 6\right]$, with the parameter $\Omega_\mathrm{m}$ for the matter density and $\Omega_\Lambda$ for the cosmological constant.}, $C_\mathrm{l}$ the comoving distance to the lens, and $D_\mathrm{r}$ is given by the expression on the right-hand side of Equation~\eqref{eq:distance_ratio}. Thus, $\Delta C_\mathrm{s} / C_\mathrm{s} = \Delta D_\mathrm{r} / (1-D_\mathrm{r})$, so that the uncertainty in $C_\mathrm{s}$ is approximately 10\%, mainly caused by the uncertainty of the lens model for typical redshifts of sources and lenses.

Alternatively, the approach developed in \cite{bib:Kneib, bib:Jullo} determines the most probable parametric lens model in a Markov-Chain-Monte-Carlo optimisation and simultaneously constrains unknown redshifts of multiple images. The user can set a uniform or a Gaussian prior on the redshift and set boundary values or a mean and a standard deviation, respectively. In this way, the redshift of the FRB can be determined to a predefined precision.

%%%%%%%%%%%%%%%%%
\section{Conclusion}
\label{sec:conclusion}

\begin{table*}[ht]
 \caption{Results of the comparison between different time-varying, multiply-imaged sources behind galaxy clusters, the number of observed cases (2nd column), the estimated detection rate for sources up to $z_\mathrm{lim} = 5$ as determined in Section~\ref{sec:comparison} (3rd column), the relative uncertainty of the time-delay measurements for cluster-scale lenses (4th column), the duration of the transient event (5th column), the possibility of a repeating time variation from the same source (6th column), and the possibility that microlensing affects the time delay (7th column).}
\label{tab:summary}
\begin{center}
\begin{tabular}{lllllll}
\hline
\noalign{\smallskip}
Probe & Observed & Detection rate & $\tau$ uncertainty & Duration & Repetition & Microlensing \\
\noalign{\smallskip}
\hline
\noalign{\smallskip}
FRB & 0 & $10^{-4} \, \mbox{deg}^{-2} \mbox{a}^{-1}$ & $10^{-8} \%$ & $\approx$ 1~ms & possible & no \\
\noalign{\smallskip}
\hline
\noalign{\smallskip}
GRB & 0 & $10^{-9} \, \mbox{deg}^{-2} \mbox{a}^{-1}$ & $0.5 \%$ & $>$ 2 s & improbable & possible \\
\noalign{\smallskip}
\hline
\noalign{\smallskip}
Quasar & 5 & $10^{-4} \, \mbox{deg}^{-2} \mbox{a}^{-1}$ & $1 - 5 \%$ & -- & yes & yes \\ 
\noalign{\smallskip}
\hline
\noalign{\smallskip}
Supernova & 1 & $10^{-5} \, \mbox{deg}^{-2} \mbox{a}^{-1}$ & $30 \%$ & $\approx$ 100 d & no & yes \\
\noalign{\smallskip}
\hline
\end{tabular}
\end{center}
\end{table*}

\noindent
Using the spectral properties of the repeating FRB121102 as a model, we estimated the rate of multiply-imaged FRBs behind galaxy-cluster scale gravitational lenses detectable by an SKA-phase-2-like telescope. We found the detection rate to be on the order of $10^{-4}$ per square degree per year. Comparing this rate with the observation rates of other multiply-imaged time-varying sources, as summarised in Table~\ref{tab:summary}, we find that FRBs could be as frequent as multiply-imaged quasars and be more frequent than multiply-imaged supernovae. The least likely time-varying source that will be found behind cluster-scale lenses are GRBs. Table~\ref{tab:summary} also lists the number of already observed multiple images, the (estimated) uncertainty of the time-delay measurement, the time scale of their duration and whether they even may repeat and may be subject to microlensing. The underlying assumptions to obtain these estimates are detailed in Sections~\ref{sec:FRBs} and \ref{sec:comparison}. From these findings, we conclude that FRBs could have a lot of advantages compared to the other time-varying sources,
\begin{itemize}
\item because they could be as frequent as quasars but they are not subject to microlensing due to their short duration time and they do not outshine the host galaxy for more than milliseconds, so that the host can be resolved more easily,
\item because they might also be calibrated to be standard candles but with a higher detection rate than supernovae and a higher precision in the time-delay measurement,
\item because the time delay could be measured several times for repeating FRBs.
\end{itemize}
These advantages are based on the assumptions stated in Section~\ref{sec:FRBs}, which, due to the limited amount of observations, may not hold. Yet, the main advantage that they are transients of short duration and subsequently allow to observe the properties of their hosts remains true in any case and makes them very valuable probes of the mass distribution in a gravitational lens.

FRBs also have disadvantages because, due to their short duration time, they require a different detection technique. In a recent wide-field observation, the Australian Square Kilometre Array Pathfinder detected 19 most probably non-repeating FRBs, \cite{bib:Shannon}. Almost doubling the amount of known FRBs so far, this survey demonstrated that the usage of a fly's-eye configuration of 5-12 antennas pointing in different directions is very efficient to locate FRBs. In this configuration, the location precision is 10 arcminutes times 10 arcminutes. An ideal configuration to detect multiply-imaged FRBs is a continuum all-sky SKA-phase-2-like survey with angular resolution of about 10 arcseconds detects the first two multiple images. Subsequently, a high-resolution follow-up observation is necessary to detect and locate further image(s). Additional optical and infra-red observations may also be of advantage to investigate the host galaxy in further detail. To match these requirements, galaxy-cluster scale gravitational lenses are more suitable than galaxy-scale lenses because cluster lenses cause angular separations between multiple images on the order of 10 arcseconds and time delays on the order of 100 days, which leaves enough time to prepare the high-resolution follow-up observation.

Given that the observational requirements could be fulfilled in the future, we could use multiply-imaged FRBs to increase our knowledge about local and global properties of galaxy clusters and increase the precision of the results by orders of magnitude. In particular, we showed that it is possible to
\begin{itemize}
\item determine differences in the gravitational lensing potential between all positions of multiple images to a greater precision than currently possible,
\item globally alleviate the mass sheet degeneracy in the reconstructions of the mass density profiles based on lens models,
\item determine cosmological parameters to a higher precision than currently possible,
\item determine the redshift or the distance of a very faint, FRB-emitting host galaxy,
\item determine changes in the cluster mass of more than $10^{5} M_\odot$ from repeating FRBs.
%\item determine differences in the electron column density between all positions of multiple images.
\end{itemize}

As found for the simulated galaxy-cluster scale lens in Section~\ref{sec:mb_information}, the overall reconstruction accuracy for the surface mass density distribution is increased when taking time-delay constraints into account, but this may come at the cost of introducing a bias towards shallower slopes in the cluster core (see Figure~\ref{fig:Jori_radial_profiles}). With one time-delay constraint coming from a pair of multiple images from opposite sides of the lens centre, approximately the same reconstruction accuracy in the surface mass density distribution map can be achieved as for the three time-delay constraints from three multiple images. 

%Combining the first and the last item, we obtain information on the X-ray gas density and the lensing potential at the same positions. Hopefully, this may contribute to calibrate both probes against each other to consistently reconstruct the gravitational potential of galaxy clusters from multi-band observations. This is useful, for instance, for mergers along the line of sight like CL0024, for which unresolved inconsistencies between the mass density profile reconstructed from lensing and X-ray measurements exist, \cite{bib:Zhang,bib:Umetsu}.

%Current estimates on the number of multiply-imaged FRBs are too few to reconstruct the gas density for the entire galaxy cluster, however, if long-term surveys with high sensitivity detect on the order of a hundred multiply-imaged FRBs behind a galaxy cluster, gas density reconstructions in the manner of \cite{bib:Konrad} become feasible.
 
Although the determination of $H_0$ from supernova Refsdal, \cite{bib:Vega, bib:Grillo}, is an important achievement, we are in favour of putting more emphasis on the analysis of the cluster lensing potential by means of the high-precision time delays. Understanding the degeneracies of the lensing formalism and the lens models, e.g. as pointed out in \cite{bib:Schneider, bib:Liesenborgs2, bib:Unruh, bib:Wertz}, \cite{bib:Wagner5}, may also contribute to resolve the 3-$\sigma$ tension in the determination of $H_0$ as found by \cite{bib:Planck, bib:Riess16,bib:Riess18}.

%%%%%%%%%%%%%%%%%
\begin{acknowledgements}
We would like to thank Claudio Grillo, Eric Jullo, Julian Merten, Adi Zitrin, and the Galaxy Cluster Group at the Institute for Theoretical Astrophysics for helpful discussions and comments. JW gratefully acknowledges the support by the Deutsche Forschungsgemeinschaft (DFG) WA3547/1-1 and WA3547/1-3. JL acknowledges the use of the computational resources and services provided by the VSC (Flemish Supercomputer Center), funded by the Research Foundation - Flanders (FWO) and the Flemish Government – department EWI. DE acknowledges support from the Israel - U.S. Binational Science Foundation, the Israel Science Foundation including an ISF-UGC grant, and the Joan and Robert Arnow Chair of Theoretical Astrophysics.
\end{acknowledgements}

\bibliographystyle{aa}
\bibliography{aa}

%%%%%%%%%%%%%%%%%%%%%%%%%%%%%%%%%%%%%%%%%%%%%%%%%%%%%%%%%%%%%%%%%%%%%%%%%%%%%
\appendix

\section{Derivations of Equations~\eqref{eq:tau_fold} and \eqref{eq:tau_opp}}
\label{app:NFW_config}

The general formula for the time delay between two points in the image plane $\vec{x}_i$ and $\vec{x}_j$ coming from the same source position $\vec{y}$, as given e.g.\ in \cite{bib:SEF} reads
\begin{equation}
\tau = \dfrac{\xi_0^2 D_\mathrm{s}}{c D_\mathrm{l} D_\mathrm{ls}} \left( 1+z_\mathrm{l} \right) \Delta \phi \left(\vec{y},\vec{x}_i,\vec{x}_j \right) \;,
\label{eq:tau_general}
\end{equation}
with
\begin{equation}
\Delta \phi \left(\vec{y},\vec{x}_i,\vec{x}_j \right) =\dfrac12 \left( \left(\vec{x}_i - \vec{y} \right)^2 - \left( \vec{x}_j - \vec{y} \right)^2 \right) - \psi(\vec{x}_i) + \psi(\vec{x}_j) \;. 
\label{eq:potential_difference}
\end{equation}
$\xi_0$ is the scaling length to convert observed angular positions $\vec{\theta}$ into $\vec{x}$, i.e. $\vec{x}~=~\vec{\theta} D_\mathrm{l}/\xi_0$ and $\psi(\vec{x})$ is the deflection potential given in Equation~\eqref{eq:NFW_potential}, as introduced in \cite{bib:Meneghetti}. For the NFW-profile, we observe the three images at positions $\vec{x}_i$, $i~=~1,2,3$, as defined in Section~\ref{sec:props_of_interest}. Employing polar coordinates and using the axisymmetry of the potential, we locate the three images of a source at $\vec{y} = (r_y, 0)$ without loss of generality at 
\begin{equation}
\vec{x}_1 = \left(r_1, \pi \right) \;, \quad \vec{x}_2 = \left(r_2, \pi \right) \;, \quad \vec{x}_3 = \left(r_3, 0 \right) \;, \quad r_1 < r_2 < r_3 \;.
\label{eq:coords}
\end{equation}
Writing Equation~\eqref{eq:potential_difference} in polar coordinates, taking into account that $\psi(\vec{x})~=~\psi(r)$ for an axisymmetric lensing profile, yields
\begin{align}
\Delta \phi \left(\vec{y},\vec{x}_i,\vec{x}_j \right) =&\dfrac12 \left( r_i^2 + r_j^2 - 2 r_y \left(r_i \cos (\eta_{iy}) - r_j \cos(\eta_{jy}) \right) \right) \nonumber \\ 
& - \psi(r_i) + \psi(r_j) \;,
\label{eq:pot_diff_polar}
\end{align}
in which $\eta_{iy}$ denotes the angle between $\vec{x}_i$ with $\vec{y}$. 

Using Equations~\eqref{eq:coords} and \eqref{eq:pot_diff_polar}, we can now derive Equation~\eqref{eq:tau_fold} as follows
\begin{align}
\Delta \phi \left(\vec{y},\vec{x}_1,\vec{x}_2 \right) =& \dfrac12 \left( r_1^2 - r_2^2 + 2 r_y \left(r_1 - r_2 \right) \right) - \psi(r_1) + \psi(r_2)  \\
=& \dfrac12 \left(r_1 - r_2 \right) \left(r_1 + r_2 + 2 r_y \right) - \psi(r_1) + \psi(r_2) \\
\approx& \left(r_1 - r_2 \right) \left(r_\mathrm{r} +  y_\mathrm{r} \right) - \psi(r_1) + \psi(r_2) \;.
\end{align}
In the last step, we used that $r_1 \approx r_2 \approx r_\mathrm{r}$, i.e.\ the radii of the two images close to the radial critical curve can be approximated by the radius of the radial critical curve and the radius of the source position is given by the radius of the caustic belonging to the radial critical curve, $r_y \approx y_\mathrm{r}$. Inserting the result into Equation~\eqref{eq:tau_general}, we arrive at Equation~\eqref{eq:tau_fold}.

A similar calculation yields Equation~\eqref{eq:tau_opp}:
\begin{align}
\Delta \phi \left(\vec{y},\vec{x}_2,\vec{x}_3 \right) =& \dfrac12 \left( r_2^2 - r_3^2 + 2 r_y \left(r_2 + r_3 \right) \right) - \psi(r_2) + \psi(r_3)  \\
=& \dfrac12 \left(r_2 + r_3 \right) \left(r_2 - r_3 + 2 r_y \right) - \psi(r_2) + \psi(r_3) \\
\approx& \dfrac12 \left(r_\mathrm{r} + r_\mathrm{t} \right) \left(r_2 - r_3 \right) - \psi(r_2) + \psi(r_3) \;.
\end{align}
In the last step, we used that $r_2 \approx r_\mathrm{r}$ and $r_3 \approx r_\mathrm{t}$, i.e.\ we replaced the radii of the images by the radii of the closest critical curve. Furthermore, $2 r_y < r_2 - r_3$, so that the last term in the second bracket can be omitted. Inserting the result into Equation~\eqref{eq:tau_general}, Equation~\eqref{eq:tau_opp} is obtained.

\section{Estimated detection rate for supernovae}
\label{app:cluster_abundance}

We assume a given rate of observable multiply-imaged supernovae per cluster per year for a given source redshift interval, $R_\mathrm{SN}$, as detailed in Section~\ref{sec:comparison_supernovae}. Summing the supernovae of all types, $R_\mathrm{SN} = 30$ per cluster per year. Then, the detection rate of multiply-imaged supernovae caused by all clusters with masses between $1.8$ to $2.0 \times 10^{15} M_\odot$ at redshift $z_\mathrm{l}$ is given by
\begin{equation}
N_\mathrm{mSN} = \int \limits_{1.8 \times 10^{15} M_\odot}^{2.0 \times 10^{15} M_\odot} R_\mathrm{SN}n(M,z_\mathrm{l}) \mathrm{d} M \;,
\end{equation}
in which $n(M,z) \mathrm{d} M$ is the number density of lenses in the mass interval d$M$ as introduced in Section~\ref{sec:lensed_FRBs}.
Using the software described in \cite{bib:Murray}, we obtain for the number density of lenses in the given mass range
\begin{align}
n(M, z_\mathrm{l}) \mathrm{d}M &= \dfrac{2.52 \times 10^{-23} }{\mathrm{Mpc}^3 M_\odot} 2 \times 10^{14} M_\odot \\
 &= \dfrac{7.81 \times 10^{-7}}{\mathrm{deg}^2} \;.
\end{align}
To arrive at this result, we assumed that the cluster extends about 1~Mpc along the line of sight. Multiplying it by $R_\mathrm{SN}$, the resulting detection rate of $2.35 \times 10^{-5}$ multiply-imaged supernovae per square degree per year is obtained.

\end{document}